\documentclass{aa}  

\usepackage{graphicx}
\usepackage{txfonts}
\usepackage[colorlinks=true,
    linkcolor=blue,
    citecolor=blue,
    filecolor=magenta,      
    urlcolor=cyan]{hyperref}
\usepackage{graphicx}	
\usepackage{amsmath}	
\usepackage{amssymb}	
\usepackage{comment}
\usepackage{bigstrut}
\usepackage{multirow}
\usepackage{xcolor}
\usepackage{pifont}
\newcommand{\labsim}[1]{\fontfamily{qcr}\selectfont \textbf{#1}}

\usepackage{geometry}
\begin{document}

   \title{Winds Against Alignment: AGN Feedback and the Spin Evolution of Massive Black Hole Binaries}
   \titlerunning{Winds Against Alignment}
   \authorrunning{F. Bollati et al.}

   \author{F. Bollati
          \inst{1} \fnmsep\thanks{fbollati@aip.de}
          \and
           M. Volonteri\inst{6}
          \and
          A. Lupi\inst{2,3,4}
          \and 
          M. Dotti\inst{5,2,4}
          \and 
          F. Haardt\inst{2,3,5}
          }

   \institute{
              Leibniz Institute for Astrophysics Potsdam (AIP), An der Sternwarte 16, 14482 Potsdam, Germany
         \and
              DiSAT, Universit\`a degli Studi dell'Insubria, via Valleggio 11, I-22100 Como, Italy       
         \and
              INFN, Sezione di Milano-Bicocca, Piazza della Scienza 3, I-20126 Milano, Italy
         \and    
              Dipartimento di Fisica G. Occhialini, Universit\`a di Milano-Bicocca, Piazza della Scienza 3, I-20126 Milano, Italy
         \and 
              INAF, Osservatorio Astronomico di Brera, Via E. Bianchi 46, I-23807 Merate, Italy
         \and Institut d’Astrophysique de Paris, Sorbonne Universit´e, CNRS, UMR 7095, 98 bis bd Arago, 75014 Paris, France  
             }
        
    \date{Received XXX; accepted YYY}

\abstract{
The interaction between massive black hole (MBH) binaries and circumbinary discs (CBDs) is expected to affect the spin orientations of merging binaries — a key observable for the future LISA mission. While previous work suggested that gas accretion can align BH spins with the orbital angular momentum via the Bardeen–Petterson effect, the impact of AGN feedback on this process has remained largely unexplored. We present a suite of hydrodynamical simulations of an equal-mass MBH binary embedded in a CBD, using the \textsc{gizmo} code with a sub-grid model that self-consistently evolves BH mass, spin, and anisotropic AGN feedback. We explore different BH spin magnitudes and orientations to identify the configurations resulting in the strongest feedback effect on the disc structure and binary evolution. We find that AGN winds substantially modify the CBD and minidisc structure, enlarging the central cavity and, after 20–25 binary orbits, destroying the minidiscs in all simulations that include feedback. 
Feedback-driven cavity excavation suppresses the gravitational torques responsible for orbital evolution, causing the binary to stall. AGN feedback has a dramatic effect on MBH accretion, reducing the Eddington ratio by one to two orders of magnitude before ultimately suppressing accretion entirely. This reduction in accretion strongly inhibits Bardeen–Petterson spin alignment, delaying it well beyond the timescale found in feedback-free simulations. Extending one simulation to $\sim$ 70 binary orbits reveals a feedback-regulated duty cycle of alternating active and quiescent phases, with the binary spending most of its time in low-density, quiescent conditions that further suppress spin alignment. These results indicate that AGN feedback provides an important, previously overlooked channel for preserving spin misalignment in MBH binaries prior to coalescence, with direct implications for the interpretation of LISA gravitational-wave observations.
}

\keywords{
Methods: numerical -- quasars: supermassive black holes
}
\maketitle






\section{Introduction}

Massive black holes are known to reside in the nuclei of nearly all galaxies hosting a substantial stellar bulge \citep{Magorrian98,Ferrarese00,Haring04,Gultekin09,Koremendy13}. Within the hierarchical framework of galaxy formation, in which galaxy mergers are ubiquitous \citep{White91,Husko22m}, this naturally implies that binaries of MBHs are expected to form frequently throughout cosmic history. Understanding the formation, cosmic evolution, and eventual coalescence of massive BH pairs and binaries has therefore become a major focus of both theoretical and observational studies.

Despite these strong theoretical expectations, direct observational evidence for massive BH binaries remains scarce. Robust detections are largely limited to dual active galactic nuclei (AGN) at projected separations of a few to a few tens of kiloparsecs \citep{Hennawi06,Foreman09,Liu11,Comerford13,Vignali18,Silverman20,Zhang21a,Zhang21b,Mannucci22,Scialpi26}, with only a handful of systems identified at separations of $\lesssim$1 kpc \citep{Komossa03,Fabbiano2011,Comerford12,Chen22}. At smaller scales, confirmed binaries are increasingly rare. A few parsec-scale candidates have been identified through very long baseline interferometry \citep{Rodriguez06,Bansal17,Kharb17}, while sub-parsec candidates are mostly inferred indirectly from periodic or quasi-periodic variability \citep{Sillanpaa88,Graham15,Chen20,Millon22}, Doppler-shifted broad emission lines \citep{Tsalmantza2011,Eracleous12}, and other spectroscopic signatures \citep{Komossa2008,Dotti2012,Dotti2022}. However, none of these observables uniquely identifies a binary, and most can be explained by alternative astrophysical processes. Consequently, the existence of a large population of bound sub-parsec binaries is still supported primarily by theoretical arguments (see \citealt{DeRosa19} for a review). Recent Pulsar Timing Array observations have nevertheless provided the first evidence of a nanohertz stochastic gravitational-wave (GW) background \citep{Agazie23,Agazie23j}, most naturally interpreted as the unresolved signal from a cosmic population of merging binaries.

From a theoretical perspective, the dynamical evolution of massive BH pairs in galaxy mergers was first outlined by \cite{Begelman1980}, who showed that coalescence is possible provided the surrounding environment can efficiently extract orbital angular momentum within a Hubble time. They identified a characteristic three-stage evolutionary sequence.
The first stage begins after a galaxy merger, when the two central BHs remain separated by kiloparsec scales and orbit within the merger remnant. In this regime, dynamical friction against stars, gas, and dark matter removes orbital energy and angular momentum, driving the BHs toward the galactic centre \citep{Chandrasekar1943,Colpi2014}. This phase is captured in both cosmological simulations \citep[e.g.,][]{Tremmel15, Pfister19, Chen22df, Genina24} and idealised galaxy simulations \citep{Mayer07,Chapon13,Capelo2015,Pfister2017,Bollati23,Liao24}. Once the enclosed mass becomes comparable to or smaller than the binary mass, the two BHs form a gravitationally bound system. From this point onward, orbital decay proceeds through the so-called binary hardening phase, during which dynamical friction becomes inefficient and other mechanisms must drive the inspiral. The dominant hardening channel depends strongly on the ambient environment. In gas-poor systems, the binary shrinks mainly through repeated three-body interactions with background stars \citep{Quinlan96,Milosavljevic01,Milosavljevic03,Sesana06,Rantala17}, and may be further accelerated by interactions with a third massive BH brought in by subsequent mergers \citep{Hoffman07,Bonetti18,Mannerkoski21}. In gas-rich mergers, tidal and hydrodynamical torques efficiently funnel gas toward the nucleus \citep{Hernquist1989, Barnes1991, Mihos1996, Hopkins2010, barnes02,CD17,BB18}, where it can form a massive circumbinary disc (CBD) whose gravitational torques extract angular momentum and accelerate orbital decay \citep{Cuadra09,Roedig12,Dorazio13,Farris14,Moody19,Munoz19,Duffell20,Franchini21,Franchini22,Bourne24}. Finally, when the separation reaches $a_{\rm GW} \sim 10^{-3}(M_{\rm BH}/10^6\,\mathrm{M}_\odot)\,\mathrm{pc}$, GW emission dominates angular momentum loss, rapidly driving the system to coalescence. 
The GW signal produced by shrinking heavy ($\gtrsim 10^8$ M$_\odot$) MBHBs thousands of years from coalescence can be observed by currently ongoing PTA analyses. Lighter massive BH binaries are instead key sources for the future Laser Interferometer Space Antenna (LISA), which will probe the millihertz band and access mergers of $10^{4}-10^{7}\,\mathrm{M}_\odot$ BHs out to high redshift \citep{Amaroseoane17,Amaro23}.

LISA detections, expected to have very high signal-to-noise ratios even at high redshift, will be crucial for constraining the magnitudes and orientations of BH spins with unprecedented precision. The structure of the GW signal carries information on spin misalignments \citep{Vecchio04, Klein09, Oshaughnessy13}, and LISA will be able to distinguish between aligned, misaligned, and anti-aligned spin configurations through characteristic spin-precession signatures imprinted on the waveform \citep{Pratten23}.
Measuring the spins of merging MBHs is of high astrophysical interest, as spin configurations play a fundamental role in determining the outcome of the merger itself: the spins at coalescence set the final spin of the remnant and the magnitude and direction of the gravitational recoil velocity \citep{Campanelli07,Gonzalez13,Lousto13,Lousto19,Sperhake}. Recoil kicks, in turn, influence the retention of MBHs in galaxies, the occupation fraction of galactic nuclei \citep{Schnittman07,Volonteri10,GerosaSesana15}, the population of off-nuclear BHs, and ultimately the expected rate of GW events \citep{Sesana07,Blecha08}.

Beyond their role in shaping the merger outcome, the spins of the progenitor BHs are also powerful tracers of their recent evolutionary history. In gas-rich environments, in the presence of coherent accretion from the surrounding gaseous disc, BH spins can align with the angular momentum of the accreting gas through the Bardeen–Petterson (BP) effect. The BP effect \citep{BP75} arises from the interplay between Lense–Thirring precession and viscous torques in a misaligned accretion disc: the inner disc aligns with the BH spin on a viscous timescale, while the outer disc retains its original orientation, and the resulting warped disc exerts a gravitomagnetic torque that gradually aligns the BH spin with the angular momentum of the outer disc. For this reason, the orientation of the spins relative to the orbital angular momentum is expected to retain memory of the binary's environment during the final stages of inspiral.

Understanding how BH spins evolve and align before merger is therefore essential for connecting future LISA observations to the astrophysical processes governing the late stages of massive BH binary evolution. To this end, spin evolution has been studied in idealised analytic and numerical models of isolated BH pairs and binaries in galactic nuclei \citep{Bogdanovic07, Dotti10, Miller13, Gerosa15, Gerosa20,Bourne24, Koudmani24}, in semi-analytical models of galaxy formation \citep{Volonteri05,Berti08,Fanidakis11,Barausse12,Sesana14,IV20}, and in cosmological hydrodynamical simulations \citep{Dubois14,Bustamante19,Dubois21,Husko22,DP23,Sala24,Beckmann25,Husko26}.
A consistent picture emerging from these studies is that gas-poor mergers tend to produce binaries with nearly isotropically distributed spin orientations, whereas gas-rich environments generally promote spin alignment through gas accretion. However, the efficiency of this alignment remains highly model dependent, varying with the coherence of the gas inflow, the accretion history, and the adopted subgrid prescriptions.
More in detail on this last point,  \citet{Bogdanovic07} first proposed that gas-rich mergers can efficiently align the MBH spins during inspiral, provided each BH accretes $\sim1$--$10\%$ of its mass, whereas gas-poor mergers preserve nearly isotropic spin orientations. This picture was subsequently refined by \citet{Dotti10}, who showed that alignment can be generally efficient but depends sensitively on the properties of the accretion disc. \cite{Lodato13} further showed that nonlinear warp propagation can delay alignment in highly misaligned systems, while \cite{Miller13} demonstrated that binary torques can instead enhance the alignment process. More recently, \cite{Gerosa15} highlighted the role of differential accretion from CBDs, showing that the secondary BH aligns more rapidly than the primary. Building on this, \cite{Gerosa20} found that the outcome may range from efficient alignment to disc breaking \citep{Nixon12,Nealon22}, depending on the initial misalignment and binary properties. Finally, \cite{Bourne24} incorporated spin alignment into hydrodynamical simulations using the sub-grid disc model of \citet{Fiacconi18}, broadly confirming the analytic picture while finding more efficient primary alignment, although nonlinear warp propagation, disc breaking, and the full back-reaction of the mini-discs remain neglected. \citet{Koudmani24} further extended this framework by incorporating a unified thin-disc/ADIOS accretion model, showing that spin alignment efficiency is strongly modulated by the disc radiative state, with solid-body precession in the radiatively inefficient regime slowing alignment relative to the classical Bardeen–Petterson case.

However, notably, none of the previous works explicitly accounted for the impact of AGN feedback from the individual BHs. More generally, only a handful of studies have investigated the role of AGN feedback during the binary hardening phase in CBDs \citep{DelValle18,Williamson22,dEtigny24}. In particular, \cite{DelValle18} found that, when the binary efficiently clears a central cavity, AGN feedback is able to escape through the low-density region and has little impact on the binary dynamics. By contrast, in configurations where no clear cavity forms and the BHs remain embedded in dense disc gas, higher accretion rates and stronger feedback lead to efficient coupling with the surrounding medium, allowing AGN feedback to carve a low-density “feedback cavity” around the binary and further suppress gas torques, thereby stalling migration. More recently, \citet{dEtigny24}, using radiation-hydrodynamic simulations, showed that faster migration can be partially restored due to inefficient coupling between AGN radiation and the surrounding gas.

In this work, we aim to investigate the spin evolution of a massive BH binary embedded in a CBD, evolving self-consistently the BHs spin, mass, accretion, and AGN feedback using the sub-grid model of \cite{Bollati24}. 
The paper is structured as follows. In Section~\ref{sec: Methods}, we present the numerical methods employed in this work. In Section~\ref{sec: setup}, we describe the simulation setup. The results are presented in Section~\ref{sec: results}, followed by discussion and conclusions in Section~\ref{sec: conclusion}.

\section{Methods} \label{sec: Methods}

In this section, we describe the numerical methods employed in this work. We performed a suite of simulations of massive BH binaries embedded in CBDs using the meshless hydrodynamics/$N$-body code \textsc{gizmo} \citep{Hopkins15}. AGN radiative feedback is modelled following \cite{Bollati24}, through a sub-grid prescription that accounts for the dependence of feedback anisotropy on the BH spin orientation and magnitude. Below, we briefly summarize the main features of the model and outline the modifications introduced for the present study. We refer the reader to \cite{Bollati24} and \cite{Cenci21} for a detailed description of the model and its implementation.

\subsection{Accretion and spin evolution}\label{sec: accretion}

The sub-grid model developed by \cite{Bollati24}, building on \cite{Cenci21}, self-consistently evolves BH growth, spin evolution, AGN feedback, and its anisotropy, enabling us to capture the non-linear interplay between gas accretion, feedback, and binary dynamics.

In our implementation, each unresolved accretion system is characterized by the masses of the BH and sub-grid accretion disc, $M_\bullet$ and $M_\alpha$, and by their corresponding angular momenta, $\mathbf{J}_\bullet$ and $\mathbf{J}_\alpha$.
The evolution of the BH is governed by the accretion rate through the sub-grid disc, $\dot{M}_\mathrm{acc}$, computed as $\dot{M}_\mathrm{acc} = f_\mathrm{Edd}\dot{M}_\mathrm{Edd}$, where $\dot{M}_\mathrm{Edd}$ is the Eddington accretion rate and $f_\mathrm{Edd}$ is the corresponding Eddington ratio. The latter is evaluated from the current disc properties using the analytic prescription of \citet[][their Eq.~2]{Fiacconi18}. This model assumes a standard $\alpha$-disc \citep{SS73}, truncated at its self-gravitating radius, and provides a self-consistent estimate of the unresolved mass transport through the disc and onto the BH. The disc is in turn fed by gas inflowing from resolved scales at a rate $\dot{M}_\mathrm{in}$.

Regarding the BH spin, our model allows $\mathbf{J}_\bullet$ and $\mathbf{J}_\alpha$ to be mutually misaligned, thus accounting for warped accretion discs. The exchange of angular momentum between the BH and the disc is followed self-consistently through both gas accretion and the Bardeen--Petterson effect \citep{BP75}. In this way, the evolution of the BH spin remains coupled to that of the sub-grid disc.

The modifications introduced in the present work with respect to the model of \cite{Bollati24} concern the computation of the radiative efficiency, $\eta$ -- which enters the evaluation of both $f_\mathrm{Edd}$ and the wind mass outflow rate (Eq.~\ref{Eq: Mw}) -- and the prescription adopted for the inflow rate onto the sub-grid disc, $\dot{M}_\mathrm{in}$. These correspond to Eqs.~1 and A.6, and Eqs.~A.7--A.8 of \cite{Bollati24}, respectively.

Regarding the radiative efficiency, when the spin evolves from a counter-rotating to a co-rotating configuration with respect to the accretion disc, the prescription based on Eqs.~B1--B5 of \cite{Fiacconi18}, adopted in \cite{Bollati24}, introduces a discontinuity in $\eta$. To avoid this unphysical jump, we smooth the dependence of $\eta$ on the angle $\delta$ between the spin and disc angular momentum vectors, $\mathbf{J}_\bullet$ and $\mathbf{J}_\alpha$, such that $\eta(\delta)$ varies continuously across the transition. Specifically, we adopt

\begin{equation}
\eta(\delta) = \eta_\mathrm{ret} + \frac{1}{2}\left(\cos\delta + 1\right)\left(\eta_\mathrm{pro} - \eta_\mathrm{ret}\right),
\label{Eq: eta}
\end{equation}

where $\eta_\mathrm{ret}$ and $\eta_\mathrm{pro}$ are the radiative efficiencies corresponding to retrograde and prograde accretion, respectively, computed according to Eqs.~B1--B5 of \cite{Fiacconi18}. This interpolation is inspired by the prescription proposed in Eq.~9 of \cite{Hugher03}.

Concerning the resolved accretion rate $\dot{M}_\mathrm{in}$, each BH particle is treated as a sink. Gas particles that enter a distance $R_\textrm{sink}$ from the BH and are gravitationally bound to it are captured into the sub-grid system. If the total mass of the captured particles during a BH timestep $dt_\bullet$ is $M_\textrm{to\_be\_captured}$, the corresponding inflow rate is $\dot{M}_\mathrm{in} = M_\textrm{to\_be\_captured}/dt_\bullet$. The same procedure is also applied to the angular momentum of the accreted gas from resolved scales, which is transferred to the sub-grid disc.

If the disc mass is zero, i.e. the BH is quiescent, the mass of the captured particles is accumulated in a sub-grid variable $M_\textrm{temp}$. Once $M_\textrm{temp}$ reaches the threshold value of $M_\textrm{tresh}$ a new sub-grid disc is created with $M_\alpha = M_\textrm{temp}$ and $f_\textrm{Edd,0}$, thereby initiating a new AGN episode. Thereafter, at each timestep, the mass and angular momentum of newly captured gas particles are directly added to $M_\alpha$ and $\mathbf{J}_\alpha$.

\subsection{AGN feedback}

The accretion rate $\dot{M}_\textrm{acc}$ through the disc determines the AGN luminosity, whose angular distribution is set by the BH spin and can be described by a function $f(\phi, a)$, where $\phi$ is the polar angle measured with respect to the spin axis and $a = \textrm{sign}(\mathbf{J}_\bullet \cdot \mathbf{J}_\alpha)\,|\mathbf{J}_\bullet|c/GM_\bullet^2$ is the dimensionless spin parameter, with $c$ the speed of light and $G$ the gravitational constant \citep{Campitiello18, Ishibashi19}. More rapidly spinning BHs produce more isotropic radiation patterns, as photons are emitted closer to the BH and their geodesics undergo stronger gravitational bending.

In our model, AGN radiation is assumed to couple to the gas at the disc (sub-grid) scale, fully transferring its momentum to the surrounding material. This launches an AGN wind that inherits the angular distribution of the radiation field. AGN feedback is implemented through a wind-spawning scheme following \cite{Torrey20}. Specifically, mass is removed from the sub-grid disc and new gas particles are spawned at a rate

\begin{equation}
\dot{M}_\textrm{w} = \eta \frac{c}{v_\textrm{w}}\dot{M}_\textrm{acc},
\label{Eq: Mw}
\end{equation}

where $v_\textrm{w}$ is the wind velocity. The newly spawned particles are distributed uniformly on a sphere centred on the BH, with radius equal to the BH softening length $\epsilon_\bullet$.

In \cite{Bollati24}, the particle mass and angular distribution followed the luminosity pattern $f(\phi,a)$ (see their Eq.~15). However, when the spin evolves from a counter-rotating to a co-rotating configuration, the sign of $a$ changes abruptly, introducing a discontinuity in $f(\phi,a)$. To avoid this behaviour, and analogously to Eq.~(\ref{Eq: eta}), we smooth the dependence of the anisotropy function by adopting

\begin{equation}
f(a,\phi; \delta) = f(-|a|, \phi) + \frac{1}{2}\Bigl(\cos\delta + 1\Bigr)\Bigl(f(|a|,\phi) - f(-|a|,\phi)\Bigr),
\label{Eq: faniso}
\end{equation}
where $\delta$ is the angle between $\mathbf{J}_\bullet$ and $\mathbf{J}_\alpha$.
At each spawning event, a number of particles $N_\textrm{spawn} = \max\bigl(N_\textrm{spawn,min},\, 2M_\textrm{spawn}/m_0\bigr)$
is generated, where $M_\textrm{spawn}$ is the mass injected into the wind, $m_0$ is the initial gas particle mass in the CBD, and $N_\textrm{spawn,min}$ ensures adequate sampling of the spherical injection region around the BH particle without spawning too many particles (see Section~\ref{sec: setup}). The second term ensures that the spawned particles do not become significantly more massive than the gas particles used to resolve the disc. The particles are launched radially outward with velocity $v_\textrm{w}$ and, through their interaction with the surrounding gas, can drive a large-scale outflow.

\subsection{Viscosity}\label{sec: viscosity}

Since the gas dynamics within the CBD, including the formation of a central cavity, results from the competition between the binary tidal torque and viscous angular momentum transport, we include viscosity in the Navier--Stokes equations following an approach similar to that of \cite{Hopkins17, Franchini22}. Specifically, we assign to each gas particle a kinematic shear viscosity
\begin{equation}
\nu = \alpha R c_\textrm{s}^2/v_\varphi,
\label{Eq: visc}
\end{equation}
where $\alpha = 0.1$ is the \cite{SS73} viscosity parameter and $c_\textrm{s}$ is the gas sound speed. Bulk viscosity is neglected.
The quantities $R$ and $v_\varphi$ are evaluated with respect to the nearest BH when the particle lies within its Hill radius, taken as half the binary separation in the equal-mass case; otherwise, they are computed with respect to the centre of mass of the system.

This viscosity prescription is intended to mimic angular momentum transport driven by the magneto-rotational instability in magnetized, differentially rotating discs \citep{Balbus91} and is therefore not applied to gas participating in feedback-driven outflows. Accordingly, Eq.~(\ref{Eq: visc}) is applied only to non-wind particles with $v_r < v_\varphi/2$, where $v_r$ is the radial velocity; otherwise, $\nu$ is set to zero.

To account for unresolved angular momentum transport in the immediate vicinity of the BHs, we additionally enhance viscosity within a distance $R_\textrm{ext} = 10 \epsilon_\bullet$ from each BH. In this region, the viscosity parameter is increased such that the viscous timescale matches the local orbital timescale, corresponding to $\alpha = (v_\varphi/c_\textrm{s})^2 (2\pi)^{-1}$. As a result, particles entering $R_\textrm{ext}$ are efficiently driven inward, allowing them to reach $R_\textrm{sink}$ and be accreted by the sub-grid model.

\section{Setup and simulations} \label{sec: setup}

We consider a massive BH binary with total mass $M_{\rm bin}=10^{6}\,{\rm M_\odot}$ embedded in a CBD. This mass scale is motivated by the population of binaries expected to be detected by LISA. Current forecasts indicate that LISA will be most sensitive to systems with total masses in the range $10^{4}-10^{7}\,{\rm M_\odot}$, with optimal signal-to-noise ratios for binaries with $M_{\rm bin}\sim10^{5}-10^{6}\,{\rm M_\odot}$ \citep{Amaroseoane17, Amaro23}. Such systems are expected to form during gas-rich galaxy mergers and may be observable up to redshift $z>12$ \citep{Colpi24}. At these epochs, mergers can efficiently funnel large amounts of gas into the nuclear regions. We assume the binary is embedded in a gaseous CBD with mass $M_{\rm d}=10^{5}\,{\rm M_\odot}$.

The initial conditions are constructed as follows: we consider a circumbinary disc profile extending from $R_\textrm{in} = 2\,r_\textrm{bin,0}$, the radius of the initial inner cavity, to $R_\textrm{out} = 10\,r_\textrm{bin,0}$, with a surface density following a Mestel profile
\begin{equation}
\Sigma(R) = \frac{M_\textrm{d}}{2 \pi R_\textrm{in} (R_\textrm{out}-R_\textrm{in})} \frac{R_\textrm{in}}{R}, \quad R_\textrm{in}\leq R \leq R_\textrm{out}.
\label{Eq: Mestel}
\end{equation}
The gas azimuthal velocity and vertical structure are generated using the code \textsc{gd\_basic} \citep{Lupi15}, assuming both vertical hydrostatic equilibrium and dynamical equilibrium in the combined gravitational potential of the binary and a surrounding stellar bulge. During the initialisation of the gas component, the binary potential is approximated as that of a single point mass with mass $M_\textrm{bin}$ located at the centre of the disc. The stellar bulge is modelled with a Plummer profile, characterised by a core radius of $65$ pc and a total mass of $3\times10^8\,\mathrm{M_\odot}$, consistent with the observed BH–bulge relation \citep{Haring04}. The gas is initialised with a uniform temperature of $6\times10^3$ K and solar metallicity.
With this setup, the disc remains stable against gravitational fragmentation, preventing the formation of clumps that could perturb the binary orbit \citep[e.g.,][]{SouzaLima17} and complicate the interpretation of AGN feedback effects.

The initial gas distribution is sampled with $N = 5\times10^5$ particles, corresponding to a mass resolution of $0.2\,\mathrm{M_\odot}$. This number increases during the simulation as wind particles are spawned from the sub-grid model. During the evolution, gas particles undergo metal-line cooling and are treated as self-gravitating. A fixed gravitational softening length of $\epsilon_\textrm{gas} = 0.005$ pc is adopted for the gas. The stellar bulge is not represented by particles but is included as a static analytic potential.

Once the disc gas particles are initialised, the initial conditions are completed by placing the equal-mass binary, with initial separation $r_\textrm{bin,0} = 2$ pc, at the centre of the system. Thus, the binary is initially placed within the CBD's initial cavity. As the simulation starts, however, the gas at the inner edge of the cavity flows towards the centre due to the unbalanced pressure forces acting on it; the subsequent presence and maintenance of the cavity is then set by the balance between the binary's tidal torque on the disc and the internal viscous torques of the gas. The two BHs are set on circular orbits co-planar with the CBD (lying in the $xy$ plane of our reference frame) and co-rotating with the gas. The initial orbital period is $T_0 = 0.271$ Myr. During the simulation, the binary is fully dynamical and evolves self-consistently under the effects of gravitational torques from the gas, as well as accretion and wind feedback, which allow its orbit to shrink or expand accordingly. The BH particles adopt a fixed gravitational softening length of $\epsilon_\bullet = 0.015$ pc. Initially, both MBHs are quiescent, i.e. the sub-grid disc mass is set to zero. The unresolved discs are then self-consistently built up through accretion from resolved scales, as particles passing close to the MBHs are progressively captured (see Sec.~\ref{sec: accretion}).

Starting from these initial conditions, we explore different BH spin configurations by varying both spin magnitudes and orientations. The adopted values for each run are summarised in Table~\ref{tab: spin}. The spin orientation is defined by the angle $\theta$ between the spin vector and the $z>0$ axis, which coincides with the angular momentum direction of both the CBD and the binary. In all simulations, the primary BH spin is aligned with the $z>0$ axis.
\begin{table}
    \centering
    \caption{Summary of the spin properties adopted in our initial conditions. For all simulations $\theta_1 = 0$.}
    \small{\begin{tabular}{l|ccccc}  
         \hline\hline
         label &  $a_1$ & $a_2$ & $\theta_2 \rm$  [deg]\\
         \hline
         {\labsim{ref}}&   0.5 & 0.5 &  0 \\
         {\labsim{a1}}  &   0.5 & 0.9982 &  0 \\
         {\labsim{aa1}}  &  0.9982  & 0.9982 &  0  \\
         {\labsim{a5pi12}}  &   0.5 & 0.5  & 75  \\
         {\labsim{a7pi12}}  &   0.5 & 0.5  & 105 \\
         {\labsim{a5pi6}}  &   0.5 & 0.5 & 150 \\
         \hline\hline
    \end{tabular}}
    \label{tab: spin}   
\end{table}
To isolate the impact of feedback, we perform additional no-feedback runs for the spin configurations {\labsim{ref}}, {\labsim{a5pi12}}, {\labsim{a7pi12}}, and {\labsim{a5pi6}} of Table \ref{tab: spin}. The full suite therefore consists of ten simulations: six with feedback, corresponding to the models listed in Table~\ref{tab: spin}, and four without feedback. The latter are denoted by the same labels with the suffix {\labsim{\_NoFB}}.

\begin{figure*}
    \centering
\includegraphics[width=1\linewidth]{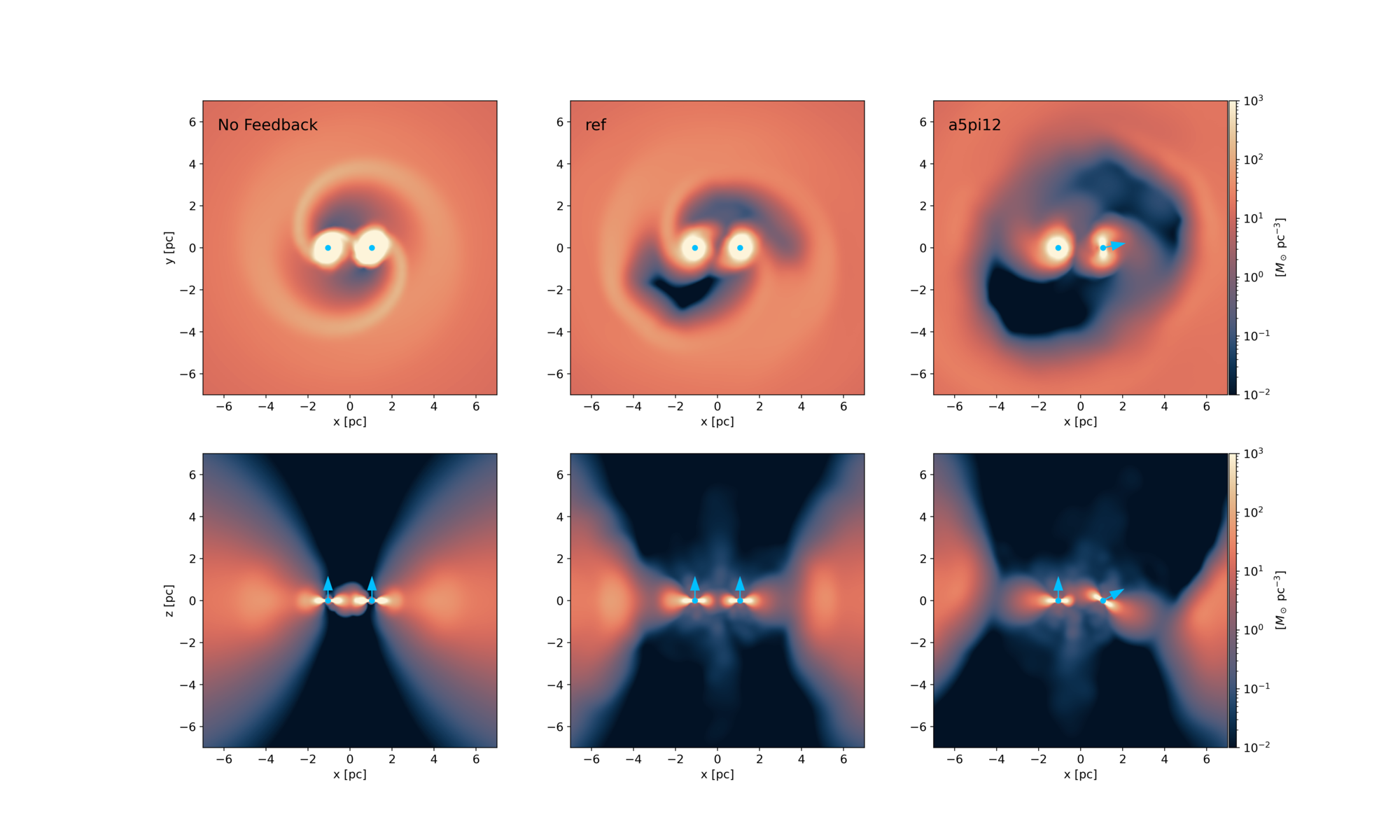}
    \caption{Density maps for runs {\labsim{ref\_NoFB}} (left column), {\labsim{ref}} (centre column), and {\labsim{a5pi12}} (right column), shown in face-on (top row) and edge-on (bottom row) projections. All maps are displayed at $t = 4.05$ Myr, corresponding to approximately 15 binary orbital periods. Blue circles mark the positions of the BHs, while arrows indicate the orientations of their spin vectors.}
    \label{fig: MapsDensity}
\end{figure*}

The simulation suite is designed to disentangle the role of spin magnitude and orientation in shaping feedback–disc coupling and the resulting dynamical evolution. Runs with misaligned spins ({\labsim{a5pi12}}, {\labsim{a7pi12}}, {\labsim{a5pi6}} and their no-feedback counterparts) probe how anisotropic feedback, collimated along the spin direction, interacts with the CBD for different inclination angles, as well as how the spin alignment process proceeds for different initial orientations.
In contrast, aligned-spin runs with different spin magnitudes ({\labsim{ref}}, {\labsim{a1}}, {\labsim{aa1}}) isolate the effect of spin magnitude on the feedback efficiency. The adopted spin values correspond to two regimes of radiative anisotropy. As shown in Figs. 1 and 2 of \cite{Bollati24}, the radiation field becomes nearly isotropic for high spins ($a \gtrsim 0.8$), while lower spins produce significantly more anisotropic emission. Our choices, $a=0.9982$ and $a=0.5$, therefore bracket these two regimes and allow us to explore qualitatively different feedback behaviours.

Each wind spawning event generates $N_\textrm{spawn,min} = 60$ wind particles, initialized with a velocity $v_{\rm wind} = 1000\,\rm km\, s^{-1}$ and a temperature $T_{\rm wind} = 2\times10^4$ K. A new sub-grid accretion disc is created once the mass of gas captured within $R_{\rm sink} = 0.035$~pc reaches the threshold value $M_\textrm{tresh} = 1500 \,\text{M}_\odot$. The newly formed disc is initialized with an initial Eddington ratio $f_{\rm Edd,0} = 0.05$.
All simulations are evolved for approximately 30 binary orbits, while the run {\labsim{a5pi12}} is extended to about 70 orbits to investigate longer-term effects.

Besides the simulations in the main suite presented above, we performed additional higher-resolution runs to test the numerical convergence of our results. These are presented in Appendix \ref{App: resolutions}.

\section{Results} \label{sec: results}

\begin{figure*}
    \centering
\includegraphics[width=1\linewidth]{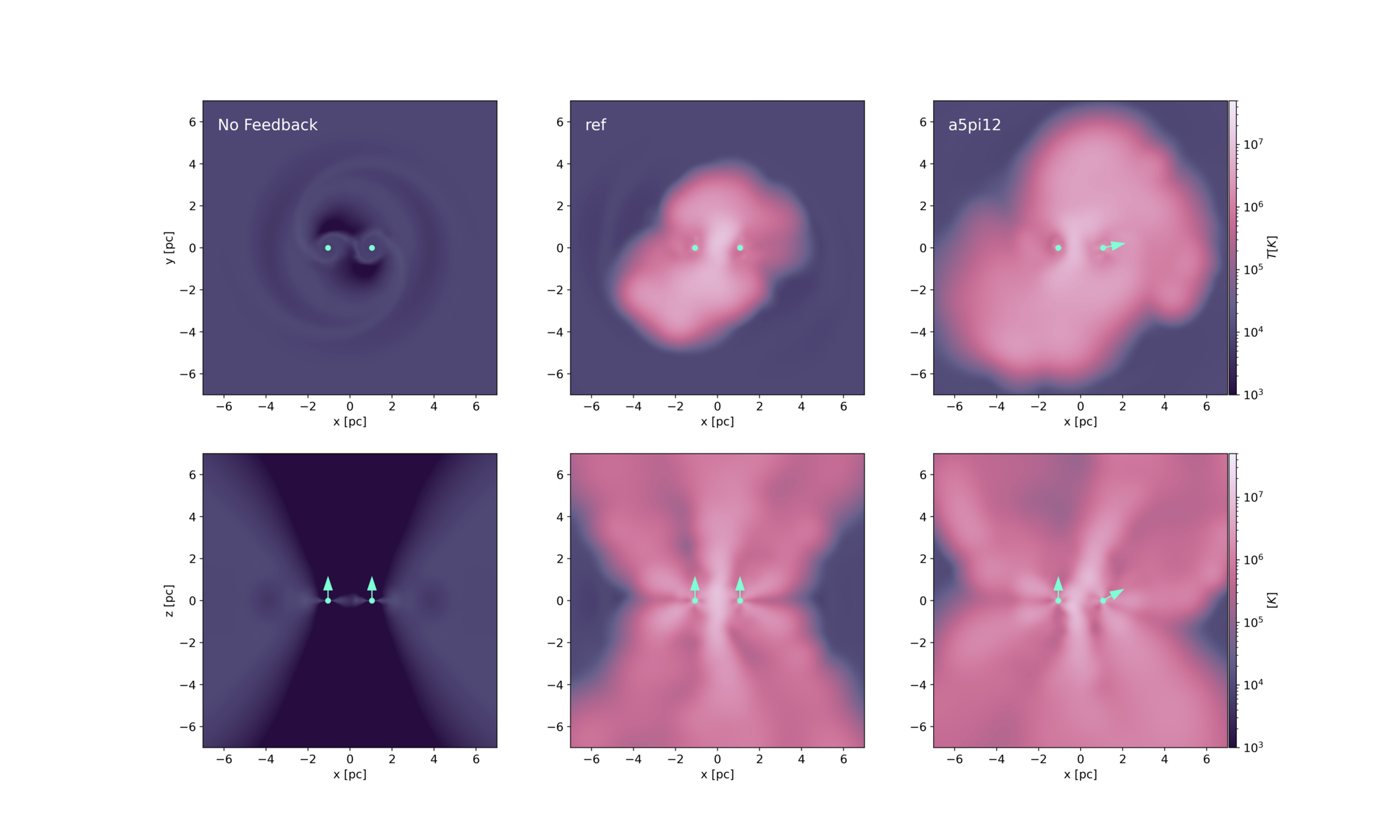}
    \caption{Temperature maps for runs {\labsim{ref\_NoFB}} (left column), {\labsim{ref}} (centre column), and {\labsim{a5pi12}} (right column), shown in face-on (top row) and edge-on (bottom row) projections. All maps are displayed at $t = 4.05$ Myr, corresponding to approximately 15 binary orbital periods. Cyan circles mark the positions of the BHs, while arrows indicate the orientations of their spin vectors.}
    \label{fig: MapsTemp}
\end{figure*}

In this section, we present the analysis of the simulations described in the previous section. Before discussing the results in detail, we first examine the density maps to gain a qualitative understanding of the system evolution and of the impact of feedback.

The first column of Figure~\ref{fig: MapsDensity} shows the face-on (top panel) and edge-on (bottom panel) density maps for the {\labsim{ref\_NoFB}} run after 15 binary orbits from the start of the simulation.
A central cavity forms around the binary, while density wakes develop in the CBD and efficiently transport angular momentum outward. Gas can nevertheless penetrate the cavity and feed minidiscs around both BHs, driven by viscous transport and binary-induced gravitational torques.  \cite{DelValle14} derived a criterion to determine whether a CBD develops and sustains a central cavity, based on the competition between the positive gravitational torque exerted by the binary and the negative angular momentum transport driven by viscous stresses within the disc. Their analysis shows that massive (relative to the binary) and geometrically thick discs are less prone to cavity excavation. Applying their criterion (Eq.~2 of \citealt{DelValle14}), and given that our simulations exhibit an aspect ratio $h/r \gtrsim 0.1$, we find that our system lies within the cavity-opening regime, in agreement with the results shown in the left column of Figure~\ref{fig: MapsDensity}. The presence of the cavity also indicates that the binary is not embedded in a dense gaseous medium capable of driving rapid inspiral over a few orbital timescales \citep[e.g.,][]{Escala05,DelValle18}. Instead, it reduces gas–binary interactions, so that orbital evolution proceeds over many orbital timescales driven by weaker gas torques, as further discussed in Section~\ref{sec: accretion}.

The second and third columns of Figure~\ref{fig: MapsDensity} show the corresponding density maps for runs {\labsim{ref}} and {\labsim{a5pi12}}, respectively. These simulations differ from {\labsim{ref\_NoFB}} by the inclusion of AGN feedback, while {\labsim{a5pi12}} additionally adopts a different orientation for the secondary BH spin (see Table \ref{tab: spin}). Despite the presence of feedback, minidiscs persist in both runs, indicating that they are not immediately disrupted by AGN activity, although they appear smaller than in the no-feedback ({\labsim{ref\_NoFB}}) case. In run {\labsim{a5pi12}}, the secondary minidisc is also tilted and partially reoriented as a consequence of anisotropic feedback originating from the misaligned secondary accretion disc.

On larger scales, feedback impacts the inner regions of the CBD, contributing to the excavation of the central cavity. This results in larger cavities compared to the no-feedback case and in a partial suppression of the density wakes generated by the binary, in agreement with \citet{DelValle18}. These effects are particularly pronounced in run {\labsim{a5pi12}}. Since AGN winds are collimated along the spin axis, the misaligned spin configuration enhances the coupling between feedback and the CBD, leading to a stronger dynamical impact on the surrounding gas. Overall, these simulations show that AGN feedback can significantly affect the gas distribution on both minidisc and CBD scales. Moreover, because the feedback geometry is directly tied to the BH spin orientation, its impact depends sensitively on the adopted spin configuration.

Figure~\ref{fig: MapsTemp} presents the same projections shown in Figure~\ref{fig: MapsDensity}, but displaying the gas temperature instead of the density. The maps show that, in the presence of feedback, hot gas fills the cavity surrounding the binary and expands preferentially along directions perpendicular to the plane of the CBD. The extent of the heated region increases with the strength of the AGN feedback--CBD coupling and is therefore most pronounced in run {\labsim{a5pi12}}.

In the following, we investigate how the different simulation setups affect the evolution of both the binary and the individual BHs. As discussed above, changes in the spin configuration modify the coupling between AGN feedback and the surrounding gas, leading to distinct alterations in the CBD structure. These differences are expected to influence the subsequent evolution of the system, since the gas properties and CBD structure regulate the torques acting on the binary, the accretion onto the BHs, and, consequently, their spin evolution.

\subsection{Dynamics and accretion}

\begin{figure}
    \centering
    \includegraphics[width=1\linewidth]{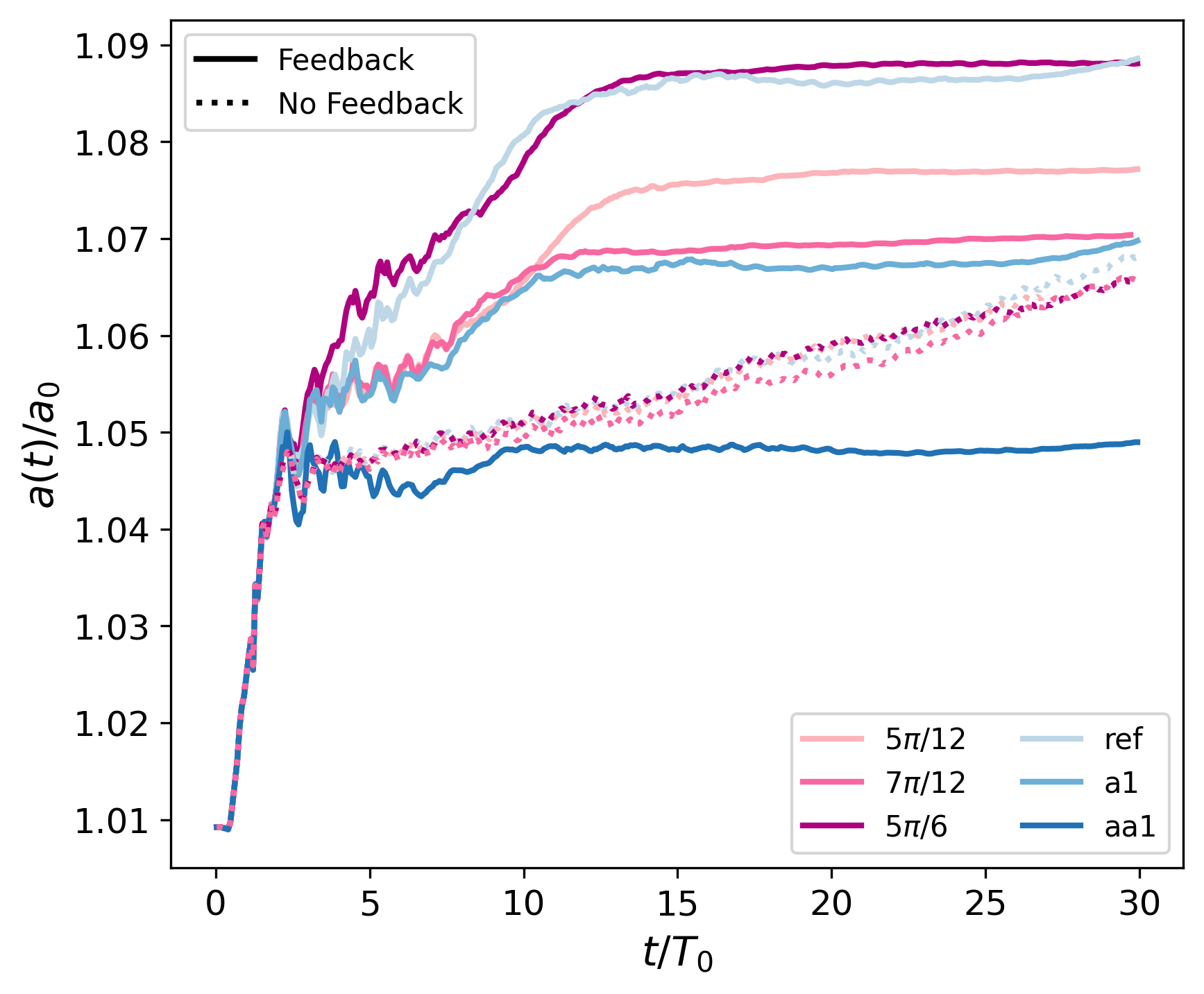}
    \caption{Evolution of the binary semi-major axis, normalized to its initial value, for all simulations in our suite. Time is expressed in units of the initial binary orbital period. Solid lines denote simulations including AGN feedback, while dotted lines correspond to their no-feedback counterparts. Different colours indicate different BH spin configurations.}
    \label{fig: Orbital}
\end{figure}

In this section, we discuss the orbital evolution of the binary and the accretion rates onto the individual BHs for all the simulations considered.
Figure~\ref{fig: Orbital} shows the time evolution of the binary semi-major axis $a(t)$, normalized to its initial value, as a function of time expressed in units of the initial orbital period. Solid lines correspond to simulations including AGN feedback, while dotted lines represent runs without feedback. Different colours identify different spin configurations: pink shades denote runs with varying inclinations of the secondary BH spin, whereas blue shades correspond to runs with different secondary spin magnitudes.

In all {\labsim{NoFb}} simulations, the semi-major axis increases throughout the duration of the runs, following nearly identical evolutionary tracks. In the absence of AGN feedback, the CBD remains essentially unchanged across the different spin configurations, resulting in very similar binary dynamics in all cases. The orbital expansion observed in these runs is consistent with previous studies (e.g. \citealt{Duffell20}, \citealt{Heath20}, \citealt{Franchini22}, \citealt{Duffell24}), which found that binaries embedded in discs with comparable viscosity and aspect ratio $h/r$ experience a net positive gravitational torque from the surrounding gas, leading to orbital expansion. 

In contrast, in simulations that include feedback, the semi-major axis initially follows the same evolution as in the {\labsim{NoFb}} runs, up to the point at which AGN activity is triggered, at $\sim 0.5$ Myr. Beyond this stage, the spin-dependent coupling between AGN feedback and the surrounding gas affects the CBD differently in each simulation, leading to distinct orbital evolutions.
In all feedback runs, the initial orbital expansion is followed by a clear stalling phase after approximately 10 binary orbits. This behaviour suggests that the torques acting on the binary become substantially weaker than during the early stages of the evolution, reducing the efficiency of the orbital expansion.
This is consistent with the orbital stalling found by \cite{DelValle18}, whereby AGN feedback excavates a cavity large enough to significantly weaken the gravitational interaction between the binary and the surrounding gas, leading to a substantial reduction in the gravitational torque. 
In all simulations, despite the different spin configurations and the corresponding differences in AGN feedback, the binary maintains a nearly circular orbit throughout the evolution, with eccentricities remaining at the level of $\sim 0.01$.

\begin{figure*}
    \centering
\includegraphics[width=1\linewidth]{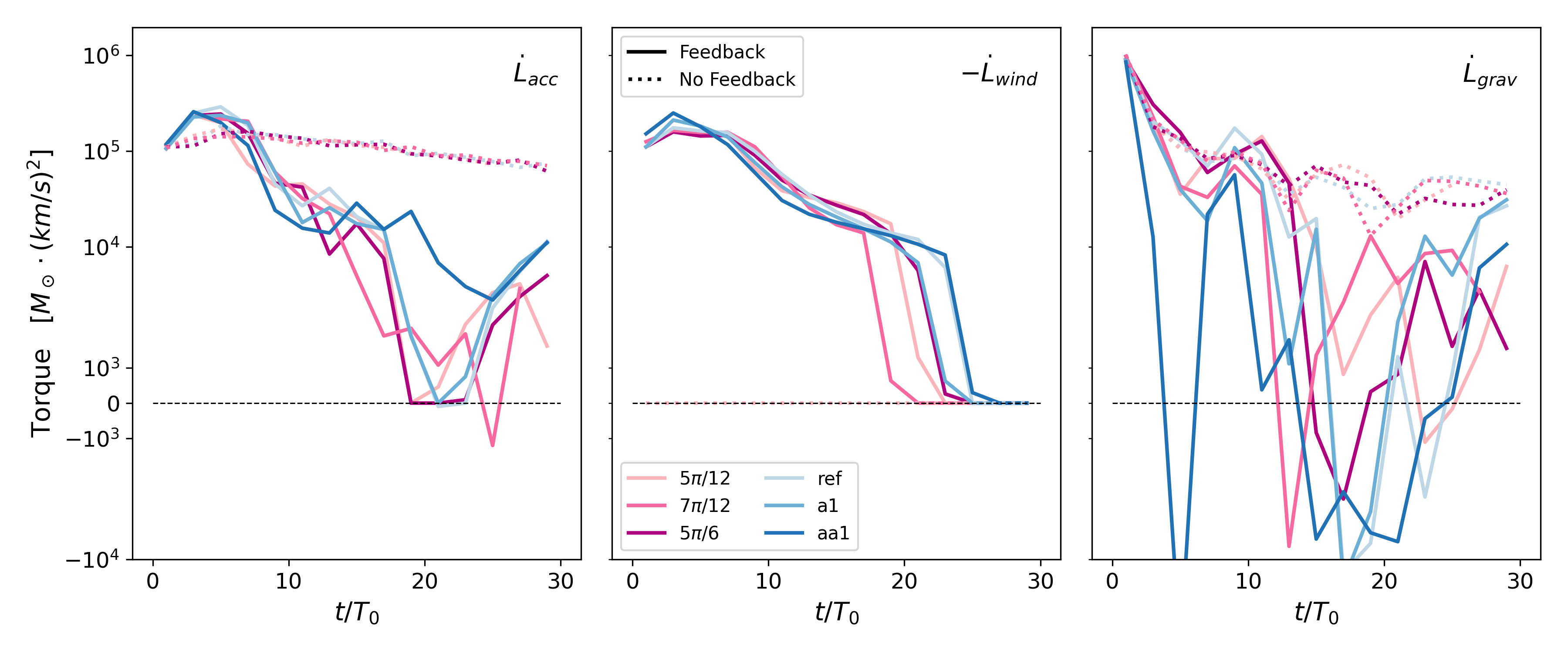}
    \caption{Time evolution of the $z$-component of the torque contributions arising from (left) accretion onto the BHs, (centre) wind launching, and (right) gravitational interaction with the CBD. Colours and line styles are as in Fig.~\ref{fig: Orbital}.}
    \label{fig: torques}
\end{figure*}

The binary evolution described above can be understood by examining the torques acting on the system, shown in Figure~\ref{fig: torques}. The three panels display the time evolution of the $z$ component of the torque contributions arising from three distinct mechanisms: (i) accretion onto the BHs, (ii) angular momentum associated with the mass loss carried away by winds launched from the accretion discs\footnote{This second term accounts for the mass lost through winds, which carries away orbital angular momentum by reducing the mass of the accretion discs attached to the BHs (Eq. \ref{Eq: Ldotwind}), thereby affecting the binary's angular momentum without directly altering the BH velocities, since the wind particles are launched isotropically in the BH rest frames. 
}, and (iii) the gravitational interaction between the CBD and the binary. A more detailed description of the numerical procedure used to compute these quantities is provided in Appendix~\ref{App: torques}. For each curve, the reported value at a given time corresponds to an average over 2 orbital periods, in order to reduce the strong short-timescale variability associated with orbital motion.

In simulations without feedback, both the accretion torque and the gravitational torque are positive. As a result, these contributions act in concert to drive orbital expansion over the entire simulated time span, consistent with the behaviour shown in Figure~\ref{fig: Orbital}. In simulations including feedback, the accretion torque is initially comparable to that measured in the {\labsim{NoFb}} runs, but subsequently decreases by one to two orders of magnitude. This trend indicates a substantial reduction in the accretion rate onto the binary. In addition, the second panel shows that the positive accretion torque is approximately balanced by a negative wind torque, which follows a similar temporal evolution but with opposite sign.
The gravitational torque in feedback runs initially exhibits positive values as large as those in the {\labsim{NoFb}} simulations; however, it decays more rapidly, settling around a smaller amplitude, but with large oscillations around zero. This behaviour indicates that feedback has excavated a larger gas cavity (as shown in Figure~\ref{fig: MapsDensity}), thereby reducing the strength of the gravitational coupling between the CBD and the binary and leading to the orbital stalling observed in Figure~\ref{fig: Orbital}. 
Finally, since different spin configurations produce feedback with varying efficiencies and geometries, they generate cavities of different sizes, as shown in Figure~\ref{fig: MapsDensity}. This is reflected in the scatter of the gravitational torque histories across spin setups (Figure~\ref{fig: torques}, third panel), which in turn leads to the diversity in semi-major axis evolution seen in Figure~\ref{fig: Orbital}.
In particular, among the misaligned-spin runs, the {\labsim{a5pi6}} run exhibits the weakest coupling between the AGN feedback and the CBD. This run has the smallest misalignment angle with respect to the $z>0$ axis and, because the spin orientation is retrograde, it also produces the most anisotropic AGN winds. Consistent with this weaker coupling, the gravitational torque in {\labsim{a5pi6}} remains comparable to that of the {\labsim{ref}} run, which experiences the weakest AGN feedback among the non-misaligned runs, and the binary consequently undergoes a similar degree of orbital expansion. In contrast, the {\labsim{a7pi12}} run exhibits the strongest feedback--CBD coupling among the misaligned cases, as its spin evolves towards an orientation nearly perpendicular to the $z$ direction (see Section~\ref{sec: spin}). As a result, this run is able to excavate the largest cavity around the binary, leading to the most rapid decline in the gravitational torque and, consequently, to the earliest onset of orbital stalling.

\begin{figure*}
    \centering
\includegraphics[width=1\linewidth]{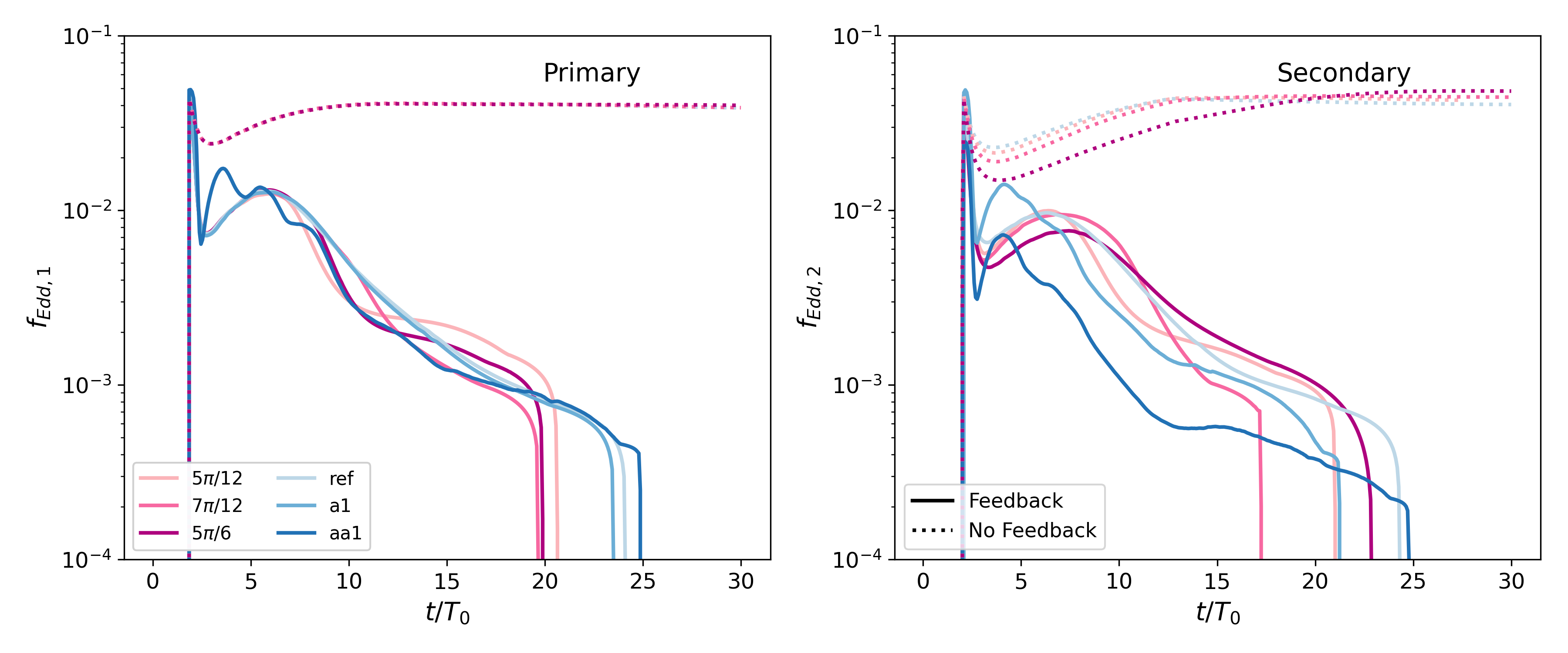}
    \caption{Time evolution of the Eddington factor $f_\textrm{Edd}$ for the primary (left) and secondary (right) BHs across all simulations in the suite. Colours and line styles are as in the previous figures.}
    \label{fig: fEdd}
\end{figure*}

In Figure~\ref{fig: fEdd}, we show the Eddington factor, $f_\mathrm{Edd}$, for the primary BH (left panel) and the secondary BH (right panel) across the full set of simulations.
In runs without feedback, the Eddington factor remains approximately constant at $f_\textrm{Edd} \sim 0.05$, indicating that both BHs accrete at a steady, nearly time-independent rate. We note that the behaviour of the secondary BH differs among the {\labsim{NoFb}} runs due to the different spin orientations. In particular, $f_\textrm{Edd} \propto \eta$  \citep[see Eq.~2 in][]{Fiacconi18}, and $\eta$, for a given spin magnitude, decreases with increasing misalignment angle $\delta$ between the BH spin and the sub-grid disc angular momentum (see Eq.~\ref{Eq: eta}). Consequently, at early times, larger spin–disc misalignments correspond to lower values of $f_\textrm{Edd}$. These differences are progressively erased during the evolution, as the BH spins align with the disc angular momentum via the Bardeen–Petterson effect.

When feedback is included, the Eddington factor decreases steadily shortly after feedback is activated. This trend indicates that feedback progressively removes gas from the immediate vicinity of the BHs, thereby suppressing accretion. This behaviour is consistent with the formation of a feedback-cavity, as shown in Figure~\ref{fig: MapsDensity}. After approximately 20--25 orbital periods, the Eddington factor drops to zero for both BHs, indicating that the system enters a quiescent phase in which the cavity has expanded sufficiently to temporarily prevent further feeding of the BHs.

\subsection{Spin evolution} \label{sec: spin}

\begin{figure*}
    \centering
\includegraphics[width=1\linewidth]{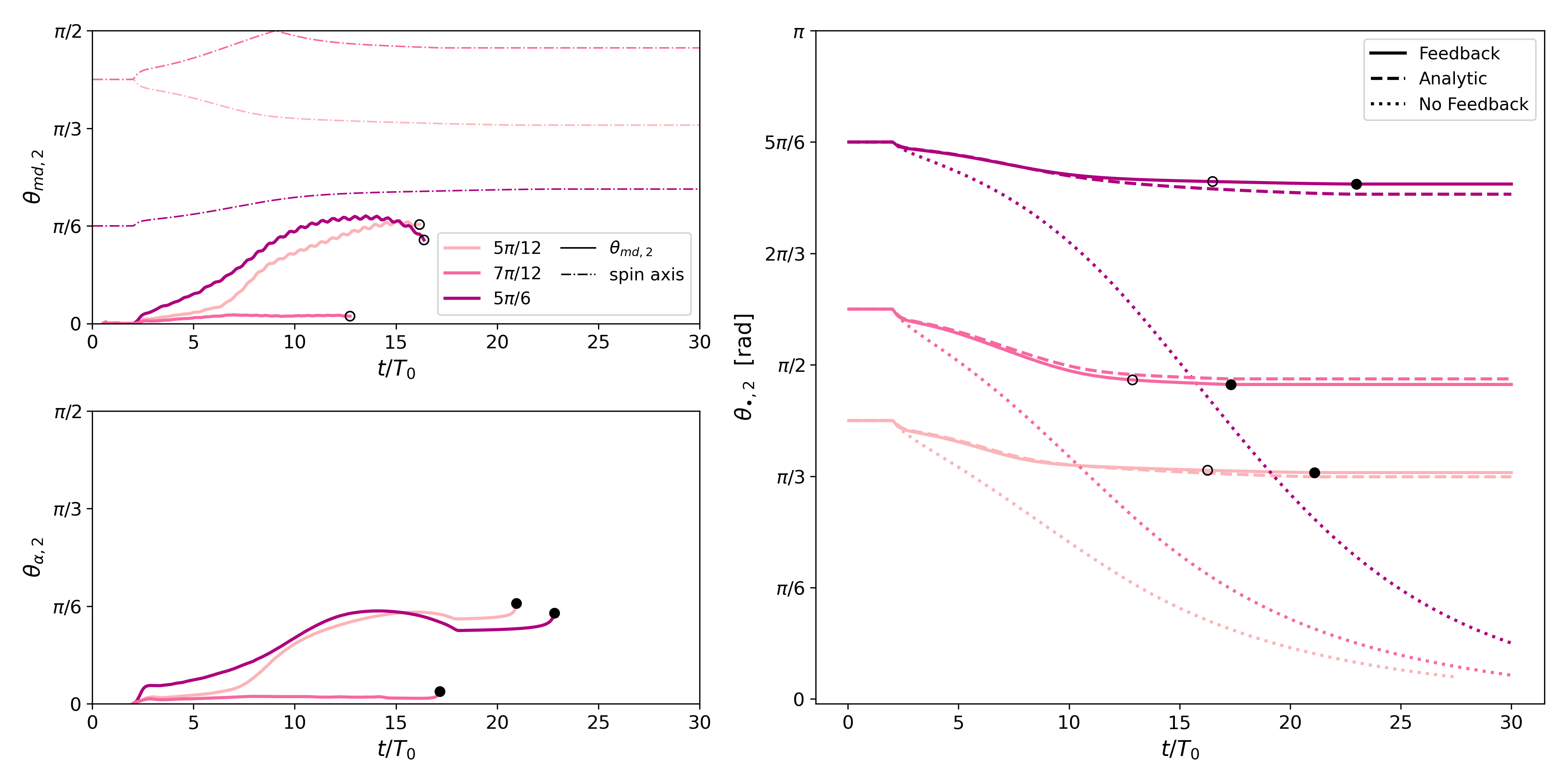}
    \caption{Time evolution of the angles between the $z>0$ axis and (i) the secondary BH minidisc angular momentum (top left), (ii) the sub-grid disc angular momentum (bottom left), and (iii) the secondary BH spin (right panel). Results are shown for the simulations {\labsim{a5pi12}}, {\labsim{a7pi12}}, and {\labsim{a5pi6}}, with solid lines in all panels. The right panel additionally includes the corresponding no-feedback runs ({\labsim{NoFb}}) as dotted lines and an analytical estimate of the spin angle evolution in dashed (see text for details). For comparison, the top-left panel also shows the spin inclination angle as a dash-dotted line. Open circles indicate the time at which minidiscs are destroyed in the feedback runs, while filled circles mark the onset of BH quiescence, when the sub-grid disc vanishes.}
    \label{fig: spin}
\end{figure*}

Figure~\ref{fig: spin} shows the evolution of the misalignment of the secondary BH minidisc (top-left) and sub-grid disc (bottom-left) for runs {\labsim{a5pi12}}, {\labsim{a7pi12}}, and {\labsim{a5pi6}}, i.e. the simulations in which the secondary spin is initially misaligned with respect to the $z$-axis.
The evolution of the secondary spin alignment is also shown (right) for the same runs (solid lines), together with their corresponding no-feedback counterparts (dotted lines).
The misalignment is quantified through the angles $\theta_{\mathrm{md},2}$, $\theta_{\alpha,2}$, and $\theta_{\bullet,2}$, which measure the inclination between the $z>0$ axis and, respectively, the angular momentum of the secondary minidisc, the sub-grid disc, and the BH spin. While the sub-grid disc angular momentum is directly evolved within the sub-grid model, the minidisc angular momentum is computed from gas particles bound to the secondary BH within a radius of 0.5 pc. The curves include markers indicating key evolutionary stages. Open circles mark the disruption of the minidisc, defined as the time when fewer than 500 gas particles remain within 0.5 pc of the BH. Filled circles indicate the disappearance of the sub-grid disc mass, marking the onset of a quiescent phase in which BH accretion ceases; these times correspond to those at which $f_\mathrm{Edd}=0$ in Figure~\ref{fig: fEdd}. 

The left panels of Figure~\ref{fig: spin} show that, in runs {\labsim{a5pi12}} and {\labsim{a5pi6}}, the secondary minidisc and, once formed, the sub-grid disc are initially aligned with the binary and CBD angular momenta (i.e. the $z>0$ axis). As the system evolves, feedback driven by the misaligned secondary spin progressively tilts the minidiscs, which in turn cause the tilting of the sub-grid accretion discs they feed.  For comparison, the dashed-dotted lines (top panel) show the evolution of the spin axis inclination angle\footnote{The angle is measured as the angular distance from the $z$ direction, hence it ranges between 0 and $\pi/2$.}, highlighting that the minidisc angular momentum tends to align with the BH spin direction. In the case of {\labsim{a5pi6}}, this corresponds to a counteralignment configuration. This behaviour, in which the minidisc aligns with the BH spin due to feedback before the spin itself can realign with the large-scale gas inflow, is also suggested by the bottom-right panel of Figure~\ref{fig: MapsDensity}. It can be understood as a consequence of anisotropic feedback: at minidisc scales, gas preferentially accumulates in directions where it experiences the weakest interaction with the outflow, namely the spin equatorial plane. As a result, feedback promotes locally coherent accretion even when the inflowing gas at CBD scales remains misaligned.
This effect is not observed in run {\labsim{a7pi12}}. This simulation corresponds to the most extreme spin configuration, with the secondary spin nearly perpendicular to the $z$-axis and therefore lying close to the CBD plane. This geometry maximises the feedback--CBD coupling, leading to the most efficient removal of gas and to the rapid disruption of the minidisc before any clear reorientation can take place.

We note that, in principle, the alignment of the minidisc and sub-grid disc with the BH equatorial plane may also affect the subsequent spin evolution. Once accretion becomes locally coherent, the sub-grid disc is expected to be less strongly warped, which in turn modifies the efficiency of the Bardeen--Petterson effect and reduces the rate at which the BH spin aligns with the large-scale angular momentum of the surrounding gas. 

We now consider the spin alignment evolution shown in the right panel. In the absence of feedback, the initially misaligned spin gradually aligns with the angular momentum of the accreting gas via the Bardeen--Petterson effect acting in the sub-grid accretion disc. As a result, all spins converge towards alignment with the $z>0$ axis within $\sim 30$ orbital periods, i.e. $\theta_{\bullet,2} \rightarrow 0$, even when initially oriented in the $z<0$ hemisphere. 

The behaviour is markedly different when feedback is included. After an initial phase of slow alignment, the evolution largely stalls, leaving the spins significantly misaligned until the end of the simulations. As in the left panel, markers indicate the times at which the minidiscs are disrupted and the BHs become quiescent. Once accretion ceases, the alignment process effectively stops.

Two mechanisms may contribute to the delayed alignment observed in the feedback runs. First, as shown in the left panels and suggested by the bottom-right panel of Figure~\ref{fig: MapsDensity}, feedback can reorient the minidisc and sub-grid disc towards the BH equatorial plane, thereby reducing the degree of disc warping and weakening the torque responsible for spin alignment. Second, Figure~\ref{fig: fEdd} shows that feedback strongly suppresses the accretion rate. Since the Bardeen--Petterson torque is mediated by the accretion flow, a lower accretion rate reduces the supply of angular momentum reaching the warp radius, where the coupling between the disc and the BH spin is strongest. This decreases the efficiency of angular momentum exchange between the disc and the spin, leading to a slower alignment process. More quantitatively \citep[see][]{Fiacconi18, Cenci21}, the Bardeen--Petterson alignment timescale can be expressed as a function of the Eddington factor, BH mass, and spin as\footnote{Note that this timescale is independent of the initial spin--disc misalignment, so the spin aligns over the same timescale in all simulations. This stems from the small-warp approximation adopted in the spin evolution model (see Eq.~10 of \citealt{Fiacconi18}). As discussed by \citet{Gerosa20}, this approximation can introduce errors of up to $\sim$50\% in the alignment timescale.}

\begin{equation}
    \tau_\textrm{BP} = 0.17 \cdot \Biggl(\frac{M_{\bullet,2}}{10^6 \mathrm{M}_\odot}\Biggr)^{-2/35}
    \Biggl(\frac{f_\textrm{Edd,2}}{\eta/0.1}\Biggr)^{-32/35}
    \Biggl(\frac{a}{0.7}\Biggr)^{5/7}.
    \label{Eq: tau}
\end{equation}
In combination, both effects act to significantly delay Bardeen--Petterson alignment in the feedback runs.

To disentangle the relative importance of these two effects, we computed the evolution of $\theta_{\bullet,2}$ for an additional idealized configuration. In this experiment, we adopt the same feedback-suppressed accretion rate measured in the feedback simulations, while assuming that the minidiscs and sub-grid accretion discs remain aligned with the $z$-axis, as in the {\labsim{NoFb}} runs. This calculation isolates the effect of the reduced accretion rate and removes the contribution associated with the feedback-driven reorientation of the gas towards the BH equatorial plane. More quantitatively, in this scenario the evolution of $\theta_{\bullet,2}$ is described by
\begin{equation}
\dot{\theta}_{\bullet,2} = - \frac{\cos(\pi/7)}{\tau_\textrm{BP}}\, \bigl|\sin(\theta_{\bullet,2})\bigr|,
\label{Eq: align}
\end{equation}
where Eq.~\ref{Eq: align} is derived from Eqs.~(8)--(10) of \cite{Cenci21}, i.e. the spin evolution model adopted in our sub-grid prescription, under the assumptions that $J_\bullet/J_\alpha \ll 1$, $|\dot{\mathbf{J}}_\mathrm{in/out}|/J_\alpha \ll 1$, and $\dot{J}_\bullet \ll 1$, where $\dot{\mathbf{J}}_\mathrm{in/out}$ denotes the net change in sub-grid disc angular momentum due to resolved gas inflow and wind ejection.

By integrating Eq.~\ref{Eq: align} using the $f_\textrm{Edd}(t)$ history measured in the feedback simulations, we obtain the expected spin evolution in the idealized regime in which both the outer regions of the sub-grid discs and the minidiscs remain coplanar with the CBD. The resulting evolution is shown by the dashed curves in the right panel. Remarkably, these curves closely reproduce the full feedback simulation results. This indicates that the reduction in accretion rate alone is sufficient to account for most of the delayed spin alignment observed in the feedback runs.

\subsection{Duty cycle}
\begin{figure*}
\includegraphics[width=1\linewidth]{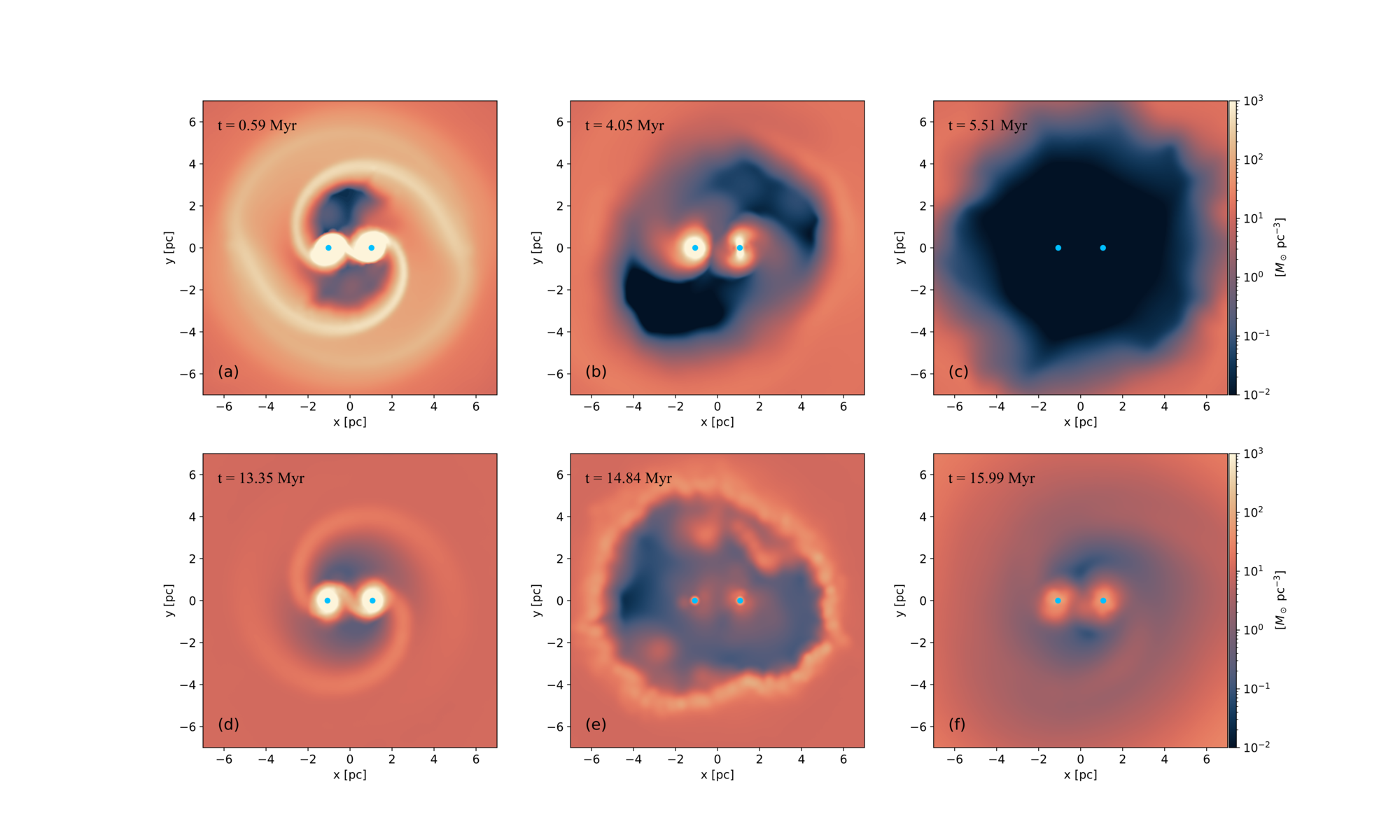}
    \caption{Face-on density maps of the prolonged {\labsim{a5pi12}} run at representative evolutionary stages. The snapshots correspond to the times marked by crosses in Figure~\ref{fig: SpinDuty}.}
    \label{fig: MapDuty}
\end{figure*}
\begin{figure}
\hspace{-0.1\linewidth}
\includegraphics[width=1.1\linewidth]{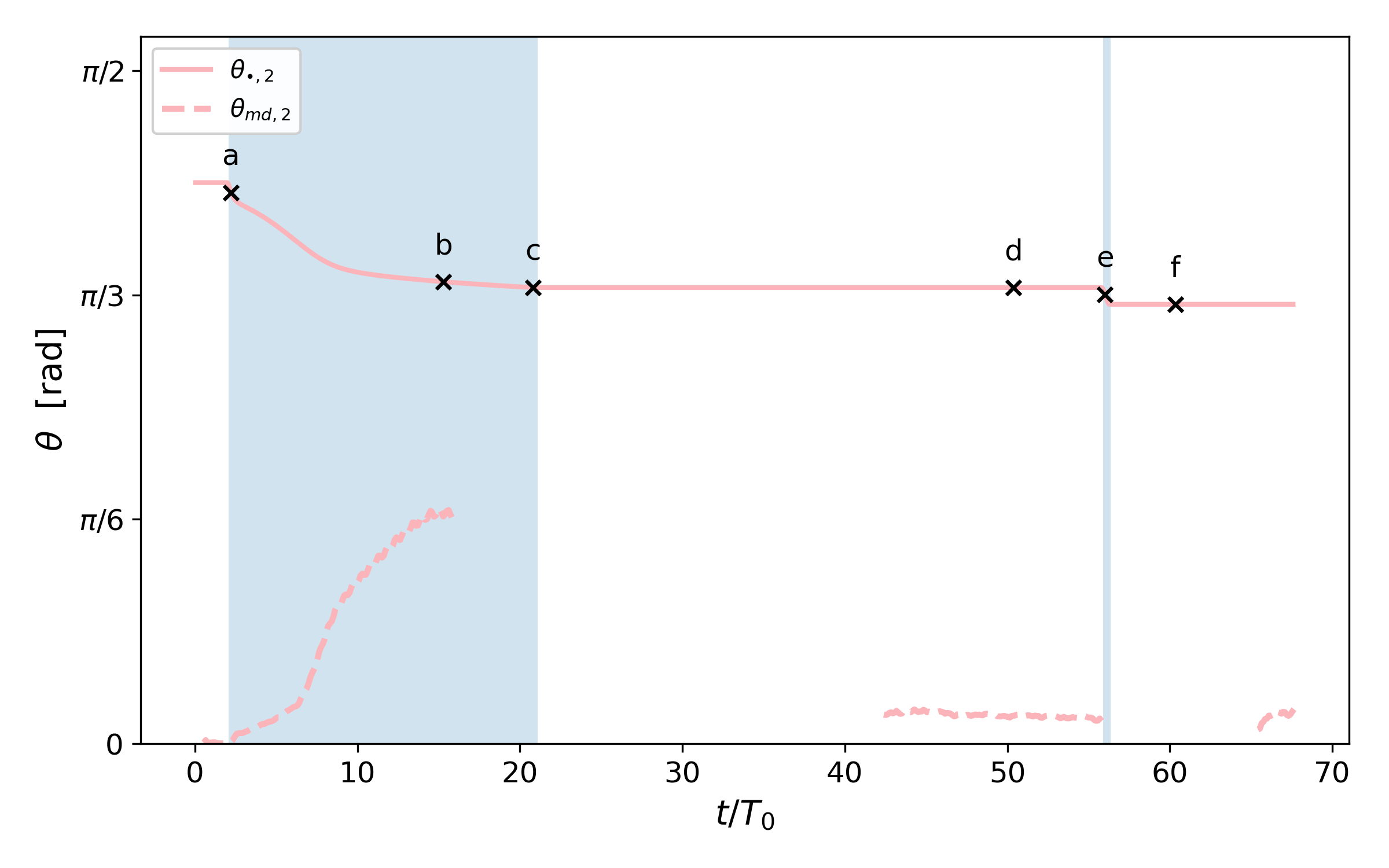}
    \caption{Time evolution of the secondary spin angle (solid line) and secondary minidisc angle (dashed) with respect to the $z>0$-axis for the prolonged {\labsim{a5pi12}} run. The minidisc curve is shown only when the minidisc is resolved with at least 500 particles, corresponding to a mass of $\sim 100\,\mathrm{M}_\odot$ or higher. Blue shaded regions indicate phases in which the secondary BH is active, and crosses mark the snapshots shown in Figure~\ref{fig: MapDuty}.}
    \label{fig: SpinDuty}
\end{figure}
To investigate the long-term evolution of the system, we extend one of our simulations, {\labsim{a5pi12}}, up to 70 initial orbital periods, corresponding to $\sim 19$ Myr. 
Figure~\ref{fig: MapDuty} shows face-on density maps at representative stages of this extended run. The first row illustrates the three main phases already discussed: (a) the formation of minidiscs and the onset of active accretion onto the BHs, (b) the development of a feedback-driven cavity accompanied by the tilting of the secondary minidisc, and (c) the subsequent depletion of the minidiscs, leaving both BHs in a quiescent state within a low-density cavity.
The second row highlights the re-emergence of gas at later times. In phase (d), gas refills the cavity and reforms minidiscs around the BHs. This is followed by a  new episode of AGN activity, accompanied by a feedback phase (e), which rapidly expels the surrounding gas, again disrupting the minidiscs and re-establishing a cleared cavity. Finally, panel (f) shows the system approaching the onset of another re-accretion episode, with mini-discs about to reform.
Overall, Figure~\ref{fig: MapDuty} suggests the presence of a duty cycle characterised by alternating active and quiescent phases. This cyclic behaviour regulates the gas-driven dynamical evolution of the binary and modulates both the accretion history and the associated spin evolution of the BHs.

In Figure~\ref{fig: SpinDuty} we show the evolution of the secondary BH spin angle, $\theta_{\bullet,2}$ (solid line), together with the secondary minidisc tilt angle, $\theta_{\mathrm{md},2}$ (dashed line), for the extended {\labsim{a5pi12}} run. The minidisc angle is reported only when the structure is resolved with more than 500 particles within a distance 0.5 pc from the BH. The shaded regions indicate intervals during which the secondary BH is active, while crosses mark the times corresponding to the snapshots shown in Figure~\ref{fig: MapDuty}.

Figure~\ref{fig: SpinDuty} shows that the system spends most of the time in a quiescent state. Feedback-driven cavity formation during active phases leads to long-lived low-density regions that delay the re-establishment of accretion, thereby prolonging the inactive phases. Since spin alignment can proceed only during active accretion phases, this extended quiescence further suppresses the overall alignment process. However, the relative duration of active and inactive phases is likely sensitive to the adopted criterion for AGN re-triggering (see Section~\ref{sec: accretion}).
As a consequence, BH quiescence induced by feedback emerges as the dominant mechanism hindering spin alignment, in addition to the reduced alignment efficiency already present during active phases due to the feedback-suppressed accretion rate (see Section~\ref{sec: spin}). 
The relative duration of the active and quiescent phases is also expected to depend on the gas feeding history. For example, strong, long-lasting, low-angular-momentum inflows that continuously refuel the circumbinary disc could reduce the quiescent intervals, thereby increasing the efficiency of spin alignment.

We also note that minidisc tilting induced by feedback occurs only during active phases, and only when these phases are sufficiently long, as in the first AGN episode of the simulation. In later cycles, newly formed minidiscs again align with the binary orbital plane, remaining quasi-perpendicular to the $z$-axis, suggesting that feedback from previous active episodes does not significantly affect the angular momentum orientation of the gas feeding subsequent minidiscs. This indicates that the impact of feedback-driven reorientation is largely confined to the immediate minidisc region.

\section{Summary and discussion} \label{sec: conclusion}

In this paper, we investigated the role of the spin-dependent anisotropy of AGN winds in shaping the evolution of MBH binaries embedded in CBDs. To this end, we employed the sub-grid model for AGN feedback, MBH accretion, and spin evolution implemented in the \textsc{gizmo} code by \cite{Bollati24}. We considered initially prograde and coplanar binary--CBD configurations in the cavity regime, according to the criterion of \cite{DelValle14}, and studied how the binary dynamics, BH accretion, and spin evolution are modified by anisotropic AGN feedback over a period of approximately 30 binary orbits. To assess the importance of the spin dependence of AGN feedback, we explored different spin configurations, varying both the spin magnitudes and inclinations, as summarised in Table~\ref{tab: spin}.

The main results of this work can be summarised as follows:
\begin{itemize}
\item AGN winds launched from the unresolved accretion discs interact with the surrounding gas in both the CBD and the minidiscs, significantly modifying their structure. As previously noted by \cite{DelValle18}, AGN feedback helps maintain the central cavity, already present in our initial conditions, and enlarges it slightly. The extent of this enlargement depends on the coupling between the feedback and the surrounding gas. In particular, misaligned spins produce winds that are partially collimated towards the CBD, leading to a more efficient coupling between the AGN outflows and the CBD and consequently to a larger cavity (Fig.~\ref{fig: MapsDensity}, top row). Moreover, after approximately $20$--$25$ orbits, the minidiscs are destroyed in all simulations including feedback. These results demonstrate that, on parsec scales, AGN feedback has a substantial impact on both the CBD and the minidiscs and therefore cannot be treated as a process coupling exclusively at larger galactic scales while leaving the circumbinary environment unaffected.

\item The impact of AGN feedback on the binary dynamics is significant. In the absence of feedback, the adopted disc viscosity and aspect ratio lead to binary expansion, in agreement with previous studies \citep{Heath20, Duffell20, Franchini22, Duffell24}. When AGN feedback is included, however, the larger cavity excavated by the outflows suppresses the positive torques responsible for the expansion, causing the orbital evolution to stall, consistently with the findings of \cite{DelValle18}. Although different spin configurations modify the CBD through different feedback efficiencies, these variations have only a minor impact on the orbital evolution (Figs.~\ref{fig: Orbital} and \ref{fig: torques}). This is because our initial conditions place the system in the cavity regime, where the coupling between winds and the CBD is already intrinsically weak, limiting the impact of differences in feedback efficiency and collimation.

\item AGN feedback has a dramatic impact on the MBH accretion rates. Without feedback, both BHs accrete at an approximately constant Eddington ratio of $f_\mathrm{Edd}\sim0.05$. In contrast, when feedback is included, the accretion rates decrease by approximately one to two orders of magnitude before eventually shutting off after $\sim20$--$25$ orbits, leaving both BHs quiescent inside the feedback-excavated cavity (Fig.~\ref{fig: fEdd}). This behaviour highlights the self-regulated nature of the feeding--feedback cycle: once feedback is accounted for, the MBH accretion rates and hence the AGN luminosities are substantially reduced, with important implications for the detectability of MBH binaries through their electromagnetic counterparts \citep{Bogdanovic22}.

\item The evolution of the BH spins is strongly affected by AGN feedback (Fig.~\ref{fig: spin}, right panel). The feedback-induced suppression of accretion dramatically reduces the efficiency of spin alignment. In the absence of feedback, Bardeen--Petterson effect aligns the spin of the secondary BH  within $\sim 30$ orbits with the angular momentum of the accreting gas, which coincides with that of the CBD and the minidiscs. This alignment is driven by the unresolved Bardeen--Petterson effect in the sub-grid disc, with a timescale inversely proportional to the accretion rate (Eq.~\ref{Eq: tau}) and governed by gas transport to the warp radius, where gravitomagnetic torques between the disc and BH spin are strongest. As AGN feedback suppresses accretion, it also weakens the coupling between the BH spin and the sub-grid disc, significantly slowing the alignment process. As a result, the BH spins retain most of their initial misalignment throughout the duration of our simulations.

\item While spin alignment is strongly suppressed, the anisotropic feedback launched along the direction of a misaligned spin alters the orientation of the surrounding gas. In particular, before being disrupted, the minidisc is progressively tilted towards the BH equatorial plane, where the interaction between the anisotropic wind and the disc is minimised (Fig.~\ref{fig: MapsDensity}, bottom row; Fig.~\ref{fig: spin}, left panel). This suggests that, on minidisc scales ($\lesssim 1$ pc), BH feeding is forced to align (or anti-align) with the MBH spin by the sole action of anisotropic AGN winds. Incidentally, this mechanism provides a plausible explanation for observations of jets being partially aligned with the axes of nuclear gas distributions, such as sub-pc megamaser discs \citep{Dotti24} and $\sim 100$ pc discs of molecular gas \citep[e.g.][]{Ruffa20}, while showing a lesser degree of alignement with the gas distribution on larger scales \citep[e.g.][]{Pringle99, Greene13}. We intend to investigate this scenario in more detail in future work.

\item Finally, our simulations suggest that the binary undergoes a feedback-regulated duty cycle, analogous to the self-regulated AGN feedback cycles found in previous studies (e.g. \citealt{Cattaneo07, Ciotti10, Gaspari16}), where feedback episodes temporarily suppress the gas supply and are followed by renewed accretion once the gas reservoir is replenished. Extending one of our runs to approximately $70$ binary orbits reveals that, after the initial feedback episode, gas gradually falls back into the cavity, reforming the minidiscs and reigniting accretion and AGN activity. The renewed feedback then expels the gas once again, repeating the cycle (Fig.~\ref{fig: MapDuty}). The duty cycle is dominated by long quiescent phases, during which the binary resides inside a low-density cavity created by previous AGN episodes. Consequently, the binary spends most of its lifetime in an environment where orbital evolution, BH accretion and spin alignment are all substantially suppressed by the effects of AGN feedback. 
This behaviour is likely more relevant for a virialized CBD with limited external gas replenishment. Stronger, long-lasting inflows could instead shorten the quiescent phases, thereby enhancing gas-driven orbital evolution, accretion, and spin alignment.

\end{itemize}
These results have important implications for predictions of MBH spin alignment at merger. Several studies investigated spin alignment during the gas-rich hardening phase of MBH binaries \citep{Bogdanovic07, Miller13, Lodato13, Gerosa15, Gerosa20, Bourne24, Koudmani24}. With the exception of \cite{Bourne24} and \cite{Koudmani24}, these studies are based on semi-analytical models. Among them, \cite{Gerosa20} provides the most complete treatment, including non-linear warp propagation, tidal torques from the companion, differential accretion, and an iterative solution of the coupled evolution equations. Crucially, it introduces a critical obliquity at which steady warped-disc solutions cease to exist, associated with disc breaking \citep{Nealon22}. In this framework, spin evolution can either lead to full alignment or stall at the critical configuration, with initially counter-aligned systems ($\theta>\pi/2$) generically driven toward this regime. This led to the conclusion that disc breaking can strongly suppress spin alignment over a significant portion of parameter space.
Building on this, \cite{Steinle23} showed that LISA may probe both spin alignment efficiency and signatures of disc breaking, although similar misalignments may also arise in gas-poor systems, introducing degeneracies. More recently, \cite{Bourne24} used hydrodynamical simulations and found more efficient alignment, broadly consistent with \cite{Gerosa15}, but without including key effects such as non-linear warps and critical obliquity, which tend to reduce alignment efficiency.  Building on \citet{Bourne24}, \citet{Koudmani24} developed a unified sub-grid accretion model spanning thin-disc and radiatively inefficient regimes. They showed that MBH spin evolution depends strongly on the accretion state, ranging from slow Bardeen--Petterson alignment to fast solid-body precession; these different regimes can significantly affect spin alignment in binaries, either delaying or facilitating alignment.

In this context, our results highlight an additional physical ingredient: AGN feedback. We find that feedback strongly suppresses Bardeen--Petterson spin alignment by reducing the accretion rate that drives the aligning torque. During active phases, accretion is reduced by up to two orders of magnitude, while a feedback-regulated duty cycle produces extended quiescent phases in which alignment is effectively halted. As a result, spin alignment can be strongly inhibited even in gas-rich conditions that would otherwise favour efficient alignment. This suggests that AGN feedback provides an additional pathway to preserve spin misalignment prior to merger, complementary to mechanisms such as disc breaking and accretion-state transitions, which can also modify the efficiency and timescale of spin alignment.

These findings have implications for models predicting MBH binary merger observables, including GW recoil, waveforms, and electromagnetic counterparts. Recent sub-grid and cosmological models have incorporated increasingly complex treatments of gas-driven evolution, accretion, and feedback in this regime (e.g. \citealt{Sayeb21,Liao23,Li25}), but a self-consistent coupling between spin evolution and the gas environment remains largely unexplored.
In this context, our results point to three main effects that may be important for next-generation models. First, AGN feedback modifies the CBD by sustaining and creating a larger central cavity, leading to a stalled binary, thus enhancing the role of other shrinking mechanisms to evolve the binary. Second, the resulting suppression of gas density lowers accretion rates, affecting BH mass growth and leading to intermittent electromagnetic emission as the system cycles between active and quiescent phases.
Third, AGN feedback can strongly suppress Bardeen--Petterson alignment by intermittently reducing or halting accretion, potentially preserving spin misalignments even in otherwise gas-rich environments.

We conclude by noting several limitations of the present work.
\begin{itemize}
\item Our treatment of sub-grid spin evolution inherits the assumptions of the model by \cite{Fiacconi18}. In particular, warp propagation is described in the linear regime\footnote{More precisely, the continuous evolution of the BH spin in \cite{Fiacconi18} is derived by linearising the Bardeen--Petterson torque at first order in the misalignment angle (their Eq.~10). This linear treatment is complemented by a separate, instantaneous (counter-)alignment prescription, applied whenever the black hole mass exceeds a critical threshold  (their Eq.~14) above which the disc cannot sustain a steady warped state; in this regime, $\mathbf{J}_\bullet$ and $\mathbf{J}_\alpha$ are instantaneously (counter-)aligned following the criterion of \citealt{King05}.}, implying that the alignment rate does not depend on the amplitude of the initial misalignment. Likewise, our model does not include the onset of critical obliquity or disc breaking \citep{Nixon12, Gerosa20, Nealon22, Steinle23},  which could significantly suppress spin alignment.
Furthermore, the structure of the warped accretion disc is computed neglecting the tidal torque exerted by the binary companion, which can shift the warp radius inwards and enhance the Bardeen--Petterson alignment torque \citep{Miller13, Gerosa20}\footnote{Following \citet{Gerosa20}, their Eq.~22 defines a dimensionless parameter $\kappa$ quantifying the relative importance of the companion torque compared to the Lense--Thirring precession, with $\kappa \gtrsim 1$ marking the onset of a significant companion contribution to the warp structure. For the black hole masses, spins, and separation considered in this work, we estimate $\kappa \sim 10^{-7}$.}. At the parsec-scale separations considered here, this effect is indeed negligible, and only becomes relevant much closer to the onset of the GW-driven regime.

\item The CBD is modelled using an effective $\alpha$-viscosity prescription \citep{SS73}, representing angular momentum transport through the Navier--Stokes equations. A more realistic treatment would require magnetohydrodynamical simulations, in which the effective viscosity naturally emerges from magnetorotational instability \citep{Balbus91, Noble12}. Since viscosity regulates cavity formation, gas inflow through the cavity, and ultimately accretion onto the individual BHs, resolving these processes may quantitatively alter both the binary evolution and the efficiency of AGN feedback.

\item The AGN feedback prescription is also idealised. Feedback is injected exclusively in kinetic form through the wind-particle spawning scheme, implicitly assuming that radiation transfers its momentum entirely to the unresolved gas before reaching resolved scales. Radiation-hydrodynamic simulations, however, have shown that photoionisation by the AGN can progressively reduce the coupling between radiation and the surrounding gas, allowing gas to flow back towards the binary more efficiently \citep[e.g.][]{dEtigny24}. In our context, such an effect could partially restore gas inflow, increasing both the torques acting on the binary and the accretion responsible for spin alignment.

\item Our accretion and feedback model assumes radiatively efficient, geometrically thin accretion discs throughout the simulations. However, after approximately $10$ orbits, the accretion rates in the feedback runs decrease below $f_\mathrm{Edd}\lesssim10^{-2}$, where accretion is expected to transition to a radiatively inefficient, geometrically thick flow. In this regime, feedback is thought to be dominated by relativistic Blandford--Znajek jets rather than radiatively driven winds \citep[e.g.][]{Husko22, Rennehan23, Koudmani24}. Since jets are generally much more collimated, they would likely couple less efficiently to the circumbinary gas unless their axis lies close to the disc plane. Consequently, the feedback-driven cavity and the associated quiescent phases may be less extended than predicted here. Moreover, spin evolution itself would proceed differently, with substantially longer alignment timescales and concurrent spin-down due to the extraction of rotational energy by the jet \citep{Husko22}.

\item Our simulations neglect the stellar component of the nuclear environment, which can contribute significantly to MBH binary hardening \citep{Bortolas21} and drive inspiral even when gas torques alone lead to stalling or expansion. In addition, stellar feedback \citep{Shin25} and torques \citep{fiacconi13,lupi15b,SouzaLima17} from self-gravitating clumps in the circumbinary disc can alter the gas distribution, introducing variability in the binary evolution, accretion history, and spin evolution.

\item Finally, our conclusions are based on a limited exploration of parameter space, focusing mainly on BH spin configurations within a single binary--disc setup. The results may vary with other binary properties, such as the mass ratio, binary-to-disc mass ratio, eccentricity, and binary--disc inclination \citep{Duffell20,DelValle14,Ragusa20,Moody19}, which regulate the gas distribution, accretion rates, feedback efficiency, and orbital torques. A broader exploration will be needed to assess the generality of our findings.

\end{itemize}

Despite these limitations, our simulations demonstrate that AGN feedback can substantially modify the evolution of MBH binaries on parsec scales primarily by suppressing accretion, and inhibiting spin alignment. Exploring the robustness of these conclusions with more realistic accretion, feedback, and environmental models will be the subject of future work.

\section*{Acknowledgements}
We acknowledge the Leibniz Institute for Astrophysics Potsdam for the availability of high-performance computing resources and support. 

\section*{Data Availability}
The data underlying this article will be shared on reasonable request to the corresponding author.



\bibliographystyle{aa}
\bibliography{example} 

@ARTICLE{Campitiello18,
       author = {{Campitiello}, Samuele and {Ghisellini}, Gabriele and {Sbarrato}, Tullia and {Calderone}, Giorgio},
        title = "{How to constrain mass and spin of supermassive black holes through their disk emission}",
      journal = {\aap},
         year = 2018,
        month = apr,
       volume = {612},
          eid = {A59},
        pages = {A59},
          doi = {10.1051/0004-6361/201731897},
archivePrefix = {arXiv},
       eprint = {1702.00011},
 primaryClass = {astro-ph.HE},
       adsurl = {https://ui.adsabs.harvard.edu/abs/2018A&A...612A..59C}
}

@ARTICLE{Hopkins15,
       author = {{Hopkins}, Philip F.},
        title = "{A new class of accurate, mesh-free hydrodynamic simulation methods}",
      journal = {\mnras},
         year = 2015,
        month = jun,
       volume = {450},
       number = {1},
        pages = {53-110},
          doi = {10.1093/mnras/stv195},
archivePrefix = {arXiv},
       eprint = {1409.7395},
 primaryClass = {astro-ph.CO},
       adsurl = {https://ui.adsabs.harvard.edu/abs/2015MNRAS.450...53H}
}

@ARTICLE{Torrey20,
       author = {{Torrey}, Paul and {Hopkins}, Philip F. and {Faucher-Gigu{\`e}re}, Claude-Andr{\'e} and {Angl{\'e}s-Alc{\'a}zar}, Daniel and {Quataert}, Eliot and {Ma}, Xiangcheng and {Feldmann}, Robert and {Keres}, Dusan and {Murray}, Norm},
        title = "{The impact of AGN wind feedback in simulations of isolated galaxies with a multiphase ISM}",
      journal = {\mnras},
         year = 2020,
        month = oct,
       volume = {497},
       number = {4},
        pages = {5292-5308},
          doi = {10.1093/mnras/staa2222},
       adsurl = {https://ui.adsabs.harvard.edu/abs/2020MNRAS.497.5292T}
}

@ARTICLE{Cenci21,
       author = {{Cenci}, Elia and {Sala}, Luca and {Lupi}, Alessandro and {Capelo}, Pedro R. and {Dotti}, Massimo},
        title = "{Black hole spin evolution in warped accretion discs}",
      journal = {\mnras},
         year = 2021,
        month = jan,
       volume = {500},
       number = {3},
        pages = {3719-3727},
          doi = {10.1093/mnras/staa3449},
archivePrefix = {arXiv},
       eprint = {2011.06596},
 primaryClass = {astro-ph.GA},
       adsurl = {https://ui.adsabs.harvard.edu/abs/2021MNRAS.500.3719C}
}

@ARTICLE{Ishibashi19,
       author = {{Ishibashi}, W. and {Fabian}, A.~C. and {Reynolds}, C.~S.},
        title = "{Radiation pattern and outflow geometry: a new probe of black hole spin?}",
      journal = {\mnras},
         year = 2019,
        month = jun,
       volume = {486},
       number = {2},
        pages = {2210-2214},
          doi = {10.1093/mnras/stz987},
archivePrefix = {arXiv},
       eprint = {1904.03203},
 primaryClass = {astro-ph.GA},
       adsurl = {https://ui.adsabs.harvard.edu/abs/2019MNRAS.486.2210I}
}

@ARTICLE{Fiacconi18,
       author = {{Fiacconi}, Davide and {Sijacki}, Debora and {Pringle}, J.~E.},
        title = "{Galactic nuclei evolution with spinning black holes: method and implementation}",
      journal = {\mnras},
         year = 2018,
        month = jul,
       volume = {477},
       number = {3},
        pages = {3807-3835},
          doi = {10.1093/mnras/sty893},
archivePrefix = {arXiv},
       eprint = {1712.00023},
 primaryClass = {astro-ph.GA},
       adsurl = {https://ui.adsabs.harvard.edu/abs/2018MNRAS.477.3807F}
}

@ARTICLE{SS73,
       author = {{Shakura}, N.~I. and {Sunyaev}, R.~A.},
        title = "{Black holes in binary systems. Observational appearance.}",
      journal = {\aap},
         year = 1973,
        month = jan,
       volume = {24},
        pages = {337-355},
       adsurl = {https://ui.adsabs.harvard.edu/abs/1973A&A....24..337S}
}

@ARTICLE{King05,
       author = {{King}, Andrew},
        title = "{The AGN-Starburst Connection, Galactic Superwinds, and M$_{BH}$-{\ensuremath{\sigma}}}",
      journal = {\apjl},
         year = 2005,
        month = dec,
       volume = {635},
       number = {2},
        pages = {L121-L123},
          doi = {10.1086/499430},
archivePrefix = {arXiv},
       eprint = {astro-ph/0511034},
 primaryClass = {astro-ph},
       adsurl = {https://ui.adsabs.harvard.edu/abs/2005ApJ...635L.121K}
}

@ARTICLE{Bollati23,
       author = {{Bollati}, Francesco and {Lupi}, Alessandro and {Dotti}, Massimo and {Haardt}, Francesco},
        title = "{Dynamical evolution of massive black hole pairs in the presence of spin-dependent radiative feedback}",
      journal = {\mnras},
         year = 2023,
        month = apr,
       volume = {520},
       number = {3},
        pages = {3696-3705},
          doi = {10.1093/mnras/stad329},
archivePrefix = {arXiv},
       eprint = {2212.08669},
 primaryClass = {astro-ph.GA},
       adsurl = {https://ui.adsabs.harvard.edu/abs/2023MNRAS.520.3696B}
}

@ARTICLE{BP75,
       author = {{Bardeen}, James M. and {Petterson}, Jacobus A.},
        title = "{The Lense-Thirring Effect and Accretion Disks around Kerr Black Holes}",
      journal = {\apjl},
         year = 1975,
        month = jan,
       volume = {195},
        pages = {L65},
          doi = {10.1086/181711},

}

@ARTICLE{Haring04,
       author = {{H{\"a}ring}, Nadine and {Rix}, Hans-Walter},
        title = "{On the Black Hole Mass-Bulge Mass Relation}",
      journal = {\apjl},
         year = 2004,
        month = apr,
       volume = {604},
       number = {2},
        pages = {L89-L92},
          doi = {10.1086/383567},
archivePrefix = {arXiv},
       eprint = {astro-ph/0402376},
 primaryClass = {astro-ph},
       adsurl = {https://ui.adsabs.harvard.edu/abs/2004ApJ...604L..89H}
}

@ARTICLE{Magorrian98,
       author = {{Magorrian}, John and {Tremaine}, Scott and {Richstone}, Douglas and {Bender}, Ralf and {Bower}, Gary and {Dressler}, Alan and {Faber}, S.~M. and {Gebhardt}, Karl and {Green}, Richard and {Grillmair}, Carl and {Kormendy}, John and {Lauer}, Tod},
        title = "{The Demography of Massive Dark Objects in Galaxy Centers}",
      journal = {\aj},
         year = 1998,
        month = jun,
       volume = {115},
       number = {6},
        pages = {2285-2305},
          doi = {10.1086/300353},
archivePrefix = {arXiv},
       eprint = {astro-ph/9708072},
 primaryClass = {astro-ph},
       adsurl = {https://ui.adsabs.harvard.edu/abs/1998AJ....115.2285M}
}

@ARTICLE{Ferrarese00,
       author = {{Ferrarese}, Laura and {Merritt}, David},
        title = "{A Fundamental Relation between Supermassive Black Holes and Their Host Galaxies}",
      journal = {\apjl},
         year = 2000,
        month = aug,
       volume = {539},
       number = {1},
        pages = {L9-L12},
          doi = {10.1086/312838},
archivePrefix = {arXiv},
       eprint = {astro-ph/0006053},
 primaryClass = {astro-ph},
       adsurl = {https://ui.adsabs.harvard.edu/abs/2000ApJ...539L...9F}
}

@ARTICLE{Fanidakis11,
       author = {{Fanidakis}, N. and {Baugh}, C.~M. and {Benson}, A.~J. and {Bower}, R.~G. and {Cole}, S. and {Done}, C. and {Frenk}, C.~S.},
        title = "{Grand unification of AGN activity in the {\ensuremath{\Lambda}}CDM cosmology}",
      journal = {\mnras},
         year = 2011,
        month = jan,
       volume = {410},
       number = {1},
        pages = {53-74},
          doi = {10.1111/j.1365-2966.2010.17427.x},
archivePrefix = {arXiv},
       eprint = {0911.1128},
 primaryClass = {astro-ph.CO},
       adsurl = {https://ui.adsabs.harvard.edu/abs/2011MNRAS.410...53F}
}

@ARTICLE{Bustamante19,
       author = {{Bustamante}, Sebastian and {Springel}, Volker},
        title = "{Spin evolution and feedback of supermassive black holes in cosmological simulations}",
      journal = {\mnras},
         year = 2019,
        month = dec,
       volume = {490},
       number = {3},
        pages = {4133-4153},
          doi = {10.1093/mnras/stz2836},
archivePrefix = {arXiv},
       eprint = {1902.04651},
 primaryClass = {astro-ph.GA},
       adsurl = {https://ui.adsabs.harvard.edu/abs/2019MNRAS.490.4133B}
}

@ARTICLE{Berti08,
       author = {{Berti}, Emanuele and {Volonteri}, Marta},
        title = "{Cosmological Black Hole Spin Evolution by Mergers and Accretion}",
      journal = {\apj},
         year = 2008,
        month = sep,
       volume = {684},
       number = {2},
        pages = {822-828},
          doi = {10.1086/590379},
archivePrefix = {arXiv},
       eprint = {0802.0025},
 primaryClass = {astro-ph},
       adsurl = {https://ui.adsabs.harvard.edu/abs/2008ApJ...684..822B}
}

@ARTICLE{Sesana14,
       author = {{Sesana}, A. and {Barausse}, E. and {Dotti}, M. and {Rossi}, E.~M.},
        title = "{Linking the Spin Evolution of Massive Black Holes to Galaxy Kinematics}",
      journal = {\apj},
         year = 2014,
        month = oct,
       volume = {794},
       number = {2},
          eid = {104},
        pages = {104},
          doi = {10.1088/0004-637X/794/2/104},
archivePrefix = {arXiv},
       eprint = {1402.7088},
 primaryClass = {astro-ph.CO},
       adsurl = {https://ui.adsabs.harvard.edu/abs/2014ApJ...794..104S}
}

@ARTICLE{Dubois21,
       author = {{Dubois}, Yohan and {Beckmann}, Ricarda and {Bournaud}, Fr{\'e}d{\'e}ric and {Choi}, Hoseung and {Devriendt}, Julien and {Jackson}, Ryan and {Kaviraj}, Sugata and {Kimm}, Taysun and {Kraljic}, Katarina and {Laigle}, Clotilde and {Martin}, Garreth and {Park}, Min-Jung and {Peirani}, S{\'e}bastien and {Pichon}, Christophe and {Volonteri}, Marta and {Yi}, Sukyoung K.},
        title = "{Introducing the NEWHORIZON simulation: Galaxy properties with resolved internal dynamics across cosmic time}",
      journal = {\aap},
         year = 2021,
        month = jul,
       volume = {651},
          eid = {A109},
        pages = {A109},
          doi = {10.1051/0004-6361/202039429},
archivePrefix = {arXiv},
       eprint = {2009.10578},
 primaryClass = {astro-ph.GA},
       adsurl = {https://ui.adsabs.harvard.edu/abs/2021A&A...651A.109D}
}

@ARTICLE{Koremendy13,
       author = {{Kormendy}, John and {Ho}, Luis C.},
        title = "{Coevolution (Or Not) of Supermassive Black Holes and Host Galaxies}",
      journal = {\araa},
         year = 2013,
        month = aug,
       volume = {51},
       number = {1},
        pages = {511-653},
          doi = {10.1146/annurev-astro-082708-101811},
archivePrefix = {arXiv},
       eprint = {1304.7762},
 primaryClass = {astro-ph.CO},
       adsurl = {https://ui.adsabs.harvard.edu/abs/2013ARA&A..51..511K}
}

@ARTICLE{Volonteri05,
       author = {{Volonteri}, Marta and {Madau}, Piero and {Quataert}, Eliot and {Rees}, Martin J.},
        title = "{The Distribution and Cosmic Evolution of Massive Black Hole Spins}",
      journal = {\apj},
         year = 2005,
        month = feb,
       volume = {620},
       number = {1},
        pages = {69-77},
          doi = {10.1086/426858},
archivePrefix = {arXiv},
       eprint = {astro-ph/0410342},
 primaryClass = {astro-ph},
       adsurl = {https://ui.adsabs.harvard.edu/abs/2005ApJ...620...69V}
}

@ARTICLE{Barausse12,
       author = {{Barausse}, Enrico},
        title = "{The evolution of massive black holes and their spins in their galactic hosts}",
      journal = {\mnras},
         year = 2012,
        month = jul,
       volume = {423},
       number = {3},
        pages = {2533-2557},
          doi = {10.1111/j.1365-2966.2012.21057.x},
archivePrefix = {arXiv},
       eprint = {1201.5888},
 primaryClass = {astro-ph.CO},
       adsurl = {https://ui.adsabs.harvard.edu/abs/2012MNRAS.423.2533B}
}

@ARTICLE{DelValle18,
       author = {{del Valle}, Luciano and {Volonteri}, Marta},
        title = "{The effect of AGN feedback on the migration time-scale of supermassive black holes binaries}",
      journal = {\mnras},
         year = 2018,
        month = oct,
       volume = {480},
       number = {1},
        pages = {439-450},
          doi = {10.1093/mnras/sty1815},
archivePrefix = {arXiv},
       eprint = {1807.03844},
 primaryClass = {astro-ph.GA},
       adsurl = {https://ui.adsabs.harvard.edu/abs/2018MNRAS.480..439D}
}

@ARTICLE{Hennawi06,
       author = {{Hennawi}, Joseph F. and {Strauss}, Michael A. and {Oguri}, Masamune and {Inada}, Naohisa and {Richards}, Gordon T. and {Pindor}, Bartosz and {Schneider}, Donald P. and {Becker}, Robert H. and {Gregg}, Michael D. and {Hall}, Patrick B. and {Johnston}, David E. and {Fan}, Xiaohui and {Burles}, Scott and {Schlegel}, David J. and {Gunn}, James E. and {Lupton}, Robert H. and {Bahcall}, Neta A. and {Brunner}, Robert J. and {Brinkmann}, Jon},
        title = "{Binary Quasars in the Sloan Digital Sky Survey: Evidence for Excess Clustering on Small Scales}",
      journal = {\aj},
         year = 2006,
        month = jan,
       volume = {131},
       number = {1},
        pages = {1-23},
          doi = {10.1086/498235},
archivePrefix = {arXiv},
       eprint = {astro-ph/0504535},
 primaryClass = {astro-ph},
       adsurl = {https://ui.adsabs.harvard.edu/abs/2006AJ....131....1H}
}

@ARTICLE{Foreman09,
       author = {{Foreman}, G. and {Volonteri}, M. and {Dotti}, M.},
        title = "{Double Quasars: Probes of Black Hole Scaling Relationships and Merger Scenarios}",
      journal = {\apj},
         year = 2009,
        month = mar,
       volume = {693},
       number = {2},
        pages = {1554-1562},
          doi = {10.1088/0004-637X/693/2/1554},
archivePrefix = {arXiv},
       eprint = {0812.1569},
 primaryClass = {astro-ph},
       adsurl = {https://ui.adsabs.harvard.edu/abs/2009ApJ...693.1554F}
}

@ARTICLE{Zhang21a,
       author = {{Zhang}, Yang-Wei and {Huang}, Yang and {Bai}, Jin-Ming and {Liu}, Xiao-Wei and {Wang}, Jian-guo and {Dong}, Xiao-bo},
        title = "{A Systematic Search for Dual AGNs in Merging Galaxies (Astro-daring): III: Results from the SDSS Spectroscopic Surveys}",
      journal = {\aj},
         year = 2021,
        month = dec,
       volume = {162},
       number = {6},
          eid = {276},
        pages = {276},
          doi = {10.3847/1538-3881/ac1ce7},
archivePrefix = {arXiv},
       eprint = {2111.00636},
 primaryClass = {astro-ph.GA},
       adsurl = {https://ui.adsabs.harvard.edu/abs/2021AJ....162..276Z}
}

@ARTICLE{Zhang21b,
       author = {{Zhang}, Yang-Wei and {Huang}, Yang and {Bai}, Jin-Ming and {Liu}, Xiao-Wei and {Wang}, Jian-guo and {Dong}, Xiao-bo},
        title = "{A Systematic Search for Dual Active Galactic Nuclei in Merging Galaxies (ASTRO-DARING) II: First Results from Long-slit Spectroscopic Observations}",
      journal = {\aj},
         year = 2021,
        month = dec,
       volume = {162},
       number = {6},
          eid = {289},
        pages = {289},
          doi = {10.3847/1538-3881/ac2deb},
archivePrefix = {arXiv},
       eprint = {2111.00635},
 primaryClass = {astro-ph.GA},
       adsurl = {https://ui.adsabs.harvard.edu/abs/2021AJ....162..289Z}
}

@ARTICLE{Silverman20,
       author = {{Silverman}, John D. and {Tang}, Shenli and {Lee}, Khee-Gan and {Hartwig}, Tilman and {Goulding}, Andy and {Strauss}, Michael A. and {Schramm}, Malte and {Ding}, Xuheng and {Riffel}, Rogemar A. and {Fujimoto}, Seiji and {Hikage}, Chiaki and {Imanishi}, Masatoshi and {Iwasawa}, Kazushi and {Jahnke}, Knud and {Kayo}, Issha and {Kashikawa}, Nobunari and {Kawaguchi}, Toshihiro and {Kohno}, Kotaro and {Luo}, Wentao and {Matsuoka}, Yoshiki and {Matsuda}, Yuichi and {Nagao}, Tohru and {Oguri}, Masamune and {Ono}, Yoshiaki and {Onoue}, Masafusa and {Ouchi}, Masami and {Shimasaku}, Kazuhiro and {Suh}, Hyewon and {Suzuki}, Nao and {Taniguchi}, Yoshiaki and {Toba}, Yoshiki and {Ueda}, Yoshihiro and {Yasuda}, Naoki},
        title = "{Dual Supermassive Black Holes at Close Separation Revealed by the Hyper Suprime-Cam Subaru Strategic Program}",
      journal = {\apj},
         year = 2020,
        month = aug,
       volume = {899},
       number = {2},
          eid = {154},
        pages = {154},
          doi = {10.3847/1538-4357/aba4a3},
archivePrefix = {arXiv},
       eprint = {2007.05581},
 primaryClass = {astro-ph.GA},
       adsurl = {https://ui.adsabs.harvard.edu/abs/2020ApJ...899..154S}
}

@ARTICLE{Liu11,
       author = {{Liu}, Xin and {Shen}, Yue and {Strauss}, Michael A. and {Hao}, Lei},
        title = "{Active Galactic Nucleus Pairs from the Sloan Digital Sky Survey. I. The Frequency on \raisebox{-0.5ex}\textasciitilde5-100 kpc Scales}",
      journal = {\apj},
         year = 2011,
        month = aug,
       volume = {737},
       number = {2},
          eid = {101},
        pages = {101},
          doi = {10.1088/0004-637X/737/2/101},
archivePrefix = {arXiv},
       eprint = {1104.0950},
 primaryClass = {astro-ph.CO},
       adsurl = {https://ui.adsabs.harvard.edu/abs/2011ApJ...737..101L}
}

@ARTICLE{Comerford13,
       author = {{Comerford}, Julia M. and {Schluns}, Kyle and {Greene}, Jenny E. and {Cool}, Richard J.},
        title = "{Dual Supermassive Black Hole Candidates in the AGN and Galaxy Evolution Survey}",
      journal = {\apj},
         year = 2013,
        month = nov,
       volume = {777},
       number = {1},
          eid = {64},
        pages = {64},
          doi = {10.1088/0004-637X/777/1/64},
archivePrefix = {arXiv},
       eprint = {1309.2284},
 primaryClass = {astro-ph.CO},
       adsurl = {https://ui.adsabs.harvard.edu/abs/2013ApJ...777...64C}
}

@ARTICLE{Vignali18,
       author = {{Vignali}, C. and {Piconcelli}, E. and {Perna}, M. and {Hennawi}, J. and {Gilli}, R. and {Comastri}, A. and {Zamorani}, G. and {Dotti}, M. and {Mathur}, S.},
        title = "{Probing black hole accretion in quasar pairs at high redshift}",
      journal = {\mnras},
         year = 2018,
        month = jun,
       volume = {477},
       number = {1},
        pages = {780-790},
          doi = {10.1093/mnras/sty682},
archivePrefix = {arXiv},
       eprint = {1803.08508},
 primaryClass = {astro-ph.GA},
       adsurl = {https://ui.adsabs.harvard.edu/abs/2018MNRAS.477..780V}
}

@ARTICLE{Rodriguez06,
       author = {{Rodriguez}, C. and {Taylor}, G.~B. and {Zavala}, R.~T. and {Peck}, A.~B. and {Pollack}, L.~K. and {Romani}, R.~W.},
        title = "{A Compact Supermassive Binary Black Hole System}",
      journal = {\apj},
         year = 2006,
        month = jul,
       volume = {646},
       number = {1},
        pages = {49-60},
          doi = {10.1086/504825},
archivePrefix = {arXiv},
       eprint = {astro-ph/0604042},
 primaryClass = {astro-ph},
       adsurl = {https://ui.adsabs.harvard.edu/abs/2006ApJ...646...49R}
}

@ARTICLE{White91,
       author = {{White}, Simon D.~M. and {Frenk}, Carlos S.},
        title = "{Galaxy Formation through Hierarchical Clustering}",
      journal = {\apj},
         year = 1991,
        month = sep,
       volume = {379},
        pages = {52},
          doi = {10.1086/170483},
       adsurl = {https://ui.adsabs.harvard.edu/abs/1991ApJ...379...52W}
}

@INPROCEEDINGS{Komossa03,
       author = {{Komossa}, Stefanie},
        title = "{Observational evidence for supermassive black hole binaries}",
    booktitle = {The Astrophysics of Gravitational Wave Sources},
         year = 2003,
       editor = {{Centrella}, Joan M.},
       series = {American Institute of Physics Conference Series},
       volume = {686},
        month = oct,
        pages = {161-174},
          doi = {10.1063/1.1629428},
archivePrefix = {arXiv},
       eprint = {astro-ph/0306439},
 primaryClass = {astro-ph},
       adsurl = {https://ui.adsabs.harvard.edu/abs/2003AIPC..686..161K}
}

@ARTICLE{Fabbiano2011,
       author = {{Fabbiano}, G. and {Wang}, Junfeng and {Elvis}, M. and {Risaliti}, G.},
        title = "{A close nuclear black-hole pair in the spiral galaxy NGC3393}",
      journal = {\nat},
         year = 2011,
        month = sep,
       volume = {477},
       number = {7365},
        pages = {431-434},
          doi = {10.1038/nature10364},
archivePrefix = {arXiv},
       eprint = {1109.0483},
 primaryClass = {astro-ph.CO},
       adsurl = {https://ui.adsabs.harvard.edu/abs/2011Natur.477..431F}
}

@ARTICLE{Comerford12,
       author = {{Comerford}, Julia M. and {Gerke}, Brian F. and {Stern}, Daniel and {Cooper}, Michael C. and {Weiner}, Benjamin J. and {Newman}, Jeffrey A. and {Madsen}, Kristin and {Barrows}, R. Scott},
        title = "{Kiloparsec-scale Spatial Offsets in Double-peaked Narrow-line Active Galactic Nuclei. I. Markers for Selection of Compelling Dual Active Galactic Nucleus Candidates}",
      journal = {\apj},
         year = 2012,
        month = jul,
       volume = {753},
       number = {1},
          eid = {42},
        pages = {42},
          doi = {10.1088/0004-637X/753/1/42},
archivePrefix = {arXiv},
       eprint = {1111.2862},
 primaryClass = {astro-ph.CO},
       adsurl = {https://ui.adsabs.harvard.edu/abs/2012ApJ...753...42C}
}

@ARTICLE{Chandrasekar1943,
       author = {{Chandrasekhar}, S.},
        title = "{Dynamical Friction. I. General Considerations: the Coefficient of Dynamical Friction.}",
      journal = {\apj},
         year = 1943,
        month = mar,
       volume = {97},
        pages = {255},
          doi = {10.1086/144517},
       adsurl = {https://ui.adsabs.harvard.edu/abs/1943ApJ....97..255C}
}

@ARTICLE{Colpi2014,
       author = {{Colpi}, Monica},
        title = "{Massive Binary Black Holes in Galactic Nuclei and Their Path to Coalescence}",
      journal = {\ssr},
         year = 2014,
        month = sep,
       volume = {183},
       number = {1-4},
        pages = {189-221},
          doi = {10.1007/s11214-014-0067-1},
archivePrefix = {arXiv},
       eprint = {1407.3102},
 primaryClass = {astro-ph.GA},
       adsurl = {https://ui.adsabs.harvard.edu/abs/2014SSRv..183..189C}
}

@ARTICLE{Amaro23,
       author = {{Amaro-Seoane}, Pau and {Andrews}, Jeff and {Arca Sedda}, Manuel and {Askar}, Abbas and {Baghi}, Quentin and {Balasov}, Razvan and {Bartos}, Imre and {Bavera}, Simone S. and {Bellovary}, Jillian and {Berry}, Christopher P.~L. and {Berti}, Emanuele and {Bianchi}, Stefano and {Blecha}, Laura and {Blondin}, St{\'e}phane and {Bogdanovi{\'c}}, Tamara and {Boissier}, Samuel and {Bonetti}, Matteo and {Bonoli}, Silvia and {Bortolas}, Elisa and {Breivik}, Katelyn and {Capelo}, Pedro R. and {Caramete}, Laurentiu and {Cattorini}, Federico and {Charisi}, Maria and {Chaty}, Sylvain and {Chen}, Xian and {Chru{\'s}li{\'n}ska}, Martyna and {Chua}, Alvin J.~K. and {Church}, Ross and {Colpi}, Monica and {D'Orazio}, Daniel and {Danielski}, Camilla and {Davies}, Melvyn B. and {Dayal}, Pratika and {De Rosa}, Alessandra and {Derdzinski}, Andrea and {Destounis}, Kyriakos and {Dotti}, Massimo and {Dutan}, Ioana and {Dvorkin}, Irina and {Fabj}, Gaia and {Foglizzo}, Thierry and {Ford}, Saavik and {Fouvry}, Jean-Baptiste and {Franchini}, Alessia and {Fragos}, Tassos and {Fryer}, Chris and {Gaspari}, Massimo and {Gerosa}, Davide and {Graziani}, Luca and {Groot}, Paul and {Habouzit}, Melanie and {Haggard}, Daryl and {Haiman}, Zoltan and {Han}, Wen-Biao and {Istrate}, Alina and {Johansson}, Peter H. and {Khan}, Fazeel Mahmood and {Kimpson}, Tomas and {Kokkotas}, Kostas and {Kong}, Albert and {Korol}, Valeriya and {Kremer}, Kyle and {Kupfer}, Thomas and {Lamberts}, Astrid and {Larson}, Shane and {Lau}, Mike and {Liu}, Dongliang and {Lloyd-Ronning}, Nicole and {Lodato}, Giuseppe and {Lupi}, Alessandro and {Ma}, Chung-Pei and {Maccarone}, Tomas and {Mandel}, Ilya and {Mangiagli}, Alberto and {Mapelli}, Michela and {Mathis}, St{\'e}phane and {Mayer}, Lucio and {McGee}, Sean and {McKernan}, Barry and {Miller}, M. Coleman and {Mota}, David F. and {Mumpower}, Matthew and {Nasim}, Syeda S. and {Nelemans}, Gijs and {Noble}, Scott and {Pacucci}, Fabio and {Panessa}, Francesca and {Paschalidis}, Vasileios and {Pfister}, Hugo and {Porquet}, Delphine and {Quenby}, John and {Ricarte}, Angelo and {R{\"o}pke}, Friedrich K. and {Regan}, John and {Rosswog}, Stephan and {Ruiter}, Ashley and {Ruiz}, Milton and {Runnoe}, Jessie and {Schneider}, Raffaella and {Schnittman}, Jeremy and {Secunda}, Amy and {Sesana}, Alberto and {Seto}, Naoki and {Shao}, Lijing and {Shapiro}, Stuart and {Sopuerta}, Carlos and {Stone}, Nicholas C. and {Suvorov}, Arthur and {Tamanini}, Nicola and {Tamfal}, Tomas and {Tauris}, Thomas and {Temmink}, Karel and {Tomsick}, John and {Toonen}, Silvia and {Torres-Orjuela}, Alejandro and {Toscani}, Martina and {Tsokaros}, Antonios and {Unal}, Caner and {V{\'a}zquez-Aceves}, Ver{\'o}nica and {Valiante}, Rosa and {van Putten}, Maurice and {van Roestel}, Jan and {Vignali}, Christian and {Volonteri}, Marta and {Wu}, Kinwah and {Younsi}, Ziri and {Yu}, Shenghua and {Zane}, Silvia and {Zwick}, Lorenz and {Antonini}, Fabio and {Baibhav}, Vishal and {Barausse}, Enrico and {Bonilla Rivera}, Alexander and {Branchesi}, Marica and {Branduardi-Raymont}, Graziella and {Burdge}, Kevin and {Chakraborty}, Srija and {Cuadra}, Jorge and {Dage}, Kristen and {Davis}, Benjamin and {de Mink}, Selma E. and {Decarli}, Roberto and {Doneva}, Daniela and {Escoffier}, Stephanie and {Gandhi}, Poshak and {Haardt}, Francesco and {Lousto}, Carlos O. and {Nissanke}, Samaya and {Nordhaus}, Jason and {O'Shaughnessy}, Richard and {Portegies Zwart}, Simon and {Pound}, Adam and {Schussler}, Fabian and {Sergijenko}, Olga and {Spallicci}, Alessandro and {Vernieri}, Daniele and {Vigna-G{\'o}mez}, Alejandro},
        title = "{Astrophysics with the Laser Interferometer Space Antenna}",
      journal = {Living Reviews in Relativity},
         year = 2023,
        month = dec,
       volume = {26},
       number = {1},
          eid = {2},
        pages = {2},
          doi = {10.1007/s41114-022-00041-y},
archivePrefix = {arXiv},
       eprint = {2203.06016},
 primaryClass = {gr-qc},
       adsurl = {https://ui.adsabs.harvard.edu/abs/2023LRR....26....2A}
}

@ARTICLE{DelValle14,
       author = {{del Valle}, Luciano and {Escala}, Andr{\'e}s},
        title = "{Binary-Disk Interaction. II. Gap-opening Criteria for Unequal-mass Binaries}",
      journal = {\apj},
         year = 2014,
        month = jan,
       volume = {780},
       number = {1},
          eid = {84},
        pages = {84},
          doi = {10.1088/0004-637X/780/1/84},
archivePrefix = {arXiv},
       eprint = {1310.4509},
 primaryClass = {astro-ph.GA},
       adsurl = {https://ui.adsabs.harvard.edu/abs/2014ApJ...780...84D}
}

@ARTICLE{Eracleous12,
       author = {{Eracleous}, Michael and {Boroson}, Todd A. and {Halpern}, Jules P. and {Liu}, Jia},
        title = "{A Large Systematic Search for Close Supermassive Binary and Rapidly Recoiling Black Holes}",
      journal = {\apjs},
         year = 2012,
        month = aug,
       volume = {201},
       number = {2},
          eid = {23},
        pages = {23},
          doi = {10.1088/0067-0049/201/2/23},
archivePrefix = {arXiv},
       eprint = {1106.2952},
 primaryClass = {astro-ph.CO},
       adsurl = {https://ui.adsabs.harvard.edu/abs/2012ApJS..201...23E}
}

@ARTICLE{Tsalmantza2011,
       author = {{Tsalmantza}, P. and {Decarli}, R. and {Dotti}, M. and {Hogg}, David W.},
        title = "{A Systematic Search for Massive Black Hole Binaries in the Sloan Digital Sky Survey Spectroscopic Sample}",
      journal = {\apj},
         year = 2011,
        month = sep,
       volume = {738},
       number = {1},
          eid = {20},
        pages = {20},
          doi = {10.1088/0004-637X/738/1/20},
archivePrefix = {arXiv},
       eprint = {1106.1180},
 primaryClass = {astro-ph.CO},
       adsurl = {https://ui.adsabs.harvard.edu/abs/2011ApJ...738...20T}
}

@ARTICLE{Dotti2022,
       author = {{Dotti}, Massimo and {Bonetti}, Matteo and {D'Orazio}, Daniel J. and {Haiman}, Zolt{\'a}n and {Ho}, Luis C.},
        title = "{Binary black hole signatures in polarized light curves}",
      journal = {\mnras},
         year = 2022,
        month = jan,
       volume = {509},
       number = {1},
        pages = {212-223},
          doi = {10.1093/mnras/stab2893},
archivePrefix = {arXiv},
       eprint = {2103.14652},
 primaryClass = {astro-ph.HE},
       adsurl = {https://ui.adsabs.harvard.edu/abs/2022MNRAS.509..212D}
}

@ARTICLE{Dotti2012,
       author = {{Dotti}, M. and {Sesana}, A. and {Decarli}, R.},
        title = "{Massive Black Hole Binaries: Dynamical Evolution and Observational Signatures}",
      journal = {Advances in Astronomy},
         year = 2012,
        month = jan,
       volume = {2012},
          eid = {940568},
        pages = {940568},
          doi = {10.1155/2012/940568},
archivePrefix = {arXiv},
       eprint = {1111.0664},
 primaryClass = {astro-ph.CO},
       adsurl = {https://ui.adsabs.harvard.edu/abs/2012AdAst2012E...3D}
}

@ARTICLE{Begelman1980,
       author = {{Begelman}, M.~C. and {Blandford}, R.~D. and {Rees}, M.~J.},
        title = "{Massive black hole binaries in active galactic nuclei}",
      journal = {\nat},
         year = 1980,
        month = sep,
       volume = {287},
       number = {5780},
        pages = {307-309},
          doi = {10.1038/287307a0},
       adsurl = {https://ui.adsabs.harvard.edu/abs/1980Natur.287..307B}
}

@ARTICLE{Komossa2008,
       author = {{Komossa}, S. and {Zhou}, H. and {Lu}, H.},
        title = "{A Recoiling Supermassive Black Hole in the Quasar SDSS J092712.65+294344.0?}",
      journal = {\apjl},
         year = 2008,
        month = may,
       volume = {678},
       number = {2},
        pages = {L81},
          doi = {10.1086/588656},
archivePrefix = {arXiv},
       eprint = {0804.4585},
 primaryClass = {astro-ph},
       adsurl = {https://ui.adsabs.harvard.edu/abs/2008ApJ...678L..81K}
}

@ARTICLE{Pfister2017,
       author = {{Pfister}, Hugo and {Lupi}, Alessandro and {Capelo}, Pedro R. and {Volonteri}, Marta and {Bellovary}, Jillian M. and {Dotti}, Massimo},
        title = "{The birth of a supermassive black hole binary}",
      journal = {\mnras},
         year = 2017,
        month = nov,
       volume = {471},
       number = {3},
        pages = {3646-3656},
          doi = {10.1093/mnras/stx1853},
archivePrefix = {arXiv},
       eprint = {1706.04010},
 primaryClass = {astro-ph.GA},
       adsurl = {https://ui.adsabs.harvard.edu/abs/2017MNRAS.471.3646P}
}

@ARTICLE{Mayer07,
       author = {{Mayer}, L. and {Kazantzidis}, S. and {Madau}, P. and {Colpi}, M. and {Quinn}, T. and {Wadsley}, J.},
        title = "{Rapid Formation of Supermassive Black Hole Binaries in Galaxy Mergers with Gas}",
      journal = {Science},
         year = 2007,
        month = jun,
       volume = {316},
       number = {5833},
        pages = {1874},
          doi = {10.1126/science.1141858},
archivePrefix = {arXiv},
       eprint = {0706.1562},
 primaryClass = {astro-ph},
       adsurl = {https://ui.adsabs.harvard.edu/abs/2007Sci...316.1874M}
}

@ARTICLE{Capelo2015,
       author = {{Capelo}, Pedro R. and {Volonteri}, Marta and {Dotti}, Massimo and {Bellovary}, Jillian M. and {Mayer}, Lucio and {Governato}, Fabio},
        title = "{Growth and activity of black holes in galaxy mergers with varying mass ratios}",
      journal = {\mnras},
         year = 2015,
        month = mar,
       volume = {447},
       number = {3},
        pages = {2123-2143},
          doi = {10.1093/mnras/stu2500},
archivePrefix = {arXiv},
       eprint = {1409.0004},
 primaryClass = {astro-ph.GA},
       adsurl = {https://ui.adsabs.harvard.edu/abs/2015MNRAS.447.2123C}
}

@ARTICLE{Hopkins17,
       author = {{Hopkins}, Philip F.},
        title = "{Anisotropic diffusion in mesh-free numerical magnetohydrodynamics}",
      journal = {\mnras},
         year = 2017,
        month = apr,
       volume = {466},
       number = {3},
        pages = {3387-3405},
          doi = {10.1093/mnras/stw3306},
archivePrefix = {arXiv},
       eprint = {1602.07703},
 primaryClass = {astro-ph.IM},
       adsurl = {https://ui.adsabs.harvard.edu/abs/2017MNRAS.466.3387H}
}

@ARTICLE{Balbus91,
       author = {{Balbus}, Steven A. and {Hawley}, John F.},
        title = "{A Powerful Local Shear Instability in Weakly Magnetized Disks. I. Linear Analysis}",
      journal = {\apj},
         year = 1991,
        month = jul,
       volume = {376},
        pages = {214},
          doi = {10.1086/170270},
       adsurl = {https://ui.adsabs.harvard.edu/abs/1991ApJ...376..214B}
}

@ARTICLE{Lupi15,
       author = {{Lupi}, Alessandro and {Haardt}, Francesco and {Dotti}, Massimo},
        title = "{Massive black hole and gas dynamics in galaxy nuclei mergers - I. Numerical implementation}",
      journal = {\mnras},
         year = 2015,
        month = jan,
       volume = {446},
       number = {2},
        pages = {1765-1774},
          doi = {10.1093/mnras/stu2223},
archivePrefix = {arXiv},
       eprint = {1410.0959},
 primaryClass = {astro-ph.GA},
       adsurl = {https://ui.adsabs.harvard.edu/abs/2015MNRAS.446.1765L}
}

@ARTICLE{Ragusa20,
       author = {{Ragusa}, Enrico and {Alexander}, Richard and {Calcino}, Josh and {Hirsh}, Kieran and {Price}, Daniel J.},
        title = "{The evolution of large cavities and disc eccentricity in circumbinary discs}",
      journal = {\mnras},
         year = 2020,
        month = dec,
       volume = {499},
       number = {3},
        pages = {3362-3380},
          doi = {10.1093/mnras/staa2954},
archivePrefix = {arXiv},
       eprint = {2009.10738},
 primaryClass = {astro-ph.EP},
       adsurl = {https://ui.adsabs.harvard.edu/abs/2020MNRAS.499.3362R}
}

@ARTICLE{Bortolas21,
       author = {{Bortolas}, Elisa and {Franchini}, Alessia and {Bonetti}, Matteo and {Sesana}, Alberto},
        title = "{The Competing Effect of Gas and Stars in the Evolution of Massive Black Hole Binaries}",
      journal = {\apjl},
         year = 2021,
        month = sep,
       volume = {918},
       number = {1},
          eid = {L15},
        pages = {L15},
          doi = {10.3847/2041-8213/ac1c0c},
archivePrefix = {arXiv},
       eprint = {2108.13436},
 primaryClass = {astro-ph.HE},
       adsurl = {https://ui.adsabs.harvard.edu/abs/2021ApJ...918L..15B}
}

@ARTICLE{Rennehan23,
       author = {{Rennehan}, Douglas and {Babul}, Arif and {Moa}, Belaid and {Dav{\'e}}, Romeel},
        title = "{Three regimes of black hole feedback}",
      journal = {arXiv e-prints},
         year = 2023,
        month = sep,
          eid = {arXiv:2309.15898},
        pages = {arXiv:2309.15898},
          doi = {10.48550/arXiv.2309.15898},
archivePrefix = {arXiv},
       eprint = {2309.15898},
 primaryClass = {astro-ph.GA},
       adsurl = {https://ui.adsabs.harvard.edu/abs/2023arXiv230915898R}
}

@ARTICLE{Husko22,
       author = {{Hu{\v{s}}ko}, Filip and {Lacey}, Cedric G. and {Schaye}, Joop and {Schaller}, Matthieu and {Nobels}, Folkert S.~J.},
        title = "{Spin-driven jet feedback in idealized simulations of galaxy groups and clusters}",
      journal = {\mnras},
         year = 2022,
        month = nov,
       volume = {516},
       number = {3},
        pages = {3750-3772},
          doi = {10.1093/mnras/stac2278},
archivePrefix = {arXiv},
       eprint = {2206.06402},
 primaryClass = {astro-ph.GA},
       adsurl = {https://ui.adsabs.harvard.edu/abs/2022MNRAS.516.3750H}
}

@ARTICLE{SouzaLima17,
       author = {{Souza Lima}, Rafael and {Mayer}, Lucio and {Capelo}, Pedro R. and {Bellovary}, Jillian M.},
        title = "{The Pairing of Accreting Massive Black Holes in Multiphase Circumnuclear Disks: the Interplay Between Radiative Cooling, Star Formation, and Feedback Processes}",
      journal = {\apj},
         year = 2017,
        month = mar,
       volume = {838},
       number = {1},
          eid = {13},
        pages = {13},
          doi = {10.3847/1538-4357/aa5d19},
archivePrefix = {arXiv},
       eprint = {1610.01600},
 primaryClass = {astro-ph.GA},
       adsurl = {https://ui.adsabs.harvard.edu/abs/2017ApJ...838...13S}
}

@ARTICLE{Bollati24,
       author = {{Bollati}, F. and {Lupi}, A. and {Dotti}, M. and {Haardt}, F.},
        title = "{Exploring the connection between AGN radiative feedback and massive black hole spin}",
      journal = {\aap},
         year = 2024,
        month = oct,
       volume = {690},
          eid = {A194},
        pages = {A194},
          doi = {10.1051/0004-6361/202348538},
archivePrefix = {arXiv},
       eprint = {2311.07576},
 primaryClass = {astro-ph.GA},
       adsurl = {https://ui.adsabs.harvard.edu/abs/2024A&A...690A.194B}
}

@ARTICLE{Hugher03,
       author = {{Hughes}, Scott A. and {Blandford}, Roger D.},
        title = "{Black Hole Mass and Spin Coevolution by Mergers}",
      journal = {\apjl},
         year = 2003,
        month = mar,
       volume = {585},
       number = {2},
        pages = {L101-L104},
          doi = {10.1086/375495},
archivePrefix = {arXiv},
       eprint = {astro-ph/0208484},
 primaryClass = {astro-ph},
       adsurl = {https://ui.adsabs.harvard.edu/abs/2003ApJ...585L.101H}
}

@ARTICLE{Duffell20,
       author = {{Duffell}, Paul C. and {D'Orazio}, Daniel and {Derdzinski}, Andrea and {Haiman}, Zoltan and {MacFadyen}, Andrew and {Rosen}, Anna L. and {Zrake}, Jonathan},
        title = "{Circumbinary Disks: Accretion and Torque as a Function of Mass Ratio and Disk Viscosity}",
      journal = {\apj},
         year = 2020,
        month = sep,
       volume = {901},
       number = {1},
          eid = {25},
        pages = {25},
          doi = {10.3847/1538-4357/abab95},
archivePrefix = {arXiv},
       eprint = {1911.05506},
 primaryClass = {astro-ph.SR},
       adsurl = {https://ui.adsabs.harvard.edu/abs/2020ApJ...901...25D}
}

@ARTICLE{Duffell24,
       author = {{Duffell}, Paul C. and {Dittmann}, Alexander J. and {D'Orazio}, Daniel J. and {Franchini}, Alessia and {Kratter}, Kaitlin M. and {Penzlin}, Anna B.~T. and {Ragusa}, Enrico and {Siwek}, Magdalena and {Tiede}, Christopher and {Wang}, Haiyang and {Zrake}, Jonathan and {Dempsey}, Adam M. and {Haiman}, Zoltan and {Lupi}, Alessandro and {Pirog}, Michal and {Ryan}, Geoffrey},
        title = "{The Santa Barbara Binary{\ensuremath{-}}disk Code Comparison}",
      journal = {\apj},
         year = 2024,
        month = aug,
       volume = {970},
       number = {2},
          eid = {156},
        pages = {156},
          doi = {10.3847/1538-4357/ad5a7e},
archivePrefix = {arXiv},
       eprint = {2402.13039},
 primaryClass = {astro-ph.SR},
       adsurl = {https://ui.adsabs.harvard.edu/abs/2024ApJ...970..156D}
}

@ARTICLE{Franchini22,
       author = {{Franchini}, Alessia and {Lupi}, Alessandro and {Sesana}, Alberto},
        title = "{Resolving Massive Black Hole Binary Evolution via Adaptive Particle Splitting}",
      journal = {\apjl},
         year = 2022,
        month = apr,
       volume = {929},
       number = {1},
          eid = {L13},
        pages = {L13},
          doi = {10.3847/2041-8213/ac63a2},
archivePrefix = {arXiv},
       eprint = {2201.05619},
 primaryClass = {astro-ph.HE},
       adsurl = {https://ui.adsabs.harvard.edu/abs/2022ApJ...929L..13F}
}

@ARTICLE{Heath20,
       author = {{Heath}, R.~M. and {Nixon}, C.~J.},
        title = "{On the orbital evolution of binaries with circumbinary discs}",
      journal = {\aap},
         year = 2020,
        month = sep,
       volume = {641},
          eid = {A64},
        pages = {A64},
          doi = {10.1051/0004-6361/202038548},
archivePrefix = {arXiv},
       eprint = {2007.11592},
 primaryClass = {astro-ph.HE},
       adsurl = {https://ui.adsabs.harvard.edu/abs/2020A&A...641A..64H}
}

@ARTICLE{Amaroseoane17,
       author = {{Amaro-Seoane}, Pau and {Audley}, Heather and {Babak}, Stanislav and {Baker}, John and {Barausse}, Enrico and {Bender}, Peter and {Berti}, Emanuele and {Binetruy}, Pierre and {Born}, Michael and {Bortoluzzi}, Daniele and {Camp}, Jordan and {Caprini}, Chiara and {Cardoso}, Vitor and {Colpi}, Monica and {Conklin}, John and {Cornish}, Neil and {Cutler}, Curt and {Danzmann}, Karsten and {Dolesi}, Rita and {Ferraioli}, Luigi and {Ferroni}, Valerio and {Fitzsimons}, Ewan and {Gair}, Jonathan and {Gesa Bote}, Lluis and {Giardini}, Domenico and {Gibert}, Ferran and {Grimani}, Catia and {Halloin}, Hubert and {Heinzel}, Gerhard and {Hertog}, Thomas and {Hewitson}, Martin and {Holley-Bockelmann}, Kelly and {Hollington}, Daniel and {Hueller}, Mauro and {Inchauspe}, Henri and {Jetzer}, Philippe and {Karnesis}, Nikos and {Killow}, Christian and {Klein}, Antoine and {Klipstein}, Bill and {Korsakova}, Natalia and {Larson}, Shane L and {Livas}, Jeffrey and {Lloro}, Ivan and {Man}, Nary and {Mance}, Davor and {Martino}, Joseph and {Mateos}, Ignacio and {McKenzie}, Kirk and {McWilliams}, Sean T and {Miller}, Cole and {Mueller}, Guido and {Nardini}, Germano and {Nelemans}, Gijs and {Nofrarias}, Miquel and {Petiteau}, Antoine and {Pivato}, Paolo and {Plagnol}, Eric and {Porter}, Ed and {Reiche}, Jens and {Robertson}, David and {Robertson}, Norna and {Rossi}, Elena and {Russano}, Giuliana and {Schutz}, Bernard and {Sesana}, Alberto and {Shoemaker}, David and {Slutsky}, Jacob and {Sopuerta}, Carlos F. and {Sumner}, Tim and {Tamanini}, Nicola and {Thorpe}, Ira and {Troebs}, Michael and {Vallisneri}, Michele and {Vecchio}, Alberto and {Vetrugno}, Daniele and {Vitale}, Stefano and {Volonteri}, Marta and {Wanner}, Gudrun and {Ward}, Harry and {Wass}, Peter and {Weber}, William and {Ziemer}, John and {Zweifel}, Peter},
        title = "{Laser Interferometer Space Antenna}",
      journal = {arXiv e-prints},
         year = 2017,
        month = feb,
          eid = {arXiv:1702.00786},
        pages = {arXiv:1702.00786},
          doi = {10.48550/arXiv.1702.00786},
archivePrefix = {arXiv},
       eprint = {1702.00786},
 primaryClass = {astro-ph.IM},
       adsurl = {https://ui.adsabs.harvard.edu/abs/2017arXiv170200786A}
}

@ARTICLE{Gultekin09,
       author = {{G{\"u}ltekin}, Kayhan and {Richstone}, Douglas O. and {Gebhardt}, Karl and {Lauer}, Tod R. and {Tremaine}, Scott and {Aller}, M.~C. and {Bender}, Ralf and {Dressler}, Alan and {Faber}, S.~M. and {Filippenko}, Alexei V. and {Green}, Richard and {Ho}, Luis C. and {Kormendy}, John and {Magorrian}, John and {Pinkney}, Jason and {Siopis}, Christos},
        title = "{The M-{\ensuremath{\sigma}} and M-L Relations in Galactic Bulges, and Determinations of Their Intrinsic Scatter}",
      journal = {\apj},
         year = 2009,
        month = jun,
       volume = {698},
       number = {1},
        pages = {198-221},
          doi = {10.1088/0004-637X/698/1/198},
archivePrefix = {arXiv},
       eprint = {0903.4897},
 primaryClass = {astro-ph.GA},
       adsurl = {https://ui.adsabs.harvard.edu/abs/2009ApJ...698..198G}
}

@ARTICLE{Husko22m,
       author = {{Hu{\v{s}}ko}, Filip and {Lacey}, Cedric G. and {Baugh}, Carlton M.},
        title = "{Statistics of galaxy mergers: bridging the gap between theory and observation}",
      journal = {\mnras},
         year = 2022,
        month = feb,
       volume = {509},
       number = {4},
        pages = {5918-5937},
          doi = {10.1093/mnras/stab3324},
archivePrefix = {arXiv},
       eprint = {2107.05601},
 primaryClass = {astro-ph.GA},
       adsurl = {https://ui.adsabs.harvard.edu/abs/2022MNRAS.509.5918H}
}

@ARTICLE{Mannucci22,
       author = {{Mannucci}, F. and {Pancino}, E. and {Belfiore}, F. and {Cicone}, C. and {Ciurlo}, A. and {Cresci}, G. and {Lusso}, E. and {Marasco}, A. and {Marconi}, A. and {Nardini}, E. and {Pinna}, E. and {Severgnini}, P. and {Saracco}, P. and {Tozzi}, G. and {Yeh}, S.},
        title = "{Unveiling the population of dual and lensed active galactic nuclei at sub-arcsec separations}",
      journal = {Nature Astronomy},
         year = 2022,
        month = aug,
       volume = {6},
        pages = {1185-1192},
          doi = {10.1038/s41550-022-01761-5},
archivePrefix = {arXiv},
       eprint = {2203.11234},
 primaryClass = {astro-ph.GA},
       adsurl = {https://ui.adsabs.harvard.edu/abs/2022NatAs...6.1185M}
}

@ARTICLE{Scialpi26,
       author = {{Scialpi}, M. and {Mannucci}, F. and {D'Amato}, Q. and {Marconcini}, C. and {Cresci}, G. and {Marconi}, A. and {Ulivi}, L. and {Fumagalli}, M. and {Rosati}, P. and {Tozzi}, G. and {Zanchettin}, M.~V. and {Battistini}, L. and {Bertola}, E. and {Bracci}, C. and {Carniani}, S. and {Cataldi}, E. and {Ceci}, M. and {Chakraborty}, A. and {Cicone}, C. and {Ciurlo}, A. and {De Rosa}, A. and {Di Rosa}, G. and {Feltre}, A. and {Ginolfi}, M. and {Lamperti}, I. and {Lusso}, E. and {Moreschini}, B. and {Nardini}, E. and {Parvatikar}, M. and {Perna}, M. and {Rubinur}, K. and {Severgnini}, P. and {Singh}, J. and {Spingola}, C. and {Venturi}, G. and {Vignali}, C. and {Volonteri}, M.},
        title = "{Cosmic Duets: I. High-spatial resolution spectroscopy of dual and lensed active galactic nuclei with MUSE-NFM}",
      journal = {\aap},
         year = 2026,
        month = may,
       volume = {709},
          eid = {A236},
        pages = {A236},
          doi = {10.1051/0004-6361/202558548},
archivePrefix = {arXiv},
       eprint = {2512.11960},
 primaryClass = {astro-ph.GA},
       adsurl = {https://ui.adsabs.harvard.edu/abs/2026A&A...709A.236S}
}

@ARTICLE{Chen22,
       author = {{Chen}, Yu-Ching and {Hwang}, Hsiang-Chih and {Shen}, Yue and {Liu}, Xin and {Zakamska}, Nadia L. and {Yang}, Qian and {Li}, Jennifer I.},
        title = "{Varstrometry for Off-nucleus and Dual Subkiloparsec AGN (VODKA): Hubble Space Telescope Discovers Double Quasars}",
      journal = {\apj},
         year = 2022,
        month = feb,
       volume = {925},
       number = {2},
          eid = {162},
        pages = {162},
          doi = {10.3847/1538-4357/ac401b},
archivePrefix = {arXiv},
       eprint = {2108.01672},
 primaryClass = {astro-ph.HE},
       adsurl = {https://ui.adsabs.harvard.edu/abs/2022ApJ...925..162C}
}

@ARTICLE{DeRosa19,
       author = {{De Rosa}, Alessandra and {Vignali}, Cristian and {Bogdanovi{\'c}}, Tamara and {Capelo}, Pedro R. and {Charisi}, Maria and {Dotti}, Massimo and {Husemann}, Bernd and {Lusso}, Elisabeta and {Mayer}, Lucio and {Paragi}, Zsolt and {Runnoe}, Jessie and {Sesana}, Alberto and {Steinborn}, Lisa and {Bianchi}, Stefano and {Colpi}, Monica and {del Valle}, Luciano and {Frey}, S{\'a}ndor and {Gab{\'a}nyi}, Krisztina {\'E}. and {Giustini}, Margherita and {Guainazzi}, Matteo and {Haiman}, Zoltan and {Herrera Ruiz}, Noelia and {Herrero-Illana}, Rub{\'e}n and {Iwasawa}, Kazushi and {Komossa}, S. and {Lena}, Davide and {Loiseau}, Nora and {Perez-Torres}, Miguel and {Piconcelli}, Enrico and {Volonteri}, Marta},
        title = "{The quest for dual and binary supermassive black holes: A multi-messenger view}",
      journal = {\nar},
         year = 2019,
        month = dec,
       volume = {86},
          eid = {101525},
        pages = {101525},
          doi = {10.1016/j.newar.2020.101525},
archivePrefix = {arXiv},
       eprint = {2001.06293},
 primaryClass = {astro-ph.GA},
       adsurl = {https://ui.adsabs.harvard.edu/abs/2019NewAR..8601525D}
}

@ARTICLE{Sillanpaa88,
       author = {{Sillanpaa}, A. and {Haarala}, S. and {Valtonen}, M.~J. and {Sundelius}, B. and {Byrd}, G.~G.},
        title = "{OJ 287: Binary Pair of Supermassive Black Holes}",
      journal = {\apj},
         year = 1988,
        month = feb,
       volume = {325},
        pages = {628},
          doi = {10.1086/166033},
       adsurl = {https://ui.adsabs.harvard.edu/abs/1988ApJ...325..628S}
}

@ARTICLE{Graham15,
       author = {{Graham}, Matthew J. and {Djorgovski}, S.~G. and {Stern}, Daniel and {Glikman}, Eilat and {Drake}, Andrew J. and {Mahabal}, Ashish A. and {Donalek}, Ciro and {Larson}, Steve and {Christensen}, Eric},
        title = "{A possible close supermassive black-hole binary in a quasar with optical periodicity}",
      journal = {\nat},
         year = 2015,
        month = feb,
       volume = {518},
       number = {7537},
        pages = {74-76},
          doi = {10.1038/nature14143},
archivePrefix = {arXiv},
       eprint = {1501.01375},
 primaryClass = {astro-ph.GA},
       adsurl = {https://ui.adsabs.harvard.edu/abs/2015Natur.518...74G}
}

@ARTICLE{Chen20,
       author = {{Chen}, Yu-Ching and {Liu}, Xin and {Liao}, Wei-Ting and {Holgado}, A. Miguel and {Guo}, Hengxiao and {Gruendl}, Robert A. and {Morganson}, Eric and {Shen}, Yue and {Zhang}, Kaiwen and {Abbott}, Tim M.~C. and {Aguena}, Michel and {Allam}, Sahar and {Avila}, Santiago and {Bertin}, Emmanuel and {Bhargava}, Sunayana and {Brooks}, David and {Burke}, David L. and {Carnero Rosell}, Aurelio and {Carollo}, Daniela and {Carrasco Kind}, Matias and {Carretero}, Jorge and {Costanzi}, Matteo and {da Costa}, Luiz N. and {Davis}, Tamara M. and {De Vicente}, Juan and {Desai}, Shantanu and {Diehl}, H. Thomas and {Doel}, Peter and {Everett}, Spencer and {Flaugher}, Brenna and {Friedel}, Douglas and {Frieman}, Joshua and {Garc{\'\i}a-Bellido}, Juan and {Gaztanaga}, Enrique and {Glazebrook}, Karl and {Gruen}, Daniel and {Gutierrez}, Gaston and {Hinton}, Samuel R. and {Hollowood}, Devon L. and {James}, David J. and {Kim}, Alex G. and {Kuehn}, Kyler and {Kuropatkin}, Nikolay and {Lewis}, Geraint F. and {Lidman}, Christopher and {Lima}, Marcos and {Maia}, Marcio A.~G. and {March}, Marisa and {Marshall}, Jennifer L. and {Menanteau}, Felipe and {Miquel}, Ramon and {Palmese}, Antonella and {Paz-Chinch{\'o}n}, Francisco and {Plazas}, Andr{\'e}s A. and {Sanchez}, Eusebio and {Schubnell}, Michael and {Serrano}, Santiago and {Sevilla-Noarbe}, Ignacio and {Smith}, Mathew and {Suchyta}, Eric and {Swanson}, Molly E.~C. and {Tarle}, Gregory and {Tucker}, Brad E. and {Norbert Varga}, Tamas and {Walker}, Alistair R.},
        title = "{Candidate periodically variable quasars from the Dark Energy Survey and the Sloan Digital Sky Survey}",
      journal = {\mnras},
         year = 2020,
        month = dec,
       volume = {499},
       number = {2},
        pages = {2245-2264},
          doi = {10.1093/mnras/staa2957},
archivePrefix = {arXiv},
       eprint = {2008.12329},
 primaryClass = {astro-ph.HE},
       adsurl = {https://ui.adsabs.harvard.edu/abs/2020MNRAS.499.2245C}
}

@ARTICLE{Millon22,
       author = {{Millon}, M. and {Dalang}, C. and {Lemon}, C. and {Sluse}, D. and {Paic}, E. and {Chan}, J.~H.~H. and {Courbin}, F.},
        title = "{Evidence for a milliparsec-separation supermassive binary black hole with quasar microlensing}",
      journal = {\aap},
         year = 2022,
        month = dec,
       volume = {668},
          eid = {A77},
        pages = {A77},
          doi = {10.1051/0004-6361/202244440},
archivePrefix = {arXiv},
       eprint = {2207.00598},
 primaryClass = {astro-ph.GA},
       adsurl = {https://ui.adsabs.harvard.edu/abs/2022A&A...668A..77M}
}

@ARTICLE{Bansal17,
       author = {{Bansal}, K. and {Taylor}, G.~B. and {Peck}, A.~B. and {Zavala}, R.~T. and {Romani}, R.~W.},
        title = "{Constraining the Orbit of the Supermassive Black Hole Binary 0402+379}",
      journal = {\apj},
         year = 2017,
        month = jul,
       volume = {843},
       number = {1},
          eid = {14},
        pages = {14},
          doi = {10.3847/1538-4357/aa74e1},
archivePrefix = {arXiv},
       eprint = {1705.08556},
 primaryClass = {astro-ph.GA},
       adsurl = {https://ui.adsabs.harvard.edu/abs/2017ApJ...843...14B}
}

@ARTICLE{Kharb17,
       author = {{Kharb}, P. and {Lal}, D.~V. and {Merritt}, D.},
        title = "{A candidate sub-parsec binary black hole in the Seyfert galaxy NGC 7674}",
      journal = {Nature Astronomy},
         year = 2017,
        month = sep,
       volume = {1},
        pages = {727-733},
          doi = {10.1038/s41550-017-0256-4},
archivePrefix = {arXiv},
       eprint = {1709.06258},
 primaryClass = {astro-ph.GA},
       adsurl = {https://ui.adsabs.harvard.edu/abs/2017NatAs...1..727K}
}

@ARTICLE{Agazie23,
       author = {{Agazie}, Gabriella and {Alam}, Md Faisal and {Anumarlapudi}, Akash and {Archibald}, Anne M. and {Arzoumanian}, Zaven and {Baker}, Paul T. and {Blecha}, Laura and {Bonidie}, Victoria and {Brazier}, Adam and {Brook}, Paul R. and {Burke-Spolaor}, Sarah and {B{\'e}csy}, Bence and {Chapman}, Christopher and {Charisi}, Maria and {Chatterjee}, Shami and {Cohen}, Tyler and {Cordes}, James M. and {Cornish}, Neil J. and {Crawford}, Fronefield and {Cromartie}, H. Thankful and {Crowter}, Kathryn and {Decesar}, Megan E. and {Demorest}, Paul B. and {Dolch}, Timothy and {Drachler}, Brendan and {Ferrara}, Elizabeth C. and {Fiore}, William and {Fonseca}, Emmanuel and {Freedman}, Gabriel E. and {Garver-Daniels}, Nate and {Gentile}, Peter A. and {Glaser}, Joseph and {Good}, Deborah C. and {G{\"u}ltekin}, Kayhan and {Hazboun}, Jeffrey S. and {Jennings}, Ross J. and {Jessup}, Cody and {Johnson}, Aaron D. and {Jones}, Megan L. and {Kaiser}, Andrew R. and {Kaplan}, David L. and {Kelley}, Luke Zoltan and {Kerr}, Matthew and {Key}, Joey S. and {Kuske}, Anastasia and {Laal}, Nima and {Lam}, Michael T. and {Lamb}, William G. and {Lazio}, T. Joseph W. and {Lewandowska}, Natalia and {Lin}, Ye and {Liu}, Tingting and {Lorimer}, Duncan R. and {Luo}, Jing and {Lynch}, Ryan S. and {Ma}, Chung-Pei and {Madison}, Dustin R. and {Maraccini}, Kaleb and {McEwen}, Alexander and {McKee}, James W. and {McLaughlin}, Maura A. and {McMann}, Natasha and {Meyers}, Bradley W. and {Mingarelli}, Chiara M.~F. and {Mitridate}, Andrea and {Ng}, Cherry and {Nice}, David J. and {Ocker}, Stella Koch and {Olum}, Ken D. and {Panciu}, Elisa and {Pennucci}, Timothy T. and {Perera}, Benetge B.~P. and {Pol}, Nihan S. and {Radovan}, Henri A. and {Ransom}, Scott M. and {Ray}, Paul S. and {Romano}, Joseph D. and {Salo}, Laura and {Sardesai}, Shashwat C. and {Schmiedekamp}, Carl and {Schmiedekamp}, Ann and {Schmitz}, Kai and {Shapiro-Albert}, Brent J. and {Siemens}, Xavier and {Simon}, Joseph and {Siwek}, Magdalena S. and {Stairs}, Ingrid H. and {Stinebring}, Daniel R. and {Stovall}, Kevin and {Susobhanan}, Abhimanyu and {Swiggum}, Joseph K. and {Taylor}, Stephen R. and {Turner}, Jacob E. and {Unal}, Caner and {Vallisneri}, Michele and {Vigeland}, Sarah J. and {Wahl}, Haley M. and {Wang}, Qiaohong and {Witt}, Caitlin A. and {Young}, Olivia and {Nanograv Collaboration}},
        title = "{The NANOGrav 15 yr Data Set: Observations and Timing of 68 Millisecond Pulsars}",
      journal = {\apjl},
         year = 2023,
        month = jul,
       volume = {951},
       number = {1},
          eid = {L9},
        pages = {L9},
          doi = {10.3847/2041-8213/acda9a},
archivePrefix = {arXiv},
       eprint = {2306.16217},
 primaryClass = {astro-ph.HE},
       adsurl = {https://ui.adsabs.harvard.edu/abs/2023ApJ...951L...9A}
}

@ARTICLE{Agazie23j,
       author = {{Agazie}, Gabriella and {Anumarlapudi}, Akash and {Archibald}, Anne M. and {Arzoumanian}, Zaven and {Baker}, Paul T. and {B{\'e}csy}, Bence and {Blecha}, Laura and {Brazier}, Adam and {Brook}, Paul R. and {Burke-Spolaor}, Sarah and {Burnette}, Rand and {Case}, Robin and {Charisi}, Maria and {Chatterjee}, Shami and {Chatziioannou}, Katerina and {Cheeseboro}, Belinda D. and {Chen}, Siyuan and {Cohen}, Tyler and {Cordes}, James M. and {Cornish}, Neil J. and {Crawford}, Fronefield and {Cromartie}, H. Thankful and {Crowter}, Kathryn and {Cutler}, Curt J. and {Decesar}, Megan E. and {Degan}, Dallas and {Demorest}, Paul B. and {Deng}, Heling and {Dolch}, Timothy and {Drachler}, Brendan and {Ellis}, Justin A. and {Ferrara}, Elizabeth C. and {Fiore}, William and {Fonseca}, Emmanuel and {Freedman}, Gabriel E. and {Garver-Daniels}, Nate and {Gentile}, Peter A. and {Gersbach}, Kyle A. and {Glaser}, Joseph and {Good}, Deborah C. and {G{\"u}ltekin}, Kayhan and {Hazboun}, Jeffrey S. and {Hourihane}, Sophie and {Islo}, Kristina and {Jennings}, Ross J. and {Johnson}, Aaron D. and {Jones}, Megan L. and {Kaiser}, Andrew R. and {Kaplan}, David L. and {Kelley}, Luke Zoltan and {Kerr}, Matthew and {Key}, Joey S. and {Klein}, Tonia C. and {Laal}, Nima and {Lam}, Michael T. and {Lamb}, William G. and {Lazio}, T. Joseph W. and {Lewandowska}, Natalia and {Littenberg}, Tyson B. and {Liu}, Tingting and {Lommen}, Andrea and {Lorimer}, Duncan R. and {Luo}, Jing and {Lynch}, Ryan S. and {Ma}, Chung-Pei and {Madison}, Dustin R. and {Mattson}, Margaret A. and {McEwen}, Alexander and {McKee}, James W. and {McLaughlin}, Maura A. and {McMann}, Natasha and {Meyers}, Bradley W. and {Meyers}, Patrick M. and {Mingarelli}, Chiara M.~F. and {Mitridate}, Andrea and {Natarajan}, Priyamvada and {Ng}, Cherry and {Nice}, David J. and {Ocker}, Stella Koch and {Olum}, Ken D. and {Pennucci}, Timothy T. and {Perera}, Benetge B.~P. and {Petrov}, Polina and {Pol}, Nihan S. and {Radovan}, Henri A. and {Ransom}, Scott M. and {Ray}, Paul S. and {Romano}, Joseph D. and {Sardesai}, Shashwat C. and {Schmiedekamp}, Ann and {Schmiedekamp}, Carl and {Schmitz}, Kai and {Schult}, Levi and {Shapiro-Albert}, Brent J. and {Siemens}, Xavier and {Simon}, Joseph and {Siwek}, Magdalena S. and {Stairs}, Ingrid H. and {Stinebring}, Daniel R. and {Stovall}, Kevin and {Sun}, Jerry P. and {Susobhanan}, Abhimanyu and {Swiggum}, Joseph K. and {Taylor}, Jacob and {Taylor}, Stephen R. and {Turner}, Jacob E. and {Unal}, Caner and {Vallisneri}, Michele and {van Haasteren}, Rutger and {Vigeland}, Sarah J. and {Wahl}, Haley M. and {Wang}, Qiaohong and {Witt}, Caitlin A. and {Young}, Olivia and {Nanograv Collaboration}},
        title = "{The NANOGrav 15 yr Data Set: Evidence for a Gravitational-wave Background}",
      journal = {\apjl},
         year = 2023,
        month = jul,
       volume = {951},
       number = {1},
          eid = {L8},
        pages = {L8},
          doi = {10.3847/2041-8213/acdac6},
archivePrefix = {arXiv},
       eprint = {2306.16213},
 primaryClass = {astro-ph.HE},
       adsurl = {https://ui.adsabs.harvard.edu/abs/2023ApJ...951L...8A}
}

@ARTICLE{Chen22df,
       author = {{Chen}, Nianyi and {Ni}, Yueying and {Tremmel}, Michael and {Di Matteo}, Tiziana and {Bird}, Simeon and {DeGraf}, Colin and {Feng}, Yu},
        title = "{Dynamical friction modelling of massive black holes in cosmological simulations and effects on merger rate predictions}",
      journal = {\mnras},
         year = 2022,
        month = feb,
       volume = {510},
       number = {1},
        pages = {531-550},
          doi = {10.1093/mnras/stab3411},
archivePrefix = {arXiv},
       eprint = {2104.00021},
 primaryClass = {astro-ph.GA},
       adsurl = {https://ui.adsabs.harvard.edu/abs/2022MNRAS.510..531C}
}

@ARTICLE{Genina24,
       author = {{Genina}, Anna and {Springel}, Volker and {Rantala}, Antti},
        title = "{A calibrated model for N-body dynamical friction acting on supermassive black holes}",
      journal = {\mnras},
         year = 2024,
        month = oct,
       volume = {534},
       number = {1},
        pages = {957-977},
          doi = {10.1093/mnras/stae2144},
archivePrefix = {arXiv},
       eprint = {2405.08870},
 primaryClass = {astro-ph.GA},
       adsurl = {https://ui.adsabs.harvard.edu/abs/2024MNRAS.534..957G}
}

@ARTICLE{Liao24,
       author = {{Liao}, Shihong and {Irodotou}, Dimitrios and {Johansson}, Peter H. and {Naab}, Thorsten and {Rizzuto}, Francesco Paolo and {Hislop}, Jessica M. and {Wright}, Ruby J. and {Rawlings}, Alexander},
        title = "{RABBITS - II. The impact of AGN feedback on coalescing supermassive black holes in disc and elliptical galaxy mergers}",
      journal = {\mnras},
         year = 2024,
        month = jun,
       volume = {530},
       number = {4},
        pages = {4058-4081},
          doi = {10.1093/mnras/stae1123},
archivePrefix = {arXiv},
       eprint = {2311.01493},
 primaryClass = {astro-ph.GA},
       adsurl = {https://ui.adsabs.harvard.edu/abs/2024MNRAS.530.4058L}
}

@ARTICLE{Pfister19,
       author = {{Pfister}, Hugo and {Volonteri}, Marta and {Dubois}, Yohan and {Dotti}, Massimo and {Colpi}, Monica},
        title = "{The erratic dynamical life of black hole seeds in high-redshift galaxies}",
      journal = {\mnras},
         year = 2019,
        month = jun,
       volume = {486},
       number = {1},
        pages = {101-111},
          doi = {10.1093/mnras/stz822},
archivePrefix = {arXiv},
       eprint = {1902.01297},
 primaryClass = {astro-ph.GA},
       adsurl = {https://ui.adsabs.harvard.edu/abs/2019MNRAS.486..101P}
}

@ARTICLE{Chapon13,
       author = {{Chapon}, Damien and {Mayer}, Lucio and {Teyssier}, Romain},
        title = "{Hydrodynamics of galaxy mergers with supermassive black holes: is there a last parsec problem?}",
      journal = {\mnras},
         year = 2013,
        month = mar,
       volume = {429},
       number = {4},
        pages = {3114-3122},
          doi = {10.1093/mnras/sts568},
archivePrefix = {arXiv},
       eprint = {1110.6086},
 primaryClass = {astro-ph.GA},
       adsurl = {https://ui.adsabs.harvard.edu/abs/2013MNRAS.429.3114C}
}

@ARTICLE{Tremmel15,
       author = {{Tremmel}, M. and {Governato}, F. and {Volonteri}, M. and {Quinn}, T.~R.},
        title = "{Off the beaten path: a new approach to realistically model the orbital decay of supermassive black holes in galaxy formation simulations}",
      journal = {\mnras},
         year = 2015,
        month = aug,
       volume = {451},
       number = {2},
        pages = {1868-1874},
          doi = {10.1093/mnras/stv1060},
archivePrefix = {arXiv},
       eprint = {1501.07609},
 primaryClass = {astro-ph.GA},
       adsurl = {https://ui.adsabs.harvard.edu/abs/2015MNRAS.451.1868T}
}

@ARTICLE{Quinlan96,
       author = {{Quinlan}, Gerald D.},
        title = "{The dynamical evolution of massive black hole binaries I. Hardening in a fixed stellar background}",
      journal = {\na},
         year = 1996,
        month = jul,
       volume = {1},
       number = {1},
        pages = {35-56},
          doi = {10.1016/S1384-1076(96)00003-6},
archivePrefix = {arXiv},
       eprint = {astro-ph/9601092},
 primaryClass = {astro-ph},
       adsurl = {https://ui.adsabs.harvard.edu/abs/1996NewA....1...35Q}
}

@ARTICLE{Milosavljevic01,
       author = {{Milosavljevi{\'c}}, Milo{\v{s}} and {Merritt}, David},
        title = "{Formation of Galactic Nuclei}",
      journal = {\apj},
         year = 2001,
        month = dec,
       volume = {563},
       number = {1},
        pages = {34-62},
          doi = {10.1086/323830},
archivePrefix = {arXiv},
       eprint = {astro-ph/0103350},
 primaryClass = {astro-ph},
       adsurl = {https://ui.adsabs.harvard.edu/abs/2001ApJ...563...34M}
}

@ARTICLE{Milosavljevic03,
       author = {{Milosavljevi{\'c}}, Milo{\v{s}} and {Merritt}, David},
        title = "{Long-Term Evolution of Massive Black Hole Binaries}",
      journal = {\apj},
         year = 2003,
        month = oct,
       volume = {596},
       number = {2},
        pages = {860-878},
          doi = {10.1086/378086},
archivePrefix = {arXiv},
       eprint = {astro-ph/0212459},
 primaryClass = {astro-ph},
       adsurl = {https://ui.adsabs.harvard.edu/abs/2003ApJ...596..860M}
}

@ARTICLE{Sesana06,
       author = {{Sesana}, Alberto and {Haardt}, Francesco and {Madau}, Piero},
        title = "{Interaction of Massive Black Hole Binaries with Their Stellar Environment. I. Ejection of Hypervelocity Stars}",
      journal = {\apj},
         year = 2006,
        month = nov,
       volume = {651},
       number = {1},
        pages = {392-400},
          doi = {10.1086/507596},
archivePrefix = {arXiv},
       eprint = {astro-ph/0604299},
 primaryClass = {astro-ph},
       adsurl = {https://ui.adsabs.harvard.edu/abs/2006ApJ...651..392S}
}

@ARTICLE{Rantala17,
       author = {{Rantala}, Antti and {Pihajoki}, Pauli and {Johansson}, Peter H. and {Naab}, Thorsten and {Lah{\'e}n}, Natalia and {Sawala}, Till},
        title = "{Post-Newtonian Dynamical Modeling of Supermassive Black Holes in Galactic-scale Simulations}",
      journal = {\apj},
         year = 2017,
        month = may,
       volume = {840},
       number = {1},
          eid = {53},
        pages = {53},
          doi = {10.3847/1538-4357/aa6d65},
archivePrefix = {arXiv},
       eprint = {1611.07028},
 primaryClass = {astro-ph.GA},
       adsurl = {https://ui.adsabs.harvard.edu/abs/2017ApJ...840...53R}
}

@ARTICLE{Hoffman07,
       author = {{Hoffman}, Loren and {Loeb}, Abraham},
        title = "{Dynamics of triple black hole systems in hierarchically merging massive galaxies}",
      journal = {\mnras},
         year = 2007,
        month = may,
       volume = {377},
       number = {3},
        pages = {957-976},
          doi = {10.1111/j.1365-2966.2007.11694.x},
archivePrefix = {arXiv},
       eprint = {astro-ph/0612517},
 primaryClass = {astro-ph},
       adsurl = {https://ui.adsabs.harvard.edu/abs/2007MNRAS.377..957H}
}

@ARTICLE{Mannerkoski21,
       author = {{Mannerkoski}, Matias and {Johansson}, Peter H. and {Rantala}, Antti and {Naab}, Thorsten and {Liao}, Shihong},
        title = "{Resolving the Complex Evolution of a Supermassive Black Hole Triplet in a Cosmological Simulation}",
      journal = {\apjl},
         year = 2021,
        month = may,
       volume = {912},
       number = {2},
          eid = {L20},
        pages = {L20},
          doi = {10.3847/2041-8213/abf9a5},
archivePrefix = {arXiv},
       eprint = {2103.16254},
 primaryClass = {astro-ph.GA},
       adsurl = {https://ui.adsabs.harvard.edu/abs/2021ApJ...912L..20M}
}

@ARTICLE{Bonetti18,
       author = {{Bonetti}, Matteo and {Haardt}, Francesco and {Sesana}, Alberto and {Barausse}, Enrico},
        title = "{Post-Newtonian evolution of massive black hole triplets in galactic nuclei - II. Survey of the parameter space}",
      journal = {\mnras},
         year = 2018,
        month = jul,
       volume = {477},
       number = {3},
        pages = {3910-3926},
          doi = {10.1093/mnras/sty896},
archivePrefix = {arXiv},
       eprint = {1709.06088},
 primaryClass = {astro-ph.GA},
       adsurl = {https://ui.adsabs.harvard.edu/abs/2018MNRAS.477.3910B}
}

@ARTICLE{barnes02,
       author = {{Barnes}, Joshua E.},
        title = "{Formation of gas discs in merging galaxies}",
      journal = {\mnras},
         year = 2002,
        month = jul,
       volume = {333},
       number = {3},
        pages = {481-494},
          doi = {10.1046/j.1365-8711.2002.05335.x},
archivePrefix = {arXiv},
       eprint = {astro-ph/0201250},
 primaryClass = {astro-ph},
       adsurl = {https://ui.adsabs.harvard.edu/abs/2002MNRAS.333..481B}
}

@ARTICLE{CD17,
       author = {{Capelo}, Pedro R. and {Dotti}, Massimo},
        title = "{Shocks and angular momentum flips: a different path to feeding the nuclear regions of merging galaxies}",
      journal = {\mnras},
         year = 2017,
        month = mar,
       volume = {465},
       number = {3},
        pages = {2643-2653},
          doi = {10.1093/mnras/stw2872},
archivePrefix = {arXiv},
       eprint = {1610.08507},
 primaryClass = {astro-ph.GA},
       adsurl = {https://ui.adsabs.harvard.edu/abs/2017MNRAS.465.2643C}
}

@ARTICLE{BB18,
       author = {{Blumenthal}, Kelly A. and {Barnes}, Joshua E.},
        title = "{Go with the Flow: Understanding inflow mechanisms in galaxy collisions}",
      journal = {\mnras},
         year = 2018,
        month = sep,
       volume = {479},
       number = {3},
        pages = {3952-3965},
          doi = {10.1093/mnras/sty1605},
archivePrefix = {arXiv},
       eprint = {1806.05132},
 primaryClass = {astro-ph.GA},
       adsurl = {https://ui.adsabs.harvard.edu/abs/2018MNRAS.479.3952B}
}

@ARTICLE{Cuadra09,
       author = {{Cuadra}, J. and {Armitage}, P.~J. and {Alexander}, R.~D. and {Begelman}, M.~C.},
        title = "{Massive black hole binary mergers within subparsec scale gas discs}",
      journal = {\mnras},
         year = 2009,
        month = mar,
       volume = {393},
       number = {4},
        pages = {1423-1432},
          doi = {10.1111/j.1365-2966.2008.14147.x},
archivePrefix = {arXiv},
       eprint = {0809.0311},
 primaryClass = {astro-ph},
       adsurl = {https://ui.adsabs.harvard.edu/abs/2009MNRAS.393.1423C}
}

@ARTICLE{Roedig12,
       author = {{Roedig}, C. and {Sesana}, A. and {Dotti}, M. and {Cuadra}, J. and {Amaro-Seoane}, P. and {Haardt}, F.},
        title = "{Evolution of binary black holes in self gravitating discs. Dissecting the torques}",
      journal = {\aap},
         year = 2012,
        month = sep,
       volume = {545},
          eid = {A127},
        pages = {A127},
          doi = {10.1051/0004-6361/201219986},
archivePrefix = {arXiv},
       eprint = {1202.6063},
 primaryClass = {astro-ph.CO},
       adsurl = {https://ui.adsabs.harvard.edu/abs/2012A&A...545A.127R}
}

@ARTICLE{Dorazio13,
       author = {{D'Orazio}, Daniel J. and {Haiman}, Zolt{\'a}n and {MacFadyen}, Andrew},
        title = "{Accretion into the central cavity of a circumbinary disc}",
      journal = {\mnras},
         year = 2013,
        month = dec,
       volume = {436},
       number = {4},
        pages = {2997-3020},
          doi = {10.1093/mnras/stt1787},
archivePrefix = {arXiv},
       eprint = {1210.0536},
 primaryClass = {astro-ph.GA},
       adsurl = {https://ui.adsabs.harvard.edu/abs/2013MNRAS.436.2997D}
}

@ARTICLE{Farris14,
       author = {{Farris}, Brian D. and {Duffell}, Paul and {MacFadyen}, Andrew I. and {Haiman}, Zoltan},
        title = "{Binary Black Hole Accretion from a Circumbinary Disk: Gas Dynamics inside the Central Cavity}",
      journal = {\apj},
         year = 2014,
        month = mar,
       volume = {783},
       number = {2},
          eid = {134},
        pages = {134},
          doi = {10.1088/0004-637X/783/2/134},
archivePrefix = {arXiv},
       eprint = {1310.0492},
 primaryClass = {astro-ph.HE},
       adsurl = {https://ui.adsabs.harvard.edu/abs/2014ApJ...783..134F}
}

@ARTICLE{Moody19,
       author = {{Moody}, Mackenzie S.~L. and {Shi}, Ji-Ming and {Stone}, James M.},
        title = "{Hydrodynamic Torques in Circumbinary Accretion Disks}",
      journal = {\apj},
         year = 2019,
        month = apr,
       volume = {875},
       number = {1},
          eid = {66},
        pages = {66},
          doi = {10.3847/1538-4357/ab09ee},
archivePrefix = {arXiv},
       eprint = {1903.00008},
 primaryClass = {astro-ph.HE},
       adsurl = {https://ui.adsabs.harvard.edu/abs/2019ApJ...875...66M}
}

@ARTICLE{Munoz19,
       author = {{Mu{\~n}oz}, Diego J. and {Miranda}, Ryan and {Lai}, Dong},
        title = "{Hydrodynamics of Circumbinary Accretion: Angular Momentum Transfer and Binary Orbital Evolution}",
      journal = {\apj},
         year = 2019,
        month = jan,
       volume = {871},
       number = {1},
          eid = {84},
        pages = {84},
          doi = {10.3847/1538-4357/aaf867},
archivePrefix = {arXiv},
       eprint = {1810.04676},
 primaryClass = {astro-ph.HE},
       adsurl = {https://ui.adsabs.harvard.edu/abs/2019ApJ...871...84M}
}

@ARTICLE{Franchini21,
       author = {{Franchini}, Alessia and {Sesana}, Alberto and {Dotti}, Massimo},
        title = "{Circumbinary disc self-gravity governing supermassive black hole binary mergers}",
      journal = {\mnras},
         year = 2021,
        month = oct,
       volume = {507},
       number = {1},
        pages = {1458-1467},
          doi = {10.1093/mnras/stab2234},
archivePrefix = {arXiv},
       eprint = {2106.13253},
 primaryClass = {astro-ph.HE},
       adsurl = {https://ui.adsabs.harvard.edu/abs/2021MNRAS.507.1458F}
}

@ARTICLE{Bourne24,
       author = {{Bourne}, Martin A. and {Fiacconi}, Davide and {Sijacki}, Debora and {Piotrowska}, Joanna M. and {Koudmani}, Sophie},
        title = "{Dynamics and spin alignment in massive, gravito-turbulent circumbinary discs around supermassive black hole binaries}",
      journal = {\mnras},
         year = 2024,
        month = nov,
       volume = {534},
       number = {4},
        pages = {3448-3477},
          doi = {10.1093/mnras/stae2143},
archivePrefix = {arXiv},
       eprint = {2311.17144},
 primaryClass = {astro-ph.HE},
       adsurl = {https://ui.adsabs.harvard.edu/abs/2024MNRAS.534.3448B}
}

@ARTICLE{Vecchio04,
       author = {{Vecchio}, Alberto},
        title = "{LISA observations of rapidly spinning massive black hole binary systems}",
      journal = {\prd},
         year = 2004,
        month = aug,
       volume = {70},
       number = {4},
          eid = {042001},
        pages = {042001},
          doi = {10.1103/PhysRevD.70.042001},
archivePrefix = {arXiv},
       eprint = {astro-ph/0304051},
 primaryClass = {astro-ph},
       adsurl = {https://ui.adsabs.harvard.edu/abs/2004PhRvD..70d2001V}
}

@ARTICLE{Klein09,
       author = {{Klein}, Antoine and {Jetzer}, Philippe and {Sereno}, Mauro},
        title = "{Parameter estimation for coalescing massive binary black holes with LISA using the full 2-post-Newtonian gravitational waveform and spin-orbit precession}",
      journal = {\prd},
         year = 2009,
        month = sep,
       volume = {80},
       number = {6},
          eid = {064027},
        pages = {064027},
          doi = {10.1103/PhysRevD.80.064027},
archivePrefix = {arXiv},
       eprint = {0907.3318},
 primaryClass = {astro-ph.CO},
       adsurl = {https://ui.adsabs.harvard.edu/abs/2009PhRvD..80f4027K}
}

@ARTICLE{Oshaughnessy13,
       author = {{O'Shaughnessy}, R. and {London}, L. and {Healy}, J. and {Shoemaker}, D.},
        title = "{Precession during merger: Strong polarization changes are observationally accessible features of strong-field gravity during binary black hole merger}",
      journal = {\prd},
         year = 2013,
        month = feb,
       volume = {87},
       number = {4},
          eid = {044038},
        pages = {044038},
          doi = {10.1103/PhysRevD.87.044038},
archivePrefix = {arXiv},
       eprint = {1209.3712},
 primaryClass = {gr-qc},
       adsurl = {https://ui.adsabs.harvard.edu/abs/2013PhRvD..87d4038O}
}

@ARTICLE{Pratten23,
       author = {{Pratten}, Geraint and {Schmidt}, Patricia and {Middleton}, Hannah and {Vecchio}, Alberto},
        title = "{Precision tracking of massive black hole spin evolution with LISA}",
      journal = {\prd},
         year = 2023,
        month = dec,
       volume = {108},
       number = {12},
          eid = {124045},
        pages = {124045},
          doi = {10.1103/PhysRevD.108.124045},
archivePrefix = {arXiv},
       eprint = {2307.13026},
 primaryClass = {gr-qc},
       adsurl = {https://ui.adsabs.harvard.edu/abs/2023PhRvD.108l4045P}
}

@ARTICLE{Campanelli07,
       author = {{Campanelli}, Manuela and {Lousto}, Carlos O. and {Zlochower}, Yosef and {Merritt}, David},
        title = "{Maximum Gravitational Recoil}",
      journal = {\prl},
         year = 2007,
        month = jun,
       volume = {98},
       number = {23},
          eid = {231102},
        pages = {231102},
          doi = {10.1103/PhysRevLett.98.231102},
archivePrefix = {arXiv},
       eprint = {gr-qc/0702133},
 primaryClass = {gr-qc},
       adsurl = {https://ui.adsabs.harvard.edu/abs/2007PhRvL..98w1102C}
}

@ARTICLE{Gonzalez13,
       author = {{Gonz{\'a}lez}, Jos{\'e} A. and {Sperhake}, Ulrich and {Br{\"u}gmann}, Bernd and {Hannam}, Mark and {Husa}, Sascha},
        title = "{Maximum Kick from Nonspinning Black-Hole Binary Inspiral}",
      journal = {\prl},
         year = 2007,
        month = mar,
       volume = {98},
       number = {9},
          eid = {091101},
        pages = {091101},
          doi = {10.1103/PhysRevLett.98.091101},
archivePrefix = {arXiv},
       eprint = {gr-qc/0610154},
 primaryClass = {gr-qc},
       adsurl = {https://ui.adsabs.harvard.edu/abs/2007PhRvL..98i1101G}
}

@ARTICLE{Lousto13,
       author = {{Lousto}, Carlos O. and {Zlochower}, Yosef},
        title = "{Nonlinear gravitational recoil from the mergers of precessing black-hole binaries}",
      journal = {\prd},
         year = 2013,
        month = apr,
       volume = {87},
       number = {8},
          eid = {084027},
        pages = {084027},
          doi = {10.1103/PhysRevD.87.084027},
archivePrefix = {arXiv},
       eprint = {1211.7099},
 primaryClass = {gr-qc},
       adsurl = {https://ui.adsabs.harvard.edu/abs/2013PhRvD..87h4027L}
}

@ARTICLE{Lousto19,
       author = {{Lousto}, Carlos O. and {Healy}, James},
        title = "{Kicking gravitational wave detectors with recoiling black holes}",
      journal = {\prd},
         year = 2019,
        month = nov,
       volume = {100},
       number = {10},
          eid = {104039},
        pages = {104039},
          doi = {10.1103/PhysRevD.100.104039},
archivePrefix = {arXiv},
       eprint = {1908.04382},
 primaryClass = {gr-qc},
       adsurl = {https://ui.adsabs.harvard.edu/abs/2019PhRvD.100j4039L}
}

@ARTICLE{Sperhake,
       author = {{Sperhake}, Ulrich and {Rosca-Mead}, Roxana and {Gerosa}, Davide and {Berti}, Emanuele},
        title = "{Amplification of superkicks in black-hole binaries through orbital eccentricity}",
      journal = {\prd},
         year = 2020,
        month = jan,
       volume = {101},
       number = {2},
          eid = {024044},
        pages = {024044},
          doi = {10.1103/PhysRevD.101.024044},
archivePrefix = {arXiv},
       eprint = {1910.01598},
 primaryClass = {gr-qc},
       adsurl = {https://ui.adsabs.harvard.edu/abs/2020PhRvD.101b4044S}
}

@ARTICLE{Schnittman07,
       author = {{Schnittman}, Jeremy D.},
        title = "{Retaining Black Holes with Very Large Recoil Velocities}",
      journal = {\apjl},
         year = 2007,
        month = oct,
       volume = {667},
       number = {2},
        pages = {L133-L136},
          doi = {10.1086/522203},
archivePrefix = {arXiv},
       eprint = {0706.1548},
 primaryClass = {astro-ph},
       adsurl = {https://ui.adsabs.harvard.edu/abs/2007ApJ...667L.133S}
}

@ARTICLE{Volonteri10,
       author = {{Volonteri}, Marta and {G{\"u}ltekin}, Kayhan and {Dotti}, Massimo},
        title = "{Gravitational recoil: effects on massive black hole occupation fraction over cosmic time}",
      journal = {\mnras},
         year = 2010,
        month = jun,
       volume = {404},
       number = {4},
        pages = {2143-2150},
          doi = {10.1111/j.1365-2966.2010.16431.x},
archivePrefix = {arXiv},
       eprint = {1001.1743},
 primaryClass = {astro-ph.CO},
       adsurl = {https://ui.adsabs.harvard.edu/abs/2010MNRAS.404.2143V}
}

@ARTICLE{GerosaSesana15,
       author = {{Gerosa}, Davide and {Sesana}, Alberto},
        title = "{Missing black holes in brightest cluster galaxies as evidence for the occurrence of superkicks in nature}",
      journal = {\mnras},
         year = 2015,
        month = jan,
       volume = {446},
       number = {1},
        pages = {38-55},
          doi = {10.1093/mnras/stu2049},
archivePrefix = {arXiv},
       eprint = {1405.2072},
 primaryClass = {astro-ph.GA},
       adsurl = {https://ui.adsabs.harvard.edu/abs/2015MNRAS.446...38G}
}

@ARTICLE{Sesana07,
       author = {{Sesana}, Alberto and {Volonteri}, Marta and {Haardt}, Francesco},
        title = "{The imprint of massive black hole formation models on the LISA data stream}",
      journal = {\mnras},
         year = 2007,
        month = jun,
       volume = {377},
       number = {4},
        pages = {1711-1716},
          doi = {10.1111/j.1365-2966.2007.11734.x},
archivePrefix = {arXiv},
       eprint = {astro-ph/0701556},
 primaryClass = {astro-ph},
       adsurl = {https://ui.adsabs.harvard.edu/abs/2007MNRAS.377.1711S}
}

@ARTICLE{Blecha08,
       author = {{Blecha}, Laura and {Loeb}, Abraham},
        title = "{Effects of gravitational-wave recoil on the dynamics and growth of supermassive black holes}",
      journal = {\mnras},
         year = 2008,
        month = nov,
       volume = {390},
       number = {4},
        pages = {1311-1325},
          doi = {10.1111/j.1365-2966.2008.13790.x},
archivePrefix = {arXiv},
       eprint = {0805.1420},
 primaryClass = {astro-ph},
       adsurl = {https://ui.adsabs.harvard.edu/abs/2008MNRAS.390.1311B}
}

@ARTICLE{Bogdanovic07,
       author = {{Bogdanovi{\'c}}, Tamara and {Reynolds}, Christopher S. and {Miller}, M. Coleman},
        title = "{Alignment of the Spins of Supermassive Black Holes Prior to Coalescence}",
      journal = {\apjl},
         year = 2007,
        month = jun,
       volume = {661},
       number = {2},
        pages = {L147-L150},
          doi = {10.1086/518769},
archivePrefix = {arXiv},
       eprint = {astro-ph/0703054},
 primaryClass = {astro-ph},
       adsurl = {https://ui.adsabs.harvard.edu/abs/2007ApJ...661L.147B}
}

@ARTICLE{Miller13,
       author = {{Miller}, M. Coleman and {Krolik}, Julian H.},
        title = "{Alignment of Supermassive Black Hole Binary Orbits and Spins}",
      journal = {\apj},
         year = 2013,
        month = sep,
       volume = {774},
       number = {1},
          eid = {43},
        pages = {43},
          doi = {10.1088/0004-637X/774/1/43},
archivePrefix = {arXiv},
       eprint = {1307.6569},
 primaryClass = {astro-ph.HE},
       adsurl = {https://ui.adsabs.harvard.edu/abs/2013ApJ...774...43M}
}

@ARTICLE{Lodato13,
       author = {{Lodato}, G. and {Gerosa}, D.},
        title = "{Black hole mergers: do gas discs lead to spin alignment?}",
      journal = {\mnras},
         year = 2013,
        month = feb,
       volume = {429},
        pages = {L30-L34},
          doi = {10.1093/mnrasl/sls018},
archivePrefix = {arXiv},
       eprint = {1211.0284},
 primaryClass = {astro-ph.CO},
       adsurl = {https://ui.adsabs.harvard.edu/abs/2013MNRAS.429L..30L}
}

@ARTICLE{Dotti10,
       author = {{Dotti}, M. and {Volonteri}, M. and {Perego}, A. and {Colpi}, M. and {Ruszkowski}, M. and {Haardt}, F.},
        title = "{Dual black holes in merger remnants - II. Spin evolution and gravitational recoil}",
      journal = {\mnras},
         year = 2010,
        month = feb,
       volume = {402},
       number = {1},
        pages = {682-690},
          doi = {10.1111/j.1365-2966.2009.15922.x},
archivePrefix = {arXiv},
       eprint = {0910.5729},
 primaryClass = {astro-ph.HE},
       adsurl = {https://ui.adsabs.harvard.edu/abs/2010MNRAS.402..682D}
}

@ARTICLE{Gerosa20,
       author = {{Gerosa}, Davide and {Rosotti}, Giovanni and {Barbieri}, Riccardo},
        title = "{The Bardeen-Petterson effect in accreting supermassive black hole binaries: a systematic approach}",
      journal = {\mnras},
         year = 2020,
        month = aug,
       volume = {496},
       number = {3},
        pages = {3060-3075},
          doi = {10.1093/mnras/staa1693},
archivePrefix = {arXiv},
       eprint = {2004.02894},
 primaryClass = {astro-ph.GA},
       adsurl = {https://ui.adsabs.harvard.edu/abs/2020MNRAS.496.3060G}
}

@ARTICLE{Steinle23,
       author = {{Steinle}, Nathan and {Gerosa}, Davide},
        title = "{The Bardeen-Petterson effect, disc breaking, and the spin orientations of supermassive black hole binaries}",
      journal = {\mnras},
         year = 2023,
        month = mar,
       volume = {519},
       number = {4},
        pages = {5031-5042},
          doi = {10.1093/mnras/stac3821},
archivePrefix = {arXiv},
       eprint = {2211.00044},
 primaryClass = {astro-ph.HE},
       adsurl = {https://ui.adsabs.harvard.edu/abs/2023MNRAS.519.5031S}
}

@ARTICLE{Nixon12,
       author = {{Nixon}, Christopher J. and {King}, Andrew R.},
        title = "{Broken discs: warp propagation in accretion discs}",
      journal = {\mnras},
         year = 2012,
        month = apr,
       volume = {421},
       number = {2},
        pages = {1201-1208},
          doi = {10.1111/j.1365-2966.2011.20377.x},
archivePrefix = {arXiv},
       eprint = {1201.1297},
 primaryClass = {astro-ph.HE},
       adsurl = {https://ui.adsabs.harvard.edu/abs/2012MNRAS.421.1201N}
}

@ARTICLE{Nealon22,
       author = {{Nealon}, Rebecca and {Ragusa}, Enrico and {Gerosa}, Davide and {Rosotti}, Giovanni and {Barbieri}, Riccardo},
        title = "{The Bardeen-Petterson effect in accreting supermassive black hole binaries: disc breaking and critical obliquity}",
      journal = {\mnras},
         year = 2022,
        month = feb,
       volume = {509},
       number = {4},
        pages = {5608-5621},
          doi = {10.1093/mnras/stab3328},
archivePrefix = {arXiv},
       eprint = {2111.08065},
 primaryClass = {astro-ph.HE},
       adsurl = {https://ui.adsabs.harvard.edu/abs/2022MNRAS.509.5608N}
}



\appendix

\section{Convergence} \label{App: resolutions}

In this section we examine the dependence of our main results on numerical resolution. To this end, we performed additional simulations for the {\labsim{ref}} and {\labsim{a5pi12}} setups at three ({\labsim{x3}}) and ten ({\labsim{x10}}) times higher resolution (corresponding to 1500000 and 5000000 particles in the CBD, respectively). For the {\labsim{x3}} runs we adopted $N_\textrm{spawn, min} = 180$, $\epsilon_\textrm{gas} = 0.035$ pc, and $\epsilon_\bullet = 0.025$ pc, while for the {\labsim{x10}} runs we used $N_\textrm{spawn, min} = 600$, $\epsilon_\textrm{gas} = 0.01$ pc, and $\epsilon_\bullet = 0.0075$ pc. All other parameters were kept unchanged.

\begin{figure}
\centering
\includegraphics[width=1\linewidth]{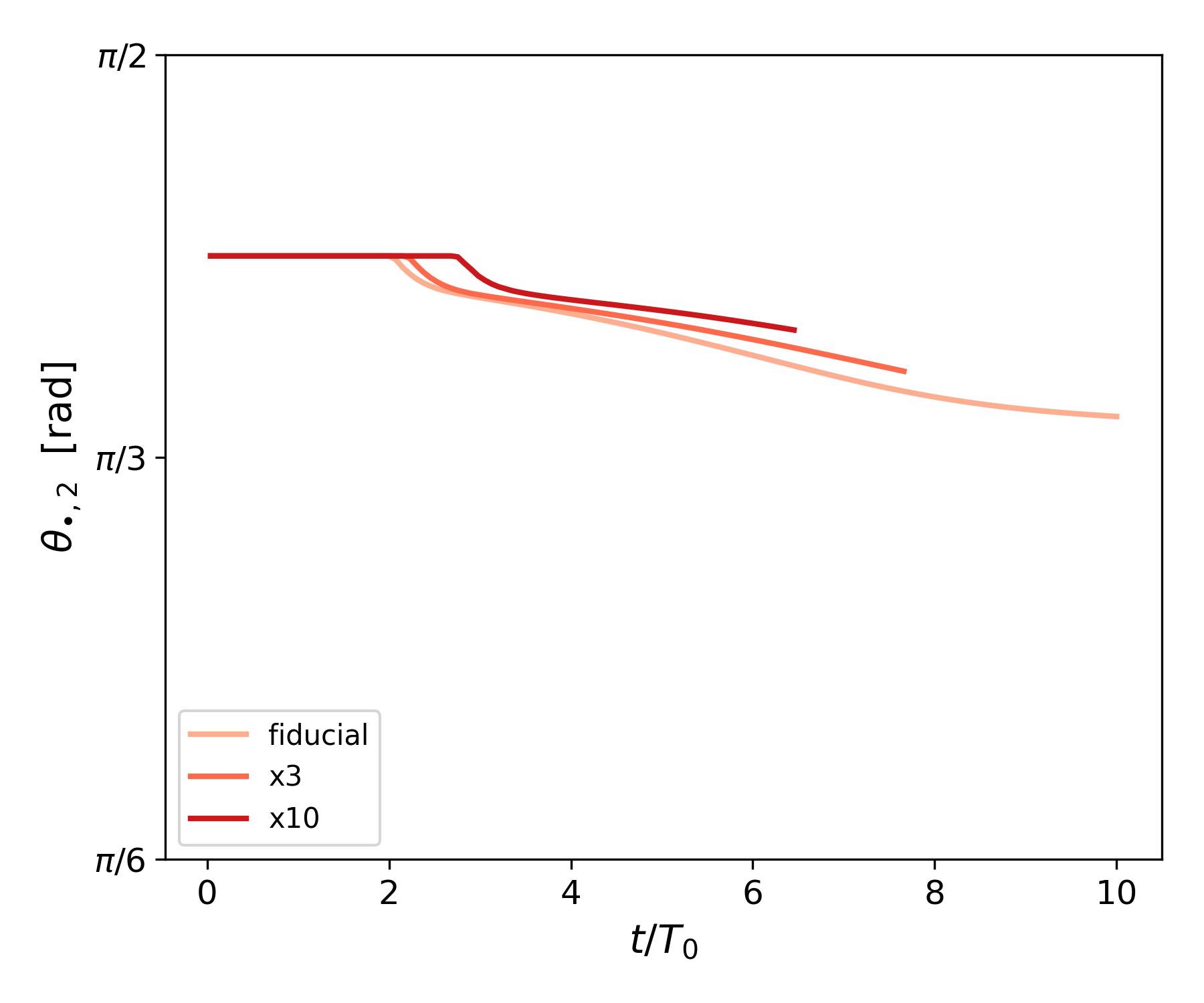}
\caption{Evolution of the angle between the secondary BH spin and the $z$-axis for the simulations {\labsim{a5pi12}}, {\labsim{a5pi12\_x3}}, and {\labsim{a5pi12\_x10}}. Time is shown in units of the initial orbital period.}
\label{fig: hres spin}
\end{figure}

Figure~\ref{fig: hres spin} shows the evolution of the secondary BH spin (corresponding to Fig.~\ref{fig: spin}, right panel) for the {\labsim{a5pi12}}, {\labsim{a5pi12\_x3}}, and {\labsim{a5pi12\_x10}} runs. Despite the limited time interval covered by the higher-resolution simulations, the spin evolution closely follows that of the fiducial resolution run (note that the $y$-axis range is reduced compared to Fig.~\ref{fig: spin}). Since the spin alignment discussed in Section~\ref{sec: spin} is driven primarily by the accretion rate onto the BH, this result indicates that both the resolved inflow and the sub-grid accretion are well converged. This supports our main conclusion that spin alignment is significantly suppressed in the presence of AGN feedback.

\begin{figure}
    \centering
    \includegraphics[width=1\linewidth]{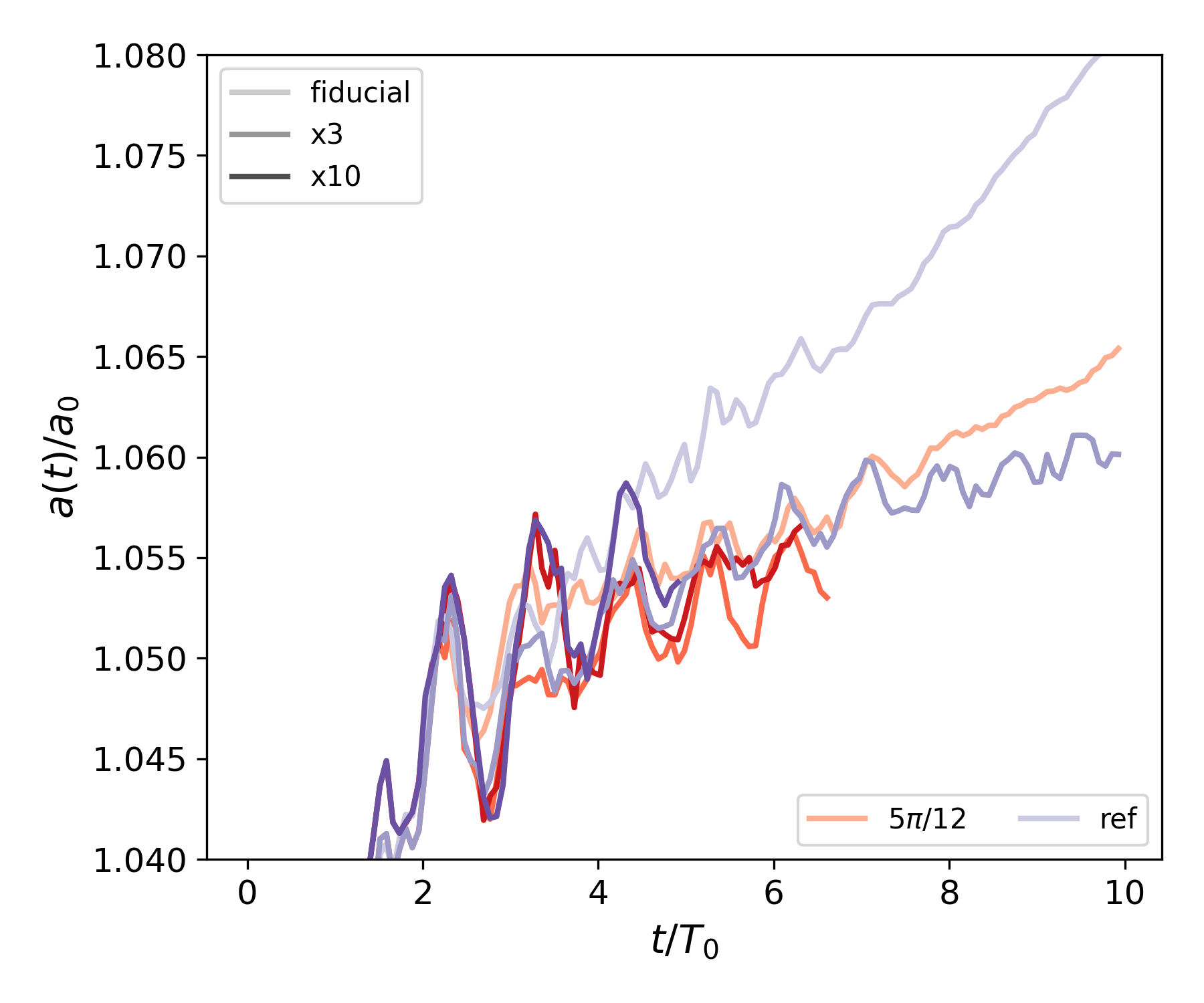}
    \caption{Evolution of the binary semi-major axis, normalised to its initial value, for the {\labsim{ref}} (blue shades) and {\labsim{a5pi12}} (red shades) simulations and their higher-resolution counterparts. Time is given in units of the initial binary orbital period. Darker shades correspond to higher resolution. }
    \label{fig: hres orbit}
\end{figure}

Figure~\ref{fig: hres orbit} shows instead the evolution of the binary semi-major axis, normalised to its initial value (analogous to Fig.~\ref{fig: Orbital}), for the {\labsim{ref}}, {\labsim{a5pi12}} simulations, and their higher-resolution counterparts. This figure indicates that, while the orbital evolution in the {\labsim{a5pi12}} setup is well converged, the {\labsim{ref}} runs are not fully converged at these resolutions (with only the lowest-resolution case showing a noticeable deviation). This suggests a dependence on how well the gas distribution in the immediate vicinity of the BHs is resolved, which in turn affects the dominant gravitational torques driving the binary evolution. Higher resolution therefore modifies the detailed evolution of these close interactions. However, despite the lack of full convergence, the largest differences remain modest (at the $\lesssim 5\%$ level).

\section{Torques} \label{App: torques}

In this section, we describe in detail how the torques presented in Section~\ref{sec: accretion} are computed.
The three $z$-components of the torques acting on the binary, namely those due to accretion, $\dot{L}_{\textrm{acc}}$, wind ejection, $\dot{L}_{\textrm{wind}}$, and gravitational interaction with the gas, $\dot{L}_{\textrm{grav}}$, can each be decomposed into the contributions from the two individual BHs, denoted by $\dot{L}_{*\textrm{source}*,\textrm{BH}}$.
Assuming that the binary centre of mass has position $(x_\textrm{cm},y_\textrm{cm},z_\textrm{cm})$ and velocity $(v_{x,\textrm{cm}}, v_{y,\textrm{cm}}, v_{z,\textrm{cm}})$, and that one of the two BHs has mass $M$, position coordinates $(x,y,z)$, velocity $(v_x, v_y, v_z)$, and acceleration $(a_x, a_y, a_z)$, the $z$-component of the accretion torque acting on this BH is given by
\begin{equation}
\begin{aligned}
    \dot{L}_{\textrm{acc},\textrm{BH}}= & \dot{M}_{\textrm{in}}\Bigl[ (x-x_{\textrm{cm}}) (v_{y} -v_{y,\textrm{cm}}) - (y-y_{\textrm{cm}}) (v_{x} -v_{x,\textrm{cm}}) \Bigr] + \\
    &M \Bigl[ \dot{x} (v_{y} -v_{y,\textrm{cm}}) - \dot{y} (v_{x} -v_{x,\textrm{cm}}) \Bigr] + \\
    &M \Bigl[ (x-x_{\textrm{cm}}) \dot{v}_{y} - (y-y_{\textrm{cm}}) \dot{v}_{x} \Bigr],
    \label{Eq: Tacc}
\end{aligned}
\end{equation}
where $\dot{M}_\textrm{in}$ is the resolved accretion rate onto the BH, and
\begin{equation}
    \begin{aligned}
    &\dot{x} = \frac{1}{M}\Biggl[ \frac{\sum_l x_l m_l}{dt} - x \dot{M}_{\textrm{in}} \Biggr], \\
    &\dot{v}_{x} = \frac{1}{M}\Biggl[ \frac{\sum_l v_{x,l} m_l}{dt} - v_{x} \dot{M}_{\textrm{in}} \Biggr],
    \end{aligned}
\end{equation}

where $dt$ is the BH timestep during which particles (indexed by $l$) are captured. The sums extend over all particles accreted by the BH during that timestep, with positions and velocities evaluated at the time of accretion. Analogous expressions hold for $\dot{y}$ and $\dot{v}_y$.

Similarly, for each BH particle, the $z$-component of the torque associated with wind ejection is given by
\begin{equation}
    \dot{L}_{\textrm{wind},\textrm{BH}} = -\dot{M}_{\textrm{w}}\Bigl[ (x-x_{\textrm{cm}}) (v_{y} -v_{y,\textrm{cm}}) - (y-y_{\textrm{cm}}) (v_{x} -v_{x,\textrm{cm}}) \Bigr],
    \label{Eq: Ldotwind}
\end{equation}

where $\dot{M}_{\textrm{w}}$ is the BH mass-loss rate due to wind ejection.

Finally, the gravitational torque is given by

\begin{equation}
    \dot{L}_{\textrm{grav},\textrm{BH}} = M\Bigl[ (x-x_{\textrm{cm}}) a_y - (y-y_{\textrm{cm}}) a_x \Bigr],
\end{equation}
and the total torque acting on the binary is obtained by summing all such contributions from the two BHs.


\label{lastpage}
\end{document}